\documentclass{elsart}
\usepackage{epsfig}
\usepackage{amsmath}
\usepackage{hhline}
\usepackage{amssymb}
\usepackage{times}
\usepackage{cite}
\usepackage{feynmp}
\usepackage{booktabs}

\usepackage{amssymb}

\begin{document}
\newcommand{\mrm}[1]{\ensuremath{\mathrm{#1}}}
\newcommand{\ttt}[1]{\texttt{#1}}
\newcommand{\tsc}[1]{\textsc{#1}}
\newcommand{\comma}{,}
\newcommand{\as}{\alpha_s}
\newcommand{\pom}{{I\!\!P}}
\newcommand{\reg}{{I\!\!R}}
\newcommand{\slowpi}{\pi_{\mathit{slow}}}
\newcommand{\fiidiii}{F_2^{D(3)}}
\newcommand{\fiidiiiarg}{\fiidiii\,(\beta,\,Q^2,\,x)}
\newcommand{\n}{1.19\pm 0.06 (stat.) \pm0.07 (syst.)}
\newcommand{\nz}{1.30\pm 0.08 (stat.)^{+0.08}_{-0.14} (syst.)}
\newcommand{\fiidiiiful}{F_2^{D(4)}\,(\beta,\,Q^2,\,x,\,t)}
\newcommand{\fiipom}{\tilde F_2^D}
\newcommand{\ALPHA}{1.10\pm0.03 (stat.) \pm0.04 (syst.)}
\newcommand{\ALPHAZ}{1.15\pm0.04 (stat.)^{+0.04}_{-0.07} (syst.)}
\newcommand{\fiipomarg}{\fiipom\,(\beta,\,Q^2)}
\newcommand{\pomflux}{f_{\pom / p}}
\newcommand{\nxpom}{1.19\pm 0.06 (stat.) \pm0.07 (syst.)}
\newcommand {\gapprox}
   {\raisebox{-0.7ex}{$\stackrel {\textstyle>}{\sim}$}}
\newcommand {\lapprox}
   {\raisebox{-0.7ex}{$\stackrel {\textstyle<}{\sim}$}}
\def\gsim{\,\lower.25ex\hbox{$\scriptstyle\sim$}\kern-1.30ex%
\raise 0.55ex\hbox{$\scriptstyle >$}\,}
\def\lsim{\,\lower.25ex\hbox{$\scriptstyle\sim$}\kern-1.30ex%
\raise 0.55ex\hbox{$\scriptstyle <$}\,}
\newcommand{\pomfluxarg}{f_{\pom / p}\,(x_\pom)}
\newcommand{\dsf}{\mbox{$F_2^{D(3)}$}}
\newcommand{\dsfva}{\mbox{$F_2^{D(3)}(\beta,Q^2,x_{I\!\!P})$}}
\newcommand{\dsfvb}{\mbox{$F_2^{D(3)}(\beta,Q^2,x)$}}
\newcommand{\dsfpom}{$F_2^{I\!\!P}$}
\newcommand{\gap}{\stackrel{>}{\sim}}
\newcommand{\lap}{\stackrel{<}{\sim}}
\newcommand{\fem}{$F_2^{em}$}
\newcommand{\tsnmp}{$\tilde{\sigma}_{NC}(e^{\mp})$}
\newcommand{\tsnm}{$\tilde{\sigma}_{NC}(e^-)$}
\newcommand{\tsnp}{$\tilde{\sigma}_{NC}(e^+)$}
\newcommand{\st}{$\star$}
\newcommand{\sst}{$\star \star$}
\newcommand{\ssst}{$\star \star \star$}
\newcommand{\sssst}{$\star \star \star \star$}
\newcommand{\tw}{\theta_W}
\newcommand{\sw}{\sin{\theta_W}}
\newcommand{\cw}{\cos{\theta_W}}
\newcommand{\sww}{\sin^2{\theta_W}}
\newcommand{\cww}{\cos^2{\theta_W}}
\newcommand{\trm}{m_{\perp}}
\newcommand{\trp}{p_{\perp}}
\newcommand{\trmm}{m_{\perp}^2}
\newcommand{\trpp}{p_{\perp}^2}
\newcommand{\alp}{\alpha_s}

\newcommand{\alps}{\alpha_s}
\newcommand{\sqrts}{$\sqrt{s}$}
\newcommand{\LO}{$O(\alpha_s^0)$}
\newcommand{\Oa}{$O(\alpha_s)$}
\newcommand{\Oaa}{$O(\alpha_s^2)$}
\newcommand{\PT}{p_{\perp}}
\newcommand{\JPSI}{J/\psi}
\newcommand{\sh}{\hat{s}}
\newcommand{\uh}{\hat{u}}
\newcommand{\MP}{m_{J/\psi}}
\newcommand{\PO}{I\!\!P}
\newcommand{\xbj}{x}
\newcommand{\xpom}{x_{\PO}}
\newcommand{\ttbs}{\char'134}
\newcommand{\xpomlo}{3\times10^{-4}}
\newcommand{\xpomup}{0.05}
\newcommand{\dgr}{^\circ}
\newcommand{\pbarnt}{\,\mbox{{\rm pb$^{-1}$}}}
\newcommand{\gev}{\,\mbox{GeV}}
\newcommand{\WBoson}{\mbox{$W$}}
\newcommand{\fbarn}{\,\mbox{{\rm fb}}}
\newcommand{\fbarnt}{\,\mbox{{\rm fb$^{-1}$}}}
\newcommand{\gevcc}{\ensuremath{{\rm GeV}\!/c^2}}%
%
\newcommand{\qsq}{\ensuremath{Q^2} }
\newcommand{\gevsq}{\ensuremath{\mathrm{GeV}^2} }
\newcommand{\et}{\ensuremath{E_t^*} }
\newcommand{\rap}{\ensuremath{\eta^*} }
\newcommand{\gp}{\ensuremath{\gamma^*}p }
\newcommand{\dsiget}{\ensuremath{{\rm d}\sigma_{ep}/{\rm d}E_t^*} }
\newcommand{\dsigrap}{\ensuremath{{\rm d}\sigma_{ep}/{\rm d}\eta^*} }
\newcommand{\GeV}{\, \mbox{GeV}}
\newcommand{\TeV}{\, \mbox{TeV}}
\newcommand{\keV}{\, \mbox{keV}}
\newcommand{\ti}{\ensuremath{\tilde}}

%
%
\def\Journal#1#2#3#4{{#1} {\bf #2} (#4) #3}
\def\NIM{Nucl. Instrum. Methods}
\def\NIMA{{Nucl. Instrum. Methods} {\bf A}}
\def\NP{{Nucl. Phys.}}
\def\NPB{{Nucl. Phys.}   {\bf B}}
\def\NPPS{Nucl. Phys. Proc. Suppl.}
\def\NPPSC{{Nucl. Phys. Proc. Suppl.} {\bf C}}
\def\PLB{{Phys. Lett.}   {\bf B}}
\def\PRL{Phys. Rev. Lett.}
\def\PR{Phys. Rev.}
\def\PRD{{Phys. Rev.}    {\bf D}}
\def\ZPC{{Z. Phys.}      {\bf C}}
\def\EJC{{Eur. Phys. J.} {\bf C}}
\def\EPC{\EJC}
\def\EPJC{\EJC}
\def\CPC{Comp. Phys. Commun.}
\def\JPG{{J. Phys.} {\bf G}}
\def\EPCD{{Eur. Phys. J.} {\bf C} {Direct}}
\def\JHEP{{JHEP} { }}
\def\PREP{{Phys. Rept.}}
\def\SJNP{{Soviet J. Nucl. Phys.}}
\def\JETP{{Soviet J. Phys. JETP}}
\def\MPA{{Mod. Phys. Lett.} {\bf A}}
%
%

\begin{frontmatter}



\title{Stable Massive Particles at Colliders}


\author[sholm]{M.~Fairbairn}
\author[penn]{A.C.~Kraan}
\author[sholm]{D.A.~Milstead}
\author[lund]{T.~Sj\"ostrand}
\author[fermi]{P.~Skands}
\author[lancaster]{T.~Sloan}

\address[sholm]{Fysikum, Stockholm University, Sweden}
\address[penn]{Department of Physics and Astronomy, University of Pennsylvania, USA; \\
Presently at: Istituto Nazionale di Fisica Nucleare, Pisa, Italy}
\address[lund]{Department of Theoretical Physics, Lund University, Sweden; also at CERN, Switzerland}
\address[fermi]{Theoretical Physics Department, Fermi National Accelerator Laboratory, USA}
\address[lancaster]{Department of Physics, Lancaster University, UK}

\begin{abstract}

We review the theoretical motivations and experimental status of
searches for stable massive particles (SMPs) which could be
sufficiently long-lived as to be directly detected at collider
experiments. The discovery of such particles would address a number
of important questions in modern physics including the origin and
composition of dark matter in the universe and the unification of the fundamental
forces. This review describes the techniques used in SMP-searches at
collider experiments and the limits so far obtained on the
production of SMPs which possess various colour, electric and
magnetic charge quantum numbers. We also describe theoretical
scenarios which predict SMPs and the phenomenology needed to model
their production at colliders and interactions with matter. In
addition, the interplay between collider searches and open questions
in cosmology is addressed.
\end{abstract}

\begin{keyword}
 review \sep experimental results \sep colliders \sep SMP \sep monopole
\sep
 SUSY \sep extra dimensions
\PACS 14.80 \sep Hv \PACS 14.80 \sep Ly  \PACS 29.40 \sep Gx
\end{keyword}
\end{frontmatter}

\section{Introduction}
An open question in modern physics is the possible existence of
heavy, exotic particles, which can be directly detected at collider
experiments through their interactions with matter. This paper
describes collider searches for these so-called Stable Massive
Particles (SMPs), together with a review of theoretical models in
which such states appear. We also discuss the astrophysical
consequences that an SMP discovery would imply. In this work, we
define an SMP as a particle which does not
decay during its passage through a detector, and which would undergo
electromagnetic and/or strong interactions with matter. Normally we would expect such particles to be heavier than a proton.

Historically, the strange, long-lived kaons heralded a revolution in
particle physics, in terms of new fundamental matter, and also in
the shape of a new (approximately) conserved quantum number. Today,
states classifiable as SMPs recur in many theoretical extensions of
the Standard Model (SM). One crucial aspect in this context is
naturally the question of dark matter in the universe, but also
charge quantisation (magnetic monopoles? millicharges? grand
unification?), the flavour question (more generations?), parity
violation (vector-like fermions? mirror fermions?), the hierarchy
problem (supersymmetric SMPs?), and more (SMPs from extra dimensions?
your favourite particle?) are relevant. We devote Section 2 to a
review of theoretical scenarios with SMP states.


Due to the symbiosis between accelerator particle physics and
astrophysics for long-lived particles, we focus on the cosmological
and astrophysical implications of SMPs in Section 3. This discussion
includes the implications of an SMP discovery at an accelerator on
topics such as dark matter and nucleosynthesis.

In most non-generic searches for SMPs, theoretical models are needed
to calculate their production cross sections and final-state
topologies. In the case of massive coloured objects, their
fragmentation into jets must be described. Section 4 contains an
overview of the techniques used to model the production of SMPs.

The detection of SMPs is only possible once the interactions in
matter are understood. In Section 5 we summarise the phenomenology
used to model the interactions of SMPs with detector material,
including a description of their electromagnetic and hadronic
interactions.

In Section 6, a range of possible search techniques which can be
used to identify SMPs are described. These include techniques based
on ionisation energy losses, Cherenkov techniques and time-of-flight
methods. Additionally, specific techniques used in magnetic monopole
searches are summarised.

A wide variety of SMP searches have been performed in lepton-hadron,
electron-positron and hadron-hadron reactions~\cite{rn}, yielding
important constraints on the parameter spaces of a number of
theories. The searches vary from the very general types, which make
minimal assumptions regarding the quantum numbers of the SMP, to
those performed within a specific theoretical scenario. Section 7
presents a summary of the most important model-independent searches,
as well as a selection of searches made within commonly studied
exotic physics scenarios.

Approaching the exciting time when the Large Hadron Collider (LHC)
will produce its first collisions, it is appropiate to discuss
prospects for the discovery of SMPs with this machine. Searches for
SMPs with masses up to several TeV will be possible,  representing
an order of magnitude increase in mass sensitivity compared with
earlier colliders. Section 8 discusses the the discovery potential
of a range of types of SMPs at the LHC. Finally, Section 9 contains
a summary of this report.

 Our report is complementary to, and
differs from, previous reviews of
SMPs~\cite{Preskill:1984gd,Perl:2001xi,Perl:2004qc,Milton:2006cp} in
a number of ways. The principal difference is that we focus on
collider searches, although non-accelerator results such as those obtained using mass spectroscopy and cosmic ray detection are discussed,
when appropriate. In view of the circa 50 accelerator searches, which
have taken place over the past two decades, an up-to-date and
detailed description of the techniques and results of these
searches, and of their theoretical motivation, is overdue. By
specialising in collider studies we are able to cover searches for
many different proposed species of SMPs and the experimental
challenges specific to each of them. For example, we provide a
detailed treatment of hadronic SMPs, which is lacking in earlier
reviews. Furthermore, although magnetic monopoles are usually
treated separately to electrically charged SMPs in this type of
article, they are included here since both types of particles are
expected to share many common experimental signatures. Magnetic
monopole searches can provide sensitivity to SMPs with values of
electric charge beyond those covered by dedicated searches for
electrically charged SMPs, and vice versa. Thus, by considering a
broader range of collider searches, it is possible to more fully
describe the types of SMPs, and their possible masses, which have
been excluded, and those which could have been produced at colliders
but which potentially remain unobserved.

\section{Theoretical scenarios for SMPs}\label{sec:scen}

%

In this section, we first give a brief introduction to theoretical
possibilities for SMP states, followed by in-depth discussions on some
of the more commonly considered scenarios. In particular we discuss
supersymmetric models in Section~\ref{ss:susymodels} and models with
universal extra dimensions in Section~\ref{ss:xdmodels}. Sections
\ref{ss:exoticmodels} and  \ref{ss:topo} concern alternative
possibilities that do not fit smoothly into either
of the above categories, such as $Z'$-induced millicharges
and magnetic monopoles.

To gain an impression of the wide range of possible SMP states,
it is instructive first to take a look at the Standard Model itself.
Protons, neutrons, electrons, and muons are all examples of stable particles which undergo interactions in a detector. So are $\pi^+$, $K^+$, and $K^0_L$.
These well-studied states also illustrate very
well the spectrum of possibilities
\emph{beyond} the Standard Model. Consider the following
well-known properties, and apply them implicitly to hypothetical new states.
 The electron is a fundamental particle. It does not decay, e.g.\ to
  neutrinos due to the conservation of a gauged quantum number,
  electric charge in this case. It is the lightest state carrying that
  quantum number.
The proton is a complicated bound state composed of a set of more
  fundamental particles which are held together by a high-strength
  short-range gauge force.
  The kinematically allowed decay of the proton to, e.g.\ $e^+\pi^0$
  does not occur due to conservation of certain global (i.e.\
  non-gauged) quantum numbers, in this case
  baryon and lepton number. The conservation of these
  numbers are accidental symmetries of the model.
The long lifetime of the neutron
owes to a combination of the weak force being involved
 and the very small decay phase space.
Though unstable in free space, there are still many of them around,
  due to the existence of stable ``composites of
  composites'', the nuclei of baryonic matter.
The muon has a comparatively long lifetime due to the
  hierarchy between the muon mass and the weak scale ($\Gamma_\mu
  \propto m_\mu^5 / m_W^4$). At low
  energies, the decay appears to proceed via a non-renormalisable
  dimension-6 operator which, in the fundamental theory, arises due to
  a  virtual massive gauge boson, the $W$.
Finally, also the atom furnishes an example. Its stability could not
 be understood in terms of Maxwell's theory. Only after a drastic
 revision, quantum mechanics, could it be accounted for. The
 corresponding case today would be the discovery of a state whose
 stability could not be accounted for within the framework of quantum
 field theory.

Thus, already the SM contains quite a varied history of stable interacting states.
 Turning now to physics beyond the SM,
there are several generic possibilities for SMP's, essentially all recurrences
of the states mentioned above, with the addition of topological
defects, like cosmic strings or magnetic monopoles.

The most obvious possibility for an SMP is that one or more new
states exist which carry a new conserved, or almost conserved,
global quantum number\footnote{ As an aside, at the most fundamental
level any such global symmetry probably has to be the remnant of a
broken gauge or space--time symmetry (e.g.\ KK parity is the remnant
of broken higher-dimensional Poincar\'e invariance), to avoid
stability problems
\cite{Gilbert:1989nq,Ibanez:1991pr,Krauss:1988zc}, but we shall here
treat them simply as discrete global symmetries, whatever their
origin.}. SUSY with $R$-parity, extra dimensions with KK-parity, and
several other models fall into this category. The lightest of the
new states will be stable, due to the conservation of this new
``parity'', and depending on quantum numbers, mass spectra, and
interaction strengths, one or more higher-lying states may also be
stable or meta-stable. In general, electrically charged stable
states are excluded by cosmology, and also coloured particles are
strongly constrained, as will be discussed in Section~\ref{sec:cosmo}. For this reason, and to obtain a solution to the
dark-matter problem, models are usually constructed to provide
\emph{un}-charged stable dark-matter candidates, most often in the
guise of weakly interacting massive particles (WIMPs). From a
motivational point of view, SMP models thus come in two categories:

\begin{enumerate}
\item Models which solve the dark-matter
problem with a WIMP-type dark-matter particle, but which also have
one or more higher-lying meta-stable SMP states.
\item Models which
have long-lived SMP states, but which either do not address dark
matter or address it with a non-WIMP dark-matter particle.
\end{enumerate}

As a quick reference,  Tabs.~\ref{tb:smpsusy} and \ref{tb:smpstates} give a
condensed overview of the SMP states discussed in the text, along
with a description of which scenario(s) give rise to each of them.

\subsection{SMP states in Supersymmetry \label{ss:susymodels}}

Among the most interesting and comprehensively studied
possibilities for observable new physics is
supersymmetry (SUSY)
--- for excellent reviews see,
e.g.\ \cite{Martin:1997ns,Tata:1997uf}.
Interesting in its own right by being the largest possible
space--time symmetry
,
it was chiefly with a number of additional theoretical
successes during the eighties that supersymmetry gained widespread
acceptance, among these successful
gauge coupling unification
,
a natural candidate for dark matter
, and an elegant solution to the so-called hierarchy problem
associated with the smallness of $M_Z^2/M_{\mrm{Planck}}^2$.

Stated briefly, supersymmetry promotes all the fields of the SM to
superfields. Each ($N=1$) superfield contains one boson and one fermion as
dynamical degrees of freedom. Hence not only should all the SM
particles (including at least one more Higgs doublet) have
`sparticle' superpartners, with spins differing by 1/2, but also the
interactions of these superpartners should to a large extent be
fixed by supersymmetry. However, due to the absence of observed
sparticles so far, there must exist a mass gap between the SM
particles and their superpartners, implying that supersymmetry must
be broken at the weak scale. This mass gap cannot be much larger
than $\sim 1$ TeV without invalidating the solution to the hierarchy
problem. Independently, both SUSY dark matter and SUSY grand
unification also point to roughly the TeV scale,  hence there is
good reason to believe that SUSY, if it exists, should manifest
itself at the next generation of colliders, and a long list of
sophisticated theory tools for SUSY phenomenology are available,
see, e.g.\
\cite{Djouadi:2002nh,Skands:2006sk,Skands:2005vi,Skands:2003cj,slha2}.

In the most general SUSY models, both lepton and baryon number are
violated. If the SUSY-breaking mass scale \cite{Weinberg:1981wj} is
$\mathcal{O}(1~\TeV)$, then bounds on proton decay
\cite{Shiozawa:1998si,rn} require the product of the $B$ and $L$
violating couplings to be less than $10^{-25}$
\cite{Dreiner:1997uz}. Hence, either one or both of the accidental
symmetries that protect $B$ and $L$ in the Standard Model are
usually promoted to explicit ones in SUSY models. The most common
way to accomplish this is through postulating the conservation of
the quantum number $R=(-1)^{3B+L+2S}$ \cite{Farrar:1978xj}, where
$B$($L$) is baryon (lepton) number and $S$ is spin. Since SM
particles carry $R=+1$ and their SUSY partners $R=-1$, the lightest
supersymmetric particle (LSP) is stable, providing a possible
solution to the dark-matter problem. If $R$-parity is violated,
either lepton or baryon number must still be conserved, to avoid
rapid proton decay. In this case, the LSP is not stable (though it
may still be long-lived) and the dark-matter problem is not
addressed. For a charged or coloured LSP, $R$-parity violation can be invoked to
obtain a lifetime in the allowed range \cite{Dreiner:1997uz, Berger:2003kc}.

\begin{table}[t]
\begin{tabular}{lllp{9.3cm}}
\toprule
SMP & LSP & Scenario & Conditions\\
\cmidrule{1-4}
$\ti{\tau}_1$ & $\ti{\chi}^0_1$ & MSSM & $\ti{\tau}_1$ mass (determined
 by $m^2_{\ti{\tau}_{L,R}}$, $\mu$, $\tan\beta$, and $A_\tau$) close to
 $\ti{\chi}^0_1$ mass. \\
  & $\ti{G}$ & GMSB & Large $N$, small
 $M$, and/or large $\tan\beta$.\\
 &   & $\ti{g}$MSB & No detailed phenomenology studies, see
 \cite{Buchmuller:2005rt}. \\
 &   & SUGRA & Supergravity with a gravitino LSP, see \cite{Feng:2004mt}. \\
 & $\ti{\tau}_1$ & MSSM &
 Small $m_{\ti{\tau}_{L,R}}$ and/or large $\tan\beta$ and/or very large
 $A_\tau$. \\
 &          & AMSB & Small $m_0$, large $\tan\beta$. \\
&  &  $\ti{g}$MSB & Generic in minimal models.  \\
$\ti{\ell}_{i1}$ & $\ti{G}$ & GMSB &
 $\ti{\tau}_1$ NLSP (see above). $\ti{e}_1$ and
 $\ti{\mu}_1$ co-NLSP and also SMP for small $\tan\beta$ and $\mu$.\\
 & $\ti{\tau}_1$ & $\ti{g}$MSB &
 $\ti{e}_1$ and $\ti{\mu}_1$ co-LSP and also SMP when stau mixing
 small. \\
$\ti{\chi}^+_1$ & $\ti{\chi}^0_1$ & MSSM &
$m_{\ti{\chi}^+_1}-m_{\ti{\chi}^0_1} \lsim m_{\pi^+}$.
Very large $M_{1,2}\gsim 2\ \mrm{TeV}\gg |\mu|$ (Higgsino region) or
non-universal gaugino masses $M_1\gsim 4M_2$, with the latter
condition relaxed to  $M_1\gsim M_2$ for $M_2 \ll |\mu|$. Natural in
 O-II models, where simultaneously
also the $\ti{g}$ can be long-lived near $\delta_{\mrm{GS}}=-3$. \\
& & AMSB & $M_1 > M_2$ natural. $m_0$ not too small. See MSSM above.\\
$\ti{g}$ & $\ti{\chi}^0_1$& MSSM & Very large $m^2_{\ti{q}}\gg M_3$,
e.g.\
split SUSY.\\
   & $\ti{G}$ & GMSB & SUSY GUT extensions \cite{Raby:1997pb,Baer:1998pg,Mafi:2000kg}.\\
         & $\ti{g}$& MSSM & Very small $M_3\ll M_{1,2}$,
 O-II models near $\delta_{\mrm{GS}}=-3$. \\
& & GMSB & SUSY GUT extensions
\cite{Raby:1997pb,Raby:1997bp,Baer:1998pg,mafi,Mafi:2000kg}.\\
$\ti{t}_1$ & $\ti{\chi}^0_1$& MSSM & Non-universal squark and gaugino
masses. Small $m^2_{\ti{q}}$ and $M_3$, small $\tan\beta$, large
$A_t$. \\
$\ti{b}_1$ & & & Small $m^2_{\ti{q}}$ and $M_3$, large $\tan\beta$
and/or large $A_b\gg A_t$. \\
\bottomrule
\end{tabular}
\caption{Brief overview of possible SUSY SMP states considered in
the literature. Classified by SMP, LSP, scenario, and typical conditions
for this case to materialise in the given scenario. See text for
details. \label{tb:smpsusy} }
\end{table}

\subsubsection{The MSSM and CMSSM}\label{peter:mssm}
At present, little is known for certain about the nature of the
SUSY-breaking mechanism. Some special cases will be discussed below,
but at the most general level we define the minimal supersymmetric
Standard Model (MSSM) to contain all possible SUSY-breaking
interactions, which are consistent with gauge and Poincar\'e
invariance, and which do not cause the hierarchy problem to reappear
(so-called soft SUSY-breaking). This implies an additional mass term
for each of the gauginos $M_{1,2,3}$, extra mass terms squared
$m_{ij}^2$ for each pair of scalars $i$ and $j$,  and also trilinear
couplings $A_{ijk}$ between combinations of three scalars $i$, $j$,
and $k$. Due to significant constraints on CP and flavour violation,
these parameters are normally taken to be real and flavour diagonal.
To further limit the number of parameters, the additional assumption
of GUT scale unification of the scalar masses $m_0$, gaugino masses
$m_{1/2}$, and trilinear parameters $A_0$ is often made, a choice
sometimes referred to as the constrained MSSM or CMSSM. In addition,
there are also the SUSY-conserving Yukawa couplings $Y_{ijk}$, gauge
couplings $g_i$, vacuum expectation values (vev's) for each of the
Higgs doublets $v_{1,2}$, the $\mu$ parameter (a SUSY-conserving
mixing term between the two Higgs doublets), and, if $R$-parity is
violated, Yukawa-like interactions between lepton and quark
superfields as well as $L$-violating mixing terms between one of the
Higgs superfields and the lepton superfields. The sum of the Higgs
vev's is fixed by the $Z$ mass, and their relative size is usually
cast in terms of the parameter $\tan\beta = v_2/v_1$, with
$\tan\beta\lsim 50$ required for perturbativity up to the GUT scale.
See \cite{Martin:1997ns,Tata:1997uf} for further details.

As Tab.~\ref{tb:smpsusy} suggests, the general MSSM allows for
essentially any sparticle to be an SMP. In models with neutralino
dark matter, the NLSP can be long-lived if its decay phase space is
small or zero, mimicking the case of the neutron in the SM. An
interesting scenario here is that of a light stop NLSP, as motivated
by electroweak baryogenesis \cite{Carena:2002wz,Balazs:2004bu}. In
this scenario non-universal squark mass terms are used to arrange a
small mass difference between the $\ti{t}_1$ and the LSP
$\ti{\chi}_1^0$, while the lightest chargino is kept too heavy for
the decay
 $\ti{t}_1\to b\ti{\chi}_1^+$ to occur. In this case, only the radiative
process $\ti{t}_1 \to \ti{c}_1 \ti{\chi}^0_1$ is open,
and the $\ti{t}_1$ can be quite long-lived.
See \cite{Carena:2002wz,Balazs:2004bu,Kraml:2005kb,lightstop} for
detailed studies.

Though not equally well motivated at present, a light NLSP
$\ti{\tau}_1$ (small slepton masses and/or large $\tan\beta$ and/or
large $A_\tau$) or less likely $\ti{b}_1$ (small squark masses and
large $\tan\beta$ and/or very large $A_b\gg A_t$) could also be
possible. For a $\ti{\tau}_1$ NLSP with small mixing (small
$\tan\beta$ and $A_\tau$), the  $\ti{e}_1$ and $\ti{\mu}_1$ could
simultaneously also be long-lived. A small mass difference
$m_{\ti{\chi}^+_1}-m_{\ti{\chi}^0_1} \lsim m_{\pi^+}$ can occur for
very large $M_{1,2}\gsim 2\ \mrm{TeV}\gg |\mu|$ (Higgsino region) or
non-universal gaugino masses
\cite{Cheng:1998hc,Chen:1999yf,Feng:1999fu} $M_1\gsim 4M_2$, with
the latter condition relaxed to  $M_1\gsim M_2$ for $M_2 \ll |\mu|$.
To obtain a long-lived gluino NLSP, the requirement is small $M_3$,
possibly in combination with very large $m_{\ti{q}}^2$, since the
gluino decay proceeds via intermediate squarks (see, e.g.\ the
discussion of split supersymmetry below).

If the LSP is not required to be a neutralino, essentially all of
the cases just mentioned have obvious extensions where the NLSP
becomes the LSP. In this case, dark matter is much more problematic,
and $R$-parity violation may be necessary to render the LSP
unstable.

When gravity is included, as in supergravity models,
also the gravitino can in principle be the LSP, depending on its mass.
If the gravitino is light, the NLSP will then often be long-lived due
to the smallness of the gravitational coupling. For supergravity models
with a long-lived  $\ti{\tau}$ NLSP see \cite{Feng:2004mt}; other
models with a gravitino LSP will be discussed below. Finally, we note that a 
similar phenomenology can also be obtained in extensions with an axino
LSP~\cite{Brandenburg:2005he}.

\subsubsection{Gauge Mediated Supersymmetry Breaking}\label{peter:gmsb}
\begin{figure}[t]
\begin{center}
\includegraphics*[scale=0.75]{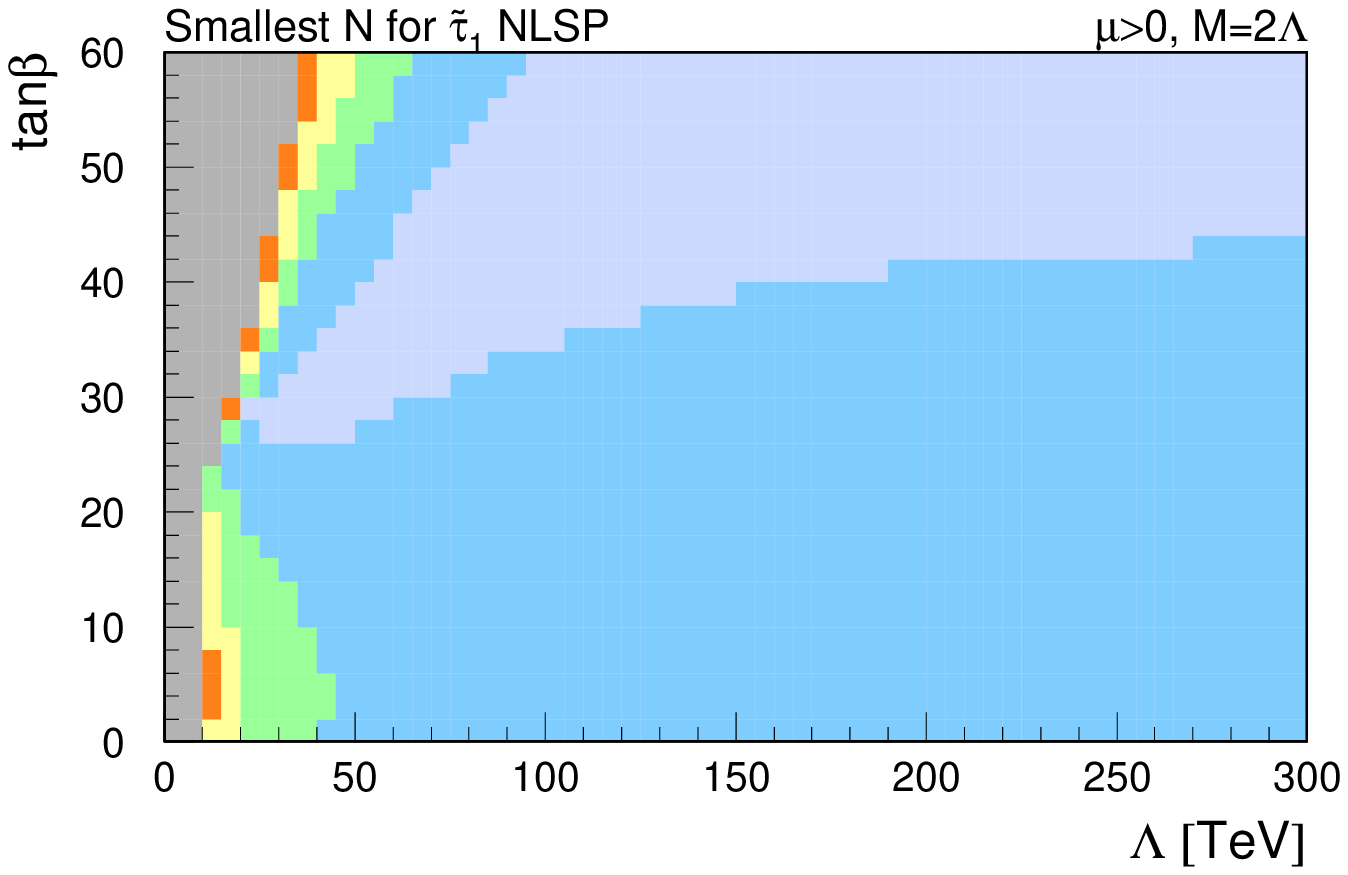}\\[-7mm]
\sl a)\\[-4mm]
\includegraphics*[scale=0.75]{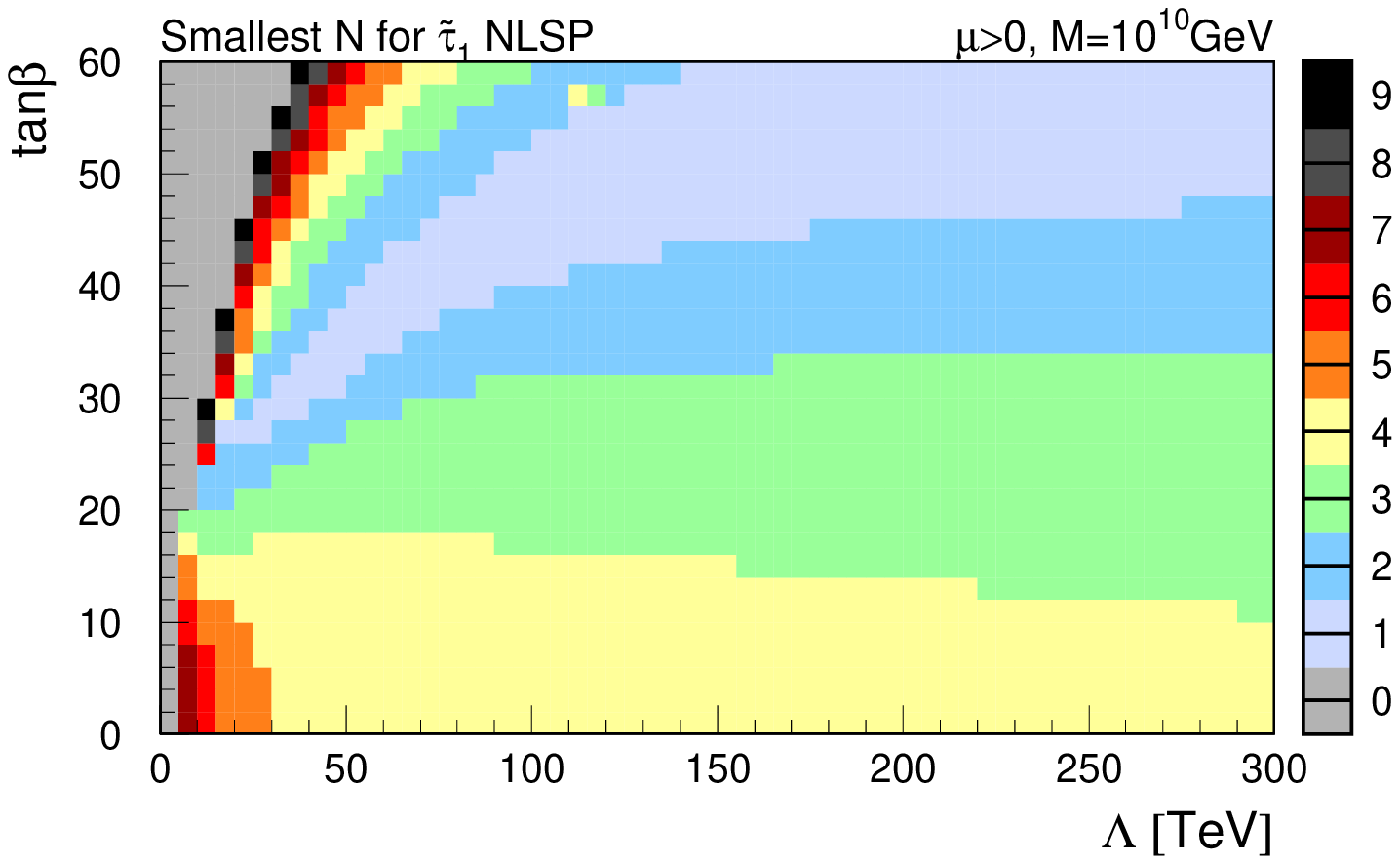}\\[-7mm]
\sl b)\\
\end{center}
\caption{
GMSB:  the smallest index number $N$ required to obtain a
$\tilde{\tau}_1$ NLSP as a function of
$\Lambda$ and $\tan\beta$ for {\sl a)} light messengers ($M=2\Lambda$)
and {\sl b)} heavy messengers ($M=10^{10}\GeV$). The colour coding is the same
for both plots and corresponds to the legend shown with {\sl b)}.
\label{fig:gmsb}}
\end{figure}

In models with gauge mediation \cite{Dine:1995ag,Dine:1994vc} (GMSB,
see \cite{Alvarez-Gaume:1981wy,Dimopoulos:1996yq,Bagger:1996ei} for
mass spectra and \cite{Giudice:1998bp} for a review), the gravitino
is very light ($m_{\ti{G}}\lsim 1\keV$) and hence the LSP for any
relevant choice of parameters.  Minimal models are cast in terms of
six parameters, typically $N$, $\Lambda$, $M$, $\tan\beta$,
$\mathrm{sgn}(\mu)$, and $c_{\mathrm{grav}}$.  Briefly stated, the
meaning of these parameters is that $N$ chiral $SU(5)$ multiplets
are added to the theory at the scale $M\lsim 10^{15}\ \mrm{GeV}$
\cite{Giudice:1998bp}. If not only $SU(5)$ multiplets are added, the
counting gets more complicated, but there is still an effective $N$.
These ``messengers'' couple directly both to the MSSM fields (via
the ordinary SM gauge interactions) and also to an unspecified
source of SUSY-breaking.  $\Lambda \sim 10 - 100$ TeV is the
effective SUSY-breaking scale, related to the fundamental
SUSY-breaking scale $\sqrt{F}$ by a relation $\Lambda = F/M$.  The
next-to-lightest sparticle (NLSP) decays only via the gravitational
coupling and can be very long-lived. For a slepton NLSP
\cite{Giudice:1998bp}:
\begin{equation}
c\tau_{\mrm{NLSP}} = 0.1 \left(\frac{100\
\mrm{GeV}}{m_{\mrm{NLSP}}}\right)^5\left(
\frac{m_{\ti{G}}}{2.4\ \mrm{eV}}
\right)\ \mrm{mm} ~~~,
\end{equation}
with the gravitino mass $m_{\ti{G}}$ controlled by $c_{\mrm{grav}}<1$
and $F$:
\begin{equation}
m_{\ti{G}} = 2.4 c_\mathrm{grav} \left(\frac{\sqrt{F}}{100\
\mrm{TeV}}\right)^2\ \mrm{eV} ~~~.
\end{equation}
Since the SUSY-breaking terms are induced by gauge interactions,
they are flavour universal and their sizes are proportional to the
amount of gauge charge carried by each field. In minimal models, the
next-to-lightest supersymmetric particle is therefore always the
``least charged'' of either the gauginos or the scalars. The latter
occurs in particular for large values of the messenger index $N$. In
a tiny and near-excluded parameter region at small values of the
model parameters
 $\Lambda$ and $M$, the NLSP is then a sneutrino, otherwise it is the
$\ti{\tau}_1$ (for a benchmark and model line, see Snowmass point 7
\cite{Allanach:2002nj}). Note, however, that $N$ cannot be chosen
arbitrarily large, perturbativity of the theory up to the GUT scale
requiring, e.g.\ $N\lsim 5$ for small $M\lsim 10^{6}\GeV$ and
$N\lsim 10$ for $M\sim 10^{10}\GeV$. To illustrate the parameter
space, Fig.~\ref{fig:gmsb} shows the smallest messenger index $N$
required to have a $\tilde{\tau}_1$ NLSP for $\mu>0$ and {\sl a)}
relatively light messengers $M=2\Lambda$ and {\sl b)} heavy
messengers $M=10^{10}\GeV$, as a function of $\tan\beta$ and
$\Lambda$ (since $c_{\mathrm{grav}}$ only affects the decay we leave
it unspecified). The numbers were obtained with \textsc{Isajet}
v.7.71 \cite{Paige:2003mg} using $m_t=175\GeV$.
The light grey areas at small $\Lambda$ are theoretically excluded due to
unstable vacua and/or non-perturbative couplings at the GUT
scale; no experimental or indirect constraints were included here.

With regard to dark matter, it is interesting to note that even in
the presence of $B$-violation, the gravitino would still be stable
and a dark-matter candidate, since no kinematically allowed decays
would be available. For other possibilities for GMSB dark matter,
see \cite{Giudice:1998bp}.

Finally, a long-lived $\ti{\tau}_1$ is not the only SMP possibility
in GMSB. If the mixing and consequently the mass splitting in the
stau sector is not too large (small $\tan\beta\lsim 8$
\cite{Ambrosanio:1997rv}), then the $\ti{e}_1$ and $\ti{\mu}_1$ may
be nearly mass-degenerate (co-NLSP) with the $\ti{\tau}_1$ and hence
can simultaneously be SMPs. As with all supersymmetric scenarios,
there is also a very large space of possible non-minimal models,
cf.\ \cite{Giudice:1998bp}. Of particular interest here are SUSY GUT
extensions of GMSB in which the coloured messengers are naturally
much heavier than their weak counterparts, resulting in a gluino
NLSP \cite{Raby:1997pb,Baer:1998pg,Mafi:2000kg} or even LSP
\cite{Raby:1997pb,Raby:1997bp,Baer:1998pg,mafi,Mafi:2000kg},
depending on the gravitino mass.

\subsubsection{Split Supersymmetry}\label{peter:split}
\begin{figure}[t]
\center\includegraphics*[scale=0.47]{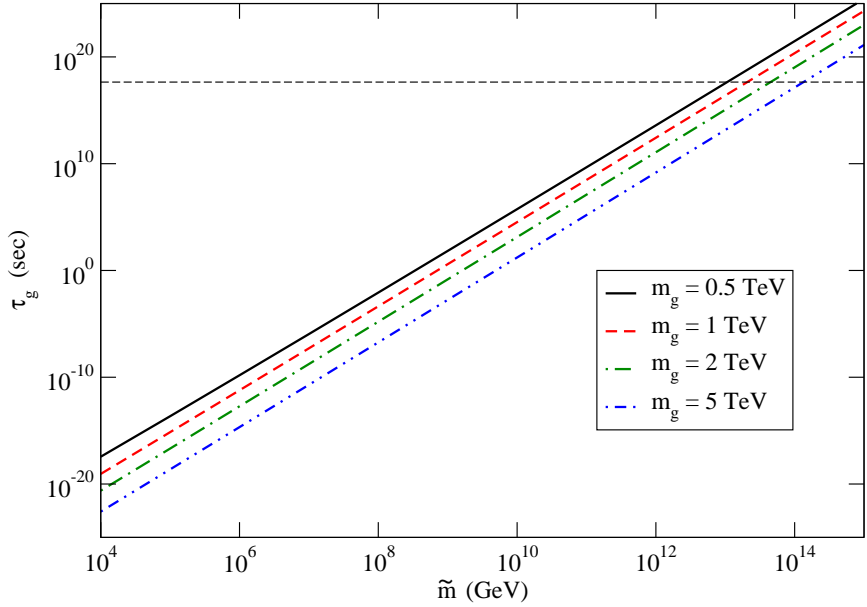}
\caption{The gluino lifetime in split SUSY as a function of the
scalar mass parameter $\tilde{m}$ for $\tan\beta=2$, $\mu >0$, and
various choices of the gluino mass, as calculated by
\cite{Gambino:2005eh}. The dashed horizontal line indicates the age
of the Universe, $\tau_U = 14$ Gyr.
\label{fig:gluinolife}}\end{figure}

Meta-stable coloured sparticles also arise in
the so-called split SUSY scenario
\cite{Arkani-Hamed:2004fb,Giudice:2004tc},
in which all the scalars
(except the ordinary Higgs) have very large masses,
 while the gaugino and higgsino masses remain at or around the weak
 scale. Though the hierarchy problem is not addressed (except
 anthropically), this
naturally suppresses both proton decay and CP and flavour violation.
Since gluinos can only decay via
squarks (independently of whether $R$-parity is conserved or not), the gluino
lifetime can be very large in this scenario, somewhat similar to the
case of the muon in the SM.
The competing channels are tree-level 3-body decays to two quarks plus a
chargino or neutralino and radiative 2-body decays into a gluon plus a
neutralino, see \cite{Gambino:2005eh} for explicit
calculations or \cite{Toharia:2005gm} for a simplified treatment.
For illustration, in Fig.~\ref{fig:gluinolife}
we include a plot from
\cite{Gambino:2005eh} showing the gluino lifetime as a function of
the scalar mass parameter $\tilde{m}$ for $\tan\beta=2$, $\mu >0$,
and various choices of the gluino mass.

Another interesting possibility in split SUSY, which is not
ordinarily viable, is that SUSY-breaking could be communicated
directly at tree-level \cite{Arkani-Hamed:2004yi,Babu:2005ui}. In
this case, the scalar masses $\tilde{m}$ could be very close to the
fundamental SUSY-breaking scale $\sqrt{F}$. Since the gluino decays
via virtual squarks discussed above are suppressed by
$\tilde{m}^{-2}$ and the coupling to the gravitino goes like $1/F$,
this would open the possibility for a large branching fraction for
gluinos to gravitinos, $\tilde{g} \to g \tilde{G}$
\cite{Gambino:2005eh}.

\subsubsection{Other SUSY-breaking scenarios: AMSB, O-II, and $\ti{g}$MSB}\label{peter:amsb}
Anomaly mediated SUSY-breaking (AMSB, see
\cite{Randall:1998uk,Giudice:1998xp,Feng:1999hg}) is a  variant of
supergravity, where the explicit SUSY-breaking terms are `switched
off' (or at least heavily suppressed), leaving a scaling (conformal)
anomaly in the supergravity Lagrangian as the sole source of
supersymmetry breaking. This ``anomaly-mediated'' contribution is
always present, but is usually much smaller than other supergravity
terms. Akin to GMSB, its virtue is that it is flavour universal by
construction. In its pure form, however, AMSB gives rise to
tachyonic sleptons (negative $m^2_{\ti{\ell}}$), hence, to obtain a
viable phenomenology, additional positive lepton mass squared terms
have to be included. The parameters of a minimal model are thus the
gravitino mass $m_{3/2}$, the usual $\tan\beta$ and
$\mrm{sgn}(\mu)$, as well as additional soft SUSY-breaking masses
for all sfermions $m_0$. The ratio of gaugino masses at the weak
scale is approximately $|M_1|:|M_2|:|M_3| \sim 2.8:1:8$
\cite{Feng:1999hg}, making the lightest chargino and neutralino
nearly mass-degenerate and wino-like. For large $m_0$, the
neutralino is the LSP, and hence the chargino can be long-lived. For
small $m_0$ the LSP can be the $\ti{\tau}_1$, except at small
$\tan\beta$ and $m_{3/2}$ where it is the $\nu_{\ti{\tau}1}$
\cite{Feng:1999hg}. For an LHC phenomenology study, see
\cite{Barr:2002ex}.

Another interesting explicit realisation of non-universal gaugino
masses is furnished by the so-called O-II orbifold model, a
string-inspired scenario in which supersymmetry breaking is
dominated by the overall `size' modulus field arising from the
orbifold compactification \cite{Brignole:1993dj,Chen:1996ap}.
Gaugino masses arise at one loop, and a large degree of
non-universality is generated naturally. The free parameters of a
minimal model are $m_0$, $\tan\beta$, $\mrm{sgn}(\mu)$, and
$\delta_{GS}$. The latter is called the Green-Schwarz mixing
parameter and preferentially lies in the range $-5 \lsim
\delta_{\mrm{GS}} < 0$, with the negative integers -4 and -5
preferred for the model studied in \cite{Chen:1996ap}. The mass
spectrum depends sensitively on $\delta_{\mrm{GS}}$ but a typical
feature, unless $|\delta_{\mrm{GS}}|$ is very large, is $|M_1|\gg
|M_2|$, resulting in a near-degeneracy between the lightest
neutralino and chargino, both of which will be wino-like as in AMSB.
Consequently, the chargino can be very long-lived. For
$\delta_{\mrm{GS}}=-3$, the gluino mass is zero at the high scale,
$M_3^0 = 0$, and hence close to this value the gluino is typically
the LSP \cite{Chen:1996ap,Baer:1998pg} or, slightly farther away
from the minimum, nearly mass-degenerate with both the
$\ti{\chi}_1^+$ and the $\ti{\chi}_1^0$. Note, however, that the
other gauginos also have minima around $\delta_{\mrm{GS}}=-3$, so
the entire gaugino spectrum can become very light in this region.

In gaugino-mediated SUSY-breaking \cite{Kaplan:1999ac,Chacko:1999mi}
($\ti{g}$MSB), the MSSM is embedded in a 5-dimensional compactified
braneworld setup. The only non-vanishing soft SUSY-breaking terms at
the compactification scale, $M_c$, are the gaugino masses. In the
original models, $M_c= M_{\mrm{GUT}}$, which gives a
$\tilde{\tau}_1$  LSP. The $\tilde{\mu}_ 1$ and $\tilde{e}_1$ are
typically
  slightly heavier, due to the smaller mixing in the first two
  generations, but they can also be SMP candidates. Due to the
  obvious constraints from cosmology, most subsequent attempts have
focussed on finding models beyond the minimal where a WIMP-like LSP
could be recovered. In particular, if $M_c>M_{\mrm{GUT}}$, then the
extra GUT running from $M_c$ (where the boundary conditions are
true) to $M_{\mrm{GUT}}$ (below which the SM running takes over) can
generate non-zero scalar masses at the GUT scale
\cite{Schmaltz:2000gy,Schmaltz:2000ei,Baer:2000gf}, which may push
the $\ti{\tau}_1$ mass above the $\ti{\chi}_1^0$ one. Another way of
modifying the spectrum is by introducing non-universal gaugino
masses, e.g.\ as in higher-dimensional GUT models where the extra
dimension is larger than the inverse GUT scale, $M_c <
M_{\mrm{GUT}}$ \cite{Baer:2002by}, but except for small parameter
regions the $\ti{\tau}_1$ LSP still dominates. A Tevatron study of
gaugino mediation can be found in \cite{Baer:2001ze}. The
interesting case of a GMSB-like phenomenology could also be
possible, with a gravitino LSP and the $\ti{\tau}_1$ the NLSP
\cite{Buchmuller:2005rt}, but detailed phenomenology studies have so
far not been carried out.

\subsection{SMP States in Universal Extra Dimensions \label{ss:xdmodels}}
In models of Universal Extra Dimensions (UED) all the fields of the
SM, including both matter and forces, are allowed to propagate in
some number of extra dimensions, usually taken to be one or two.
These models \cite{Appelquist:2000nn,Macesanu:2002db,Cheng:2002iz}
provide an interesting scenario for TeV scale physics and are
consistent with low-energy constraints \cite{Appelquist:2002wb}. The
compactification  is constructed such that momentum conservation in
the extra dimensions is preserved at tree level, leading to a
discrete Kaluza-Klein (KK) quantum number in the effective 4D
theory. This quantum number is then broken to a KK parity at the
loop level \cite{Cheng:2002iz}. At least the lightest of the KK
excitations of SM particles will thus be stable and can be a dark-matter candidate \cite{Servant:2002aq,servLH05,Shah:2006gs}.

In the simplest analysis, all the KK modes of the light SM particles
(the photon, gluon, and first generation fermions) may be
sufficiently long-lived to be SMP candidates as well (they are
stable at tree level and are almost mass-degenerate level by level),
but note that the presence of non-zero boundary terms in the extra
dimension(s) can change this picture drastically
\cite{Carena:2002me}. If small, these boundary terms can be treated
perturbatively \cite{delAguila:2006kj}, but they are not calculable
from fundamental principles and should be considered free parameters
of the theory.

The Minimal UED (MUED) model with one extra dimension can be
specified in terms of three parameters, $R$, $\Lambda$, and $m_H$,
where $R$ is the size of the extra dimension (an $S^1/Z_2$ orbifold
stretched from $0$ to $\pi R$), $\Lambda>R^{-1}$ is a scale at which
the boundary terms mentioned above are assumed to vanish, and $m_H$
is the mass of the (ground state) Higgs boson. The viable range of
$R$ found in \cite{Appelquist:2000nn} is $300\GeV < R^{-1} <
10\TeV$. For two extra dimensions \cite{Burdman:2006gy} the lower
bound becomes more dependent on the compactification and ranges
between 300 and 700 GeV, while for more extra dimensions a bound
cannot be reliably estimated. See \cite{uedpythia} for a Monte Carlo
implementation of this model.

Finally, we note that a GMSB-like phenomenology with a stable KK
graviton and a meta-stable next-to-lightest KK particle, e.g.\
$\tau_1$, has also been considered, see
\cite{Feng:2003nr,Shah:2006gs}.

\subsection{SMP States in other scenarios beyond the Standard Model (BSM) \label{ss:exoticmodels}}
We now turn to a brief overview of more exotic possibilities for SMP
states in BSM physics. In particular, it is interesting to
investigate the extent to which other models which have been
proposed to address the dark-matter problem can also give rise to
SMP states. We do not claim to be complete --- model space is in
principle infinite
--- nor do we give a detailed discussion of
each scenario, but
we hope to illustrate the
spectrum of ideas, and where possible point
the reader to relevant literature where further
information can be found. Also keep in mind that for essentially all
these scenarios, with the exception of leptoquarks and to a lesser
extent Little Higgs models,
no convenient package of collider phenomenology tools
yet exists.

\subsubsection{Models with parity-like symmetries}\label{peter:fourth}
A class of models which have recently attracted attention is Little
Higgs with $T$-parity (for recent reviews, see, e.g.\
\cite{perelLH05,Perelstein:2005ka}). As with several of the other
proposals discussed here, $T$-parity (not to be confused with time
reversal) serves the dual purpose of simultaneously suppressing
contributions to electroweak precision observables
\cite{Hubisz:2005tx} and providing a WIMP--like dark-matter
candidate \cite{Birkedal:2006fz}, the ``Lightest $T$-odd particle''
(LTP). In the context of SMPs, however, minimal models do not have
much to offer. The only corner of parameter space where the LTP can
be charged and/or coloured \cite{Birkedal:2006fz}, at large values
of the symmetry-breaking scale $f$ and small  $T$-odd fermion masses
$\tilde{m}$, roughly $f\gsim 1\TeV$ and $\tilde{m}\lsim 300\GeV$, is
excluded from Tevatron squark searches. It should be possible to
construct more viable models with SMPs using non-universal
$\tilde{m}_{\mathrm{coloured}} > \tilde{m}_{\mathrm{leptons}}$, but
so far no explicit models in this direction have been constructed.

More exotic possibilities for supersymmetric SMP candidates also
exist, in particular in models where the MSSM gauge groups and
particle content are extended to include more superfields. A recent
example is a variant of the so-called Fat Higgs model, based on the
MSSM with an extra confining $SU(3)$ symmetry \cite{Harnik:2003rs},
the ``Fat Higgs with a Fat Top'' \cite{Delgado:2005fq}. In this
model, quasi-stable exotic chiral superfields (i.e.\ a complex
scalar and a fermion) appear, which are charged under a global $Z_2$
symmetry, which makes them approximately stable. The strongly
interacting ones are probably outside the range of colliders, but
there is a weak-scale electrically charged multiplet whose members
have SMP properties \cite{Delgado:2005fq}.

In warped extra dimensions with ``GUT parity''
\cite{Goldberger:2002pc,Nomura:2003qb,Nomura:2004is,Nomura:2004it},
the combination of extra dimensions
and an effective TeV scale  supersymmetric
grand unification results in KK towers
not only of the SM gauge and Higgs fields but also of their
SUSY-GUT partners, including XY bosons and coloured Higgs
multiplets. In models where the SM fermions also propagate in the
bulk, add KK-towers for these and their SUSY-GUT partners as well.
A parity  can be chosen such that the MSSM
particles are even and their GUT partners odd, hence the lightest
``GUT-odd'' particle (LGP) is stable or long-lived if this quantum
number is conserved or approximately conserved, respectively.
In the earliest scenario of
\cite{Goldberger:2002pc}, the LGP is typically a light isospin-up
(-down) colour triplet XY gaugino, with electric charge -1/3 (-4/3),
but in the more recent models \cite{Nomura:2003qb,Nomura:2004is,Nomura:2004it}
a wide range of possibilities are open.

In a recent 5D model \cite{Nomura:2005qg} of dynamical SUSY-breaking
(DSB),  TeV scale exotic scalars with the quantum numbers of GUT XY
bosons appear, so-called xyons. If a condition similar to $R$-parity
holds in the DSB sector, these states are long-lived. Their precise
quantum numbers depend on the details of the DSB --- in general they
are both coloured and charged. In the simplest $SU(5)$ case they lie
in a colour triplet isospin doublet with electric charges
$Q=-1/3,-4/3$, but also $SO(10)$ assignments are possible, for example an
additional doublet with $Q=1/3,-2/3$, could easily be possible. See
\cite{Nomura:2005qg}  for mass spectra and phenomenology.

Finally, there also exist a few more general ideas for possible
(quasi-)stable BSM particles, including long-lived leptoquarks (see,
e.g.\ \cite{Friberg:1997nn}) and additional (generations of)
fermions \cite{Fishbane:1983hf,Fishbane:1984zv}. The latter can
either be straightforward additions to the SM generations, e.g.\ a
4th generation with 4th flavour approximately conserved
\cite{Frampton:1997up}, or they can have a non-SM like structure.
Mirror fermions (see, e.g.\ \cite{Banks:1986cg,He:2001tp,Barbieri:2005ri}) are
extra fermions whose right-chiral members lie in $SU(2)$ doublets
while the left-handed ones are singlets, i.e.\ opposite to the SM. A
`vector-like' generation is comprised of an extra SM generation
together with its mirror, as e.g.\ in $N=2$ supersymmetric models
\cite{Polonsky:2000zt,He:2001tp}.

\begin{table}[t]
\begin{tabular}{rcrp{9.5cm}}
\toprule
$Q_{\mathrm{em}}$ & $C_{\mathrm{QCD}}$ & $S$ & Model(s)\\
\cmidrule{1-4}
$0$ & \bf 8     & 1         & Universal Extra Dimensions (KK gluon) \\
$\pm 1$ & \bf 1 & $\frac12$ & Universal Extra Dimensions (KK lepton) \\
& &   & Fat Higgs with a fat top ($\psi$ fermions)\\
& & & 4th generation (chiral) fermions \\
& & & Mirror and/or vector-like fermions \\
& & 0 & Fat Higgs with a fat top ($\psi$ scalars)\\
$\pm\frac43$ & \bf 3 & $\frac12$ & Warped Extra Dimensions with GUT
parity (XY gaugino) \\
& & 0 & 5D Dynamical SUSY-breaking (xyon)\\
$-\frac13$,$\frac23$ & \bf 3 & $\frac12$ &
Universal Extra Dimensions (KK down, KK up)\\
& & & 4th generation (chiral) fermions\\
& & & Mirror and/or vector-like fermions \\
 & & & Warped Extra Dimensions with GUT
parity (XY gaugino) \\
$\epsilon<1$ & \bf 1 & $\frac12$ & GUT with $U(1)-U(1)'$ mixing\\
 & & & Extra singlets with hypercharge $Y=2\epsilon$\\
 & & & Millicharged neutrinos\\
? & ? & 0/$\frac12$/1 & ``Technibaryons'' \\
\bottomrule
\end{tabular}
\caption{ Examples of possible SMP states in a variety of models
beyond the MSSM (for MSSM SMPs, see Tab.~\ref{tb:smpsusy}).
Classified by electric charge $Q$, colour representation
$C{\mathrm{QCD}}$, spin $S$, and scenario. \label{tb:smpstates}}
\end{table}

\subsection{Models with gauged symmetries}\label{ss:topo} When
moving from global symmetries to local ones, several new
possibilities open up. Even though explicit models are scarce, we
shall still try to provide a reasonable overview and discussion.

Consider first the case of a new $U(1)_X$ gauge group. If $X$ is
conserved then a corresponding new massless exotic photon exists,
$\gamma'$ (or ``paraphoton'' \cite{Holdom:1985ag}), which mediates a
new long-range gauge force between particles charged under $X$. If
any SM field is charged under $X$, then $X$ must be $B-L$
\cite{Appelquist:2002mw}, and the coupling is constrained to be
extremely tiny $g_{B-L} \lsim 10^{-19}$~\cite{Dobrescu:2004wz}.
Furthermore, only a state carrying a non-SM-like combination of
quantum numbers could be stable, since it would otherwise decay to
SM particles via prompt photon (or $W/Z$) emission. The stability
would hence not rely directly on the smallness of $g_{B-L}$ but
rather on the conservation of other quantum numbers. We shall
therefore not dwell on this possibility further.

The simplest is thus to postulate a genuinely new $U(1)_X$, with all
SM particles having $X=0$. The lightest $U(1)_X$ charged state would
be absolutely stable (any instability would violate $X$), similarly
to the electron, and if also charged under $U(1)_Q$ and/or
$SU(3)_C$, an SMP. As above, even if a lighter electrical- and/or
colour-neutral state with $X\ne 0$ also exists, the lightest charged
one may still be long-lived, depending on what decay mechanisms are
available.

A second possibility arises from kinetic mixing between the photon
and paraphoton which implies that a particle charged only under
$U(1)_X$ will appear to also have a (small) coupling to the photon
\cite{Holdom:1985ag}. Experimental constraints on millicharged
particles \cite{Davidson:2000hf,Perl:2001xi,Dubovsky:2003yn} leave a
significant parameter space open, including a region of interest to
accelerator searches, with relatively large charges $\epsilon \gsim
10^{-4}$ in units of the electron charge and masses $m\sim 0.1 -
1000\GeV$, with the lower mass bound increasing rapidly above
$\epsilon\sim 10^{-2}$. Recently, a 5-dimensional variant of this
model has been proposed \cite{Batell:2005wa} in which the kinetic
mixing, and hence the observable millicharge, is enhanced as
compared to the 4D case.

In the Standard Model without an additional $U(1)_X$, two additional
possibilities for millicharged states in principle exist. Firstly,
additional $SU(3)\times SU(2)$ singlets with hypercharge
$Y=2\epsilon$ are not forbidden, but would be
difficult to reconcile with grand unified theories
\cite{Okun:1983vw} and suffer from much tighter
experimental constraints than in the $U(1)_X$ case
\cite{Dubovsky:2003yn}.
A small region of interest to accelerator searches still
exists at large $\epsilon = 10^{-3} - 10^{-1}$ and moderate masses $m=
0.1 - 10\GeV$. Secondly, though massless millicharged neutrinos
could be generated by a redefinition of the SM hypercharge
coupling, this is significantly more difficult in the massive case and
also implies an unobserved proton-electron charge difference unless
$\epsilon < 10^{-21}$ \cite{Foot:1989fh}.

For a conserved non-Abelian gauge symmetry the phenomenon of
millicharge is excluded and the only possibility is that the
lightest state itself carries charge and/or colour, in addition to
its exotic charge, $X$. The running coupling constant with $N_{fX}$
exoflavours in $SU(N)_X$ is given by the $\beta$ function (see,
e.g.\ \cite{Muta:1998vi}):
\begin{equation}
\beta_0 = \frac{1}{(4\pi)^2}\frac{11 N - 2 N_{fX}}{3} ~~,
\end{equation}
with $\beta<0$ corresponding to a non-confining theory and $\beta>0$
to an asymptotically free one. For example, QCD is asymptotically
free for $N_f \leq 16$. Thus there are two distinct cases, one
$U(1)$-like in which the exotic particles are not confined, and the
second technicolour-like, in which the exotic particles first
``hadronise'' among themselves into composite states of zero total
$X$. Mesonic $X$-hadrons would normally decay rapidly, but the
lightest totally antisymmetric $SU(N)$ state (the $X$-proton) could
be stable if an analogue of Baryon Number conservation holds in the
new sector (see, e.g.~\cite{Gudnason:2006ug,Gudnason:2006yj}). A QCD
hadronisation would then follow, neutralizing any leftover colour.
Some similarity to this picture is found in the hidden-valley models
of \cite{Strassler:2006im}, though mainly neutral exotics are
considered there.

Finally, we note that if the new gauge symmetry is broken, then it
is hard to see how the new states charged under it could be stable,
since there should be interactions which violate the conservation of
the corresponding charge. For the state to be long-lived \emph{and}
light, it would have to be an analogue of the muon, which has a
small mass due to a small coupling with the relevant
symmetry-breaking sector (the Higgs), but simultaneously a long
lifetime due to the gauge boson (the $W$) having a much larger mass.


\subsubsection{Magnetic Monopoles}~\label{mmtheory}
All particles so far observed possess values of magnetic charge
consistent with zero. The magnetic charge of the electron is
constrained by experiment to be
$q_e^m<10^{-24}g_D$\cite{Vant-Hull:1969ds}, where $g_D$ is the
elementary (Dirac) magnetic charge. However, in spite of a plethora
of experimental evidence to the contrary, strong theoretical
arguments continue to motivate searches for magnetic monopoles.


A potent motivation for searching for monopoles was given by Dirac
in 1931\cite{Dirac:1931kp,Dirac:1948um}. Dirac demonstrated that the
existence of only one monopole is necessary to accommodate electric
charge quantisation within quantum electrodynamics (QED). In a
modern form of Dirac's argument\cite{Griffiths}, the quantisation of
the total angular momentum in the field of a system of an electric
charge $q$ and a monopole with magnetic charge $g$ leads to the
quantisation condition $qg=\frac{nhc}{4\pi}$. Here $h$ is Planck's
constant, $c$ is the speed of light in vacuum, and $n$ is a quantum
number. Taking $q=e$ as the elementary electric charge, $n=1$ sets
the theoretical minimum magnetic charge which can be possessed by a
particle, $g_D=n\frac{137}{2}e$. A particle with magnetic charge
$g_D$ is known as a Dirac monopole. The value of the minimum
magnetic charge has profound implications for the construction of a
theory of high-energy monopole scattering. The magnetic fine
structure constant for a Dirac monopole is $\alpha_m
\approx\frac{137}{4}$, rendering perturbative field theory
inapplicable. This has implications for the reliability of models of
monopole production at colliders and for the exclusion limits, as
described in Sections~\ref{monocal} and \ref{monsearch},
respectively. A further consequence is that a Dirac monopole will be
largely equivalent to an electrically charged particle with charge
$\sim \frac{137e}{2}$ in terms of the electromagnetic force it
exercises and experiences. This implies that a Dirac monopole will
suffer a huge electromagnetic energy loss in matter compared with a
minimum ionising particle (MIP). This is discussed further in
Section~\ref{dedxMM}.

Although experiments typically search for Dirac monopoles there are
a number of reasons why alternative values of the minimum charge may
be favoured. If the elementary electric charge is considered to be
held by the down quark then the Dirac condition implies that the
minimum magnetic charge could be $3g_D$. However, it has been argued
that the Dirac condition is not appropriate for a confined
quark\cite{'tHooft:1975zi,Corrigan:1976wk,Preskill:1984gd} and an
observed minimum charge of $3g_D$ may suggest the existence of
isolated particles with charge $\frac{1}{3}e$ (free quarks). The
existence of isolated millicharged particles could imply a minimum
charge which is higher still. The value of the fundamental charge is
also affected if the Dirac argument is applied to a particle
possessing both electric and magnetic charge. Such a particle is
known as a
dyon\cite{Si,Schwinger:1968rq,Schwinger:1969ib,Yock:1969yv,DeRujula:1977rk,
Fryberger:1980sn}\footnote{Unless made clear by the context, the
term monopole is used to refer to both magnetic monopoles and dyons
in this paper.}. It has been speculated that a dyon could exist
either as a fundamental particle or as a composite of two particles,
one of which possesses electric and the other magnetic charge.
Schwinger argued that generalising the Dirac condition for a dyon
restricts values of $n$ to be
even\cite{Si,Schwinger:1968rq,Schwinger:1969ib}. It is therefore
important that searches are sensitive to as wide a range of magnetic
charge as is experimentally possible.

 While the Dirac argument provides strong motivation for the existence of
monopoles it gives no prediction of the likely monopole mass. Naive
arguments based on the classical radius of a Dirac monopole would
suggest a mass of the order of a few
GeV\cite{Klapdor-Kleingrothaus:1995as}. However, as long as the
production cross section is not vanishingly small, such a low-mass
monopole would long ago have been discovered at colliders. The
available window in mass for Dirac monopoles is largely determined
not by theoretical arguments but the results of the searches which
are described later in Section~\ref{monsearch}.

A further motivation for monopole hunting is provided by their
presence in grand unified theories (GUTs). 't Hooft and Polyakov
showed that monopoles possessing the Dirac charge, or multiples of
it, arise as topological defects of space--time. They occur when a
simple gauge group is spontaneously broken into an exact $U(1)$
subgroup\cite{'tHooft:1974qc,Polyakov:1974ek}. This would occur, for
example, in the phase transition
\begin{equation}
SU(5) \rightarrow SU(3) \otimes SU(2) \otimes U(1) \rightarrow SU(3)
\otimes U(1)
\end{equation}
 The spontaneous symmetry-breaking mechanism generates vector bosons $X$ with masses, $m_X$.
The monopole size $R_m$ can be related to the boson mass via $R_m
\sim m_X^{-1}$ and the monopole mass is $m_m \sim \frac{g_m^2}{R_m}
\sim \frac{m_w}{\alpha}$, where $\alpha$ is the common gauge
coupling at unification energy. Here, $m_X$ is typically of the
order of $10^{15}$ GeV, implying a super-heavy monopole of around
$10^{15}-10^{16}$ GeV.

In addition to massive GUT monopoles, it has also been postulated
that lighter monopoles can be produced through other
symmetry-breaking schemes. So-called intermediate-mass monopoles
(IMMs) with masses between $10^{7}$ and $10^{14}$ GeV have been
proposed \cite{Huguet:1999bu,Wick:2000yc}. These could occur in
models containing a more complicated gauge group than $SU(5)$, such
as $SO(10)$.
Both GUT monopoles and IMMs are beyond the reach of accelerator
searches.  Such monopoles are sought as primordial relics, which
could be be bound in matter or found in cosmic
rays\cite{Rhode:2002ut,Ambrosio:2004ub,Giacomelli:2005xz,Cecchini:2005ix}.

To obtain lower-mass gauge monopoles, to which accelerators could be
sensitive, requires their production via the electroweak
symmetry-breaking mechanism. Contrary to earlier work, which
asserted that the Weinberg-Salam model could not admit
monopoles\cite{Vachaspati:1992pi,Barriola:1993fy}, it has been
established that monopole solutions are
possible\cite{Cho:1996qd,Yang:1998sh,Yang:2001bb}. Furthermore, it
has been proposed that monopoles may possess masses as low as $~\sim
1$ TeV if the coupling strength of the quartic self-coupling of
W-bosons was modified, or the group $SU(2)\otimes U(1)$ is embedded
into a larger
 gauge group\cite{Bae:2002bm}. However, as
the authors of this work point out, there are
 theoretical difficulties with this approach.
  For example, the
modification of the Lagrangian may spoil the renormalisability of
the Weinberg-Salam model.

In higher-dimensional theories such as string theory, space--time is
often compactified on a topologically non-trivial manifold.
This naturally leads to a vacuum structure suitable for many
different
 kinds of defects, including monopoles.
 For example, in Ref.~\cite{Gauntlett:1992nn} a
 magnetic monopole solution is presented in the
 context of heterotic string theory.  The mass of
 this monopole is related to the size of the compact
  space, the string length and coupling. Since these
  parameters combine to determine the low-energy Planck
   mass and gauge coupling, the mass is constrained
  to be rather large, within a few orders of magnitude of the Planck scale.

Given the large amount of freedom in higher-dimensional theories
with regard to the size,
 dimension and topology of the compact space
 it is not unrealistic to believe that there
 may be monopole solutions with correspondingly
  low masses, so that they may be discovered at the LHC.

It should be noted that Dirac monopoles differ from monopoles
expected from gauge-symmetry breaking. Whilst Dirac monopoles are
considered point-like, gauge monopoles are expected to have a size
$R_m \sim \frac{1}{m_X} \sim 10^{-31}$m (for GUTs) and to possess a
complicated structure of vacuum particles surrounding it. It has
been argued that, compared to point-like monopoles, the production
of gauge monopoles from particle collisions will be suppressed by a
factor of $\geq 10^{30}$ due to form factors\cite{Drukier:1981fq}.

\subsubsection{$Q$-balls}

$Q$-balls represent a further possibility of producing topological
SMPs. The fastest way to understand what they are is to consider a theory of charged scalars with SO(2) internal symmetry, in other words a two dimensional internal space of scalar fields $\phi_1$ and $\phi_2$ with a potential which is only a function of $\phi=\sqrt{\phi_1^2+\phi_2^2}$.  We assume that $\phi$ is constant within a (real space) sphere of radius $R$ and zero outside that sphere.  If $\phi_i$ is rotating around the internal SO(2) symmetry with a specific angular frequency $\omega$, then the conserved charge $Q$ is given by
\begin{equation}
Q=\int d^3 x
\left[\phi_1\partial_0\phi_2-\phi_2\partial_0\phi_1\right]=\frac{4\pi}{3},
R^3\omega\phi^2 \label{qcharge}
\end{equation}
whereas the energy $E$ is given by
\begin{equation}
E=\frac{4\pi}{3}R^3\left(\frac{1}{2}\omega^2\phi^2+V\right),
\label{qenergy}
\end{equation}
where $V$ is the potential of $\phi$.  Using Eq.~\ref{qcharge} to
replace $\omega$ in Eq.~\ref{qenergy}, it is clear that there is a
certain radius $R$ at which
 the energy is minimised, and at
 this minimum $E=Q\sqrt{2V/\phi^2}$.
 This field configuration or ``$Q$-ball'', could
 decay by emitting the basic scalar
 particle associated with the equations
 which has unit charge and
 mass $m^2=\left.\partial^2V/\partial\phi^2\right|_{\phi=0}$.  However
 the energy per unit charge of the configuration will be $E/Q=\sqrt{2V/\phi^2}$, so
 that if the mass of
  the scalar $m> \sqrt{2V/\phi^2}$, the Q-ball will not be able to decay into 
 these scalar particles~\cite{Lee:1974ma,Friedberg:1976me,Friedberg:1976ay,Friedberg:1976eg,Coleman:1985ki}.   
Many
  potentials have characteristic logarithmic
  one-loop corrections which allow $m>\sqrt{2V/\phi^2}$, not
   least in SUSY. There are examples of $Q$-balls
   in both gravity mediated \cite{Kasuya:2000wx} and
   gauge mediated \cite{Kasuya:2000sc} SUSY-breaking models.
    For generic values of the parameters in these models,
     $Q$-balls must have rather large masses, much higher than the electroweak scale, in
     order to be stable \cite{Kusenko:1997si,Kusenko:1997vp}.Although they may be created at the end of inflation, it is less likely that they
      would appear at the LHC. However, if present, they could
      manifest themselves as very highly ionising particles. Coupling the scalar field
 to fermions can make the
  $Q$-ball unstable, since decay into
  those fermions would be possible~\cite{Cohen:1986ct}.

\section{Cosmological implications of SMPs at Colliders}\label{sec:cosmo}

 If stable or quasi-stable particles are
produced at the LHC, they will also have been produced in the early
universe.  If they are unstable, the lifetime of those particles
then determines at which cosmological epoch they will decay.  Such
decays generically involve the creation of energetic SM particles,
which cause a variety of problems that we outline below.  In this
way one can put constraints upon the regions of parameter space
where quasi-stable objects might be created.

If there are relic stable or quasi-stable particles which are
produced in the early universe then they fall into three broad
categories
\begin{itemize}
\item particles which serve as dark-matter candidates.
\item particles which are ruled out since their presence or decay
is in one way or another incompatible with what is observed (see below).
\item particles which exist in such small quantities that they do not serve as dark matter candidates and have not yet been detected.
\end{itemize}

Of particles which appear in extensions of the SM, very few fulfill
the criteria necessary to fall into the first category whereas very
many fall into the second or third.

``Incompatible with what is observed" can have several meanings, the
first being due to general relativity.  The
Friedman-Robertson-Walker solutions of Einstein's equations tell us
that the rate at which the universe expands depends upon the energy
density of the contents of the universe. One way a relic particle
could be incompatible with what is seen is that the presence of that
particle would change, via gravity, the expansion history of the
universe in a way which would be contrary to the expansion history derived from observations.  The second possible meaning is
that the particle would decay and its decay products would destroy
the observed light elements (which are formed in the first few
minutes after the Big Bang) or be visible today, for example in the
form of gamma rays. A third possible meaning is that the particle
should have already been detected experimentally by magnetic
monopole detectors, searches for anomalous isotopes or other
experiments of that nature. Finally the particle might exist in such
an abundance that it gives the correct expansion history
gravitationally, but its interaction with itself or SM particles is
such as to rule it out as a dark-matter candidate.

We will introduce the current state of cosmology before outlining
the known properties of dark matter.  We will then discuss the
constraints upon two supersymmetric candidates for quasi-stable
objects, which could be produced at the LHC, namely gluinos and
sleptons.  Then we will summarise the constraints from cosmology and
astrophysics upon magnetic monopoles. Finally we will discuss the
implications for cosmology if a charged massive stable object is
produced at the LHC.

\subsection{Cosmology Overview}

Here we will describe briefly how the energy of the universe is
divided between normal matter, dark matter, and dark energy in a way
relevant for the subsequent discussion of SMPs at colliders.  For
more in depth reviews of cosmology see e.g. \cite{bergstromgoobar}.

The ratios between the abundances of light elements such as
hydrogen, helium, and lithium in the universe depend upon the baryon
to photon ratio roughly one minute after the Big Bang
\cite{steigman}. By comparing their observed abundances with the
number of photons we observe in the cosmic microwave background
(CMB) today, we can obtain the number of baryons in the universe.
One can also measure the abundance of baryons using the CMB alone
\cite{Spergel:2006hy}.

General relativity provides us with the relationship between the
expansion of the universe and the density of matter inside it
\cite{bergstromgoobar}. The density in the form of baryons is not
enough to explain the observed expansion measured by, for example,
the Hubble Space Telescope~\cite{Freedman:2000cf}. Furthermore,
there appears to be invisible matter in galaxies and clusters of
galaxies, the presence of which is deduced from the motion of gas
and stars in those objects \cite{zwicky,galrotationcurves}. Studies
of the clustering of galaxies predict that there is 6-8 times as
much dark matter as there is baryonic matter \cite{2df,sdss}.
Observations of type 1a supernovae tell us that the universe
recently started to accelerate \cite{astier,riess}.

The conclusions drawn from these different observations are the
following - around $95\% $ of the energy density in the universe is
non-baryonic and it couples at most weakly to SM particles so
that it cannot be seen. Roughly one third of it is dark matter while
the other two thirds does not dilute quickly as the universe expands, i.e. it is dark energy rather than dark matter.  At the time of writing, dark energy does not appear
to be related to particles produced at the LHC in any direct way
and we shall not discuss it further.  The identity of the
dark matter however is obviously closely related to the rest of this
review. Dark-matter particles may be produced at the LHC, so the
more we can learn about their behaviour and properties from
cosmology, the easier it will be to find them.

If stable massive particles are discovered at the LHC, their
couplings and masses should not be such that they are over-produced in the early universe.
If there is too much matter in the universe, then the
radiation-dominated regime would finish earlier and the universe
would be younger by the time the CMB had been red-shifted to the
temperature that we see today.  The universe needs to be at least
around $12\times 10^{9}$ years old since observed objects such as
globular clusters are at least that old.  Together with structure
formation, this limit provides strong constraints upon the
over-production of new forms of matter (see e.g.
\cite{Fairbairn:2001ab}).

\subsection{Constraints on the basic properties and interactions of dark matter.}

Before we look at possible particle candidates for the dark matter
in the universe, it is interesting to summarise the information we
have about its properties.
\subsubsection{Cold dark matter}
By definition, dark matter is at most weakly coupled to Standard
Model fields so it can only have been produced in the early
universe.  The simplest mechanism for dark-matter production is to
have the relics be in thermal equilibrium with the rest of the (standard model)
plasma at early times.   We can use crossing symmetry to relate the annihilation rate to the production rate.
The criterion for thermal equilibrium is
that this annihilation rate, which depends upon the coupling and mass
of the dark-matter candidate, is comparable to the inverse timescale
of the expansion of the universe, which depends only, via gravity, upon the energy density of the universe. This means that the universe will
not change its size or temperature much over the timescale
of the particle interaction or decay rate. The dark matter observed
in the universe today is probably cold, meaning that it was
non-relativistic at the temperature at which it came out of thermal
equilibrium or 'froze' out of the plasma.  In contrast, the upper
bound on the neutrino mass \cite{neutrinoupperbound} tells us that
neutrinos were relativistic until rather late times, at most just starting to slow down at matter-radiation equality.  Indeed, depending upon the mass hierarchy, some of the neutrino species may still be
relativistic today. Such fast-moving or hot dark-matter candidates
will be able to escape all but the largest collapsing structure in
the universe, so structure formation in a universe with hot dark
matter would start with the largest structures forming first.

Observations suggest that the opposite is true, and that the
smallest structures form first \cite{sdss} indicating that the
dark matter was non-relativistic in the early universe.  This
provides us with information about the coupling of the particle to
the Standard Model (annihilation rate) and the mass of the particle:
the annihilation rate should still be large at temperatures when the
particle is not extremely relativistic.

Strictly speaking however, the fact that dark matter is cold does
not mean that it had to be non-relativistic at freeze-out; its
kinetic energy may have been red-shifted away by the expansion of
the universe between freeze-out and the epoch at which structure
formation begins when the universe becomes matter-dominated, a few
thousand years after the Big Bang. Because of this, one can obtain
cold dark matter candidates with masses at least as low as the MeV
range \cite{MeV} and probably even lower.  Most models of cold dark
matter under consideration today involve particles with masses in
the GeV to TeV range, and the combination of this mass range with
the coupling that will give good relic abundance means that it is
possible that they could be created at the LHC \cite{darkmatterLHC}
(one notable exception is the axion \cite{Sikivie:2005zz}).  Cold
dark-matter candidates which interact weakly with Standard Model
particles are called weakly interacting massive particles or WIMPS.

\subsubsection{Warm dark matter}

Dark-matter haloes in computer simulations have steeply varying
density profiles in the core, so that typically one would expect
that the density of dark matter $\rho\propto r^{-\gamma}$ where $r$
is the radius from the centre of the halo, and $\gamma$ typically has
values between 1 and 1.5 \cite{nfw,moore}.  However, reconstructions
of the distribution of dark matter based upon observations of low
surface brightness galaxies suggest a much less steep density
profile.  Such galaxies seem to possess an approximately constant
density core of dark matter~\cite{LMSB}.  Another discrepancy is
that there are many more satellite galaxies (smaller galaxies like
the Magellanic clouds) produced in $N$-body simulations of cold dark-matter universes than are observed in nature (for instance, in the
vicinity of the Milky Way) \cite{Klypin:1999uc}.  There are a number
of people who have suggested that these discrepancies can be
explained if the dark-matter candidate is 'warm', in other words
half way between cold and hot dark matter \cite{wdm}. This latent
thermal energy would smooth out the central density cusps.

A warm dark-matter candidate produced thermally would usually have a
much smaller mass than a GeV, so the prospects for dark-matter
physics at the LHC would change considerably.  There is the
possibility that weakly interacting neutralinos are produced
non-thermally just before nucleosynthesis \cite{Lin:2000qq} so that
they could be fast moving during structure formation, but producing
a relic density of neutralinos in this way ignores the fact that one
can obtain the correct relic density thermally without too much
fine-tuning.  There are equally a number of people who think that
these discrepancies between observations and computer simulations
may be explained using astrophysics rather than changing the
properties of cold dark matter~\cite{Kravtsov:2004cm}.

\subsubsection{Dark-matter self-interaction}

Constraints upon the self-interaction of dark matter can
be obtained by looking at colliding galaxy clusters.  The bullet
cluster is an example of two galaxy clusters colliding, the hot gas
forming shock waves at the interface \cite{Markevitch}.
Gravitational lensing shows that the two dark-matter populations
have moved through each other without forming shocks as far as can
be observed.  This allows one to place (mass dependent) constraints
upon the self-interaction cross section of dark matter of
$\sigma/M<2$ barn/GeV. This is actually rather a large cross section
in terms of what we are familiar with in the Standard Model, for
comparison $\sigma_{QCD}\sim$ barn (at MeV energies). Some
authors have in fact suggested that the dark matter does interact
with itself, and that this feature solves some of the same problems
which people try to solve using warm dark matter~\cite{Wandelt:2000ad} (see the previous subsection). One might
therefore think that a colour-singlet particle with a typical QCD
cross section $\sigma\sim \Lambda_{QCD}^{-2}$ (for example, perhaps
some kind of glueball) might be a viable cold dark-matter candidate.
However, the constraints on the interaction between dark matter and
baryonic matter are much stronger (as we shall see in the next section).

\subsubsection{Dark-matter coupling to Standard Model particles}

\begin{figure}
\begin{center}
\includegraphics[height=9cm,width=11cm]{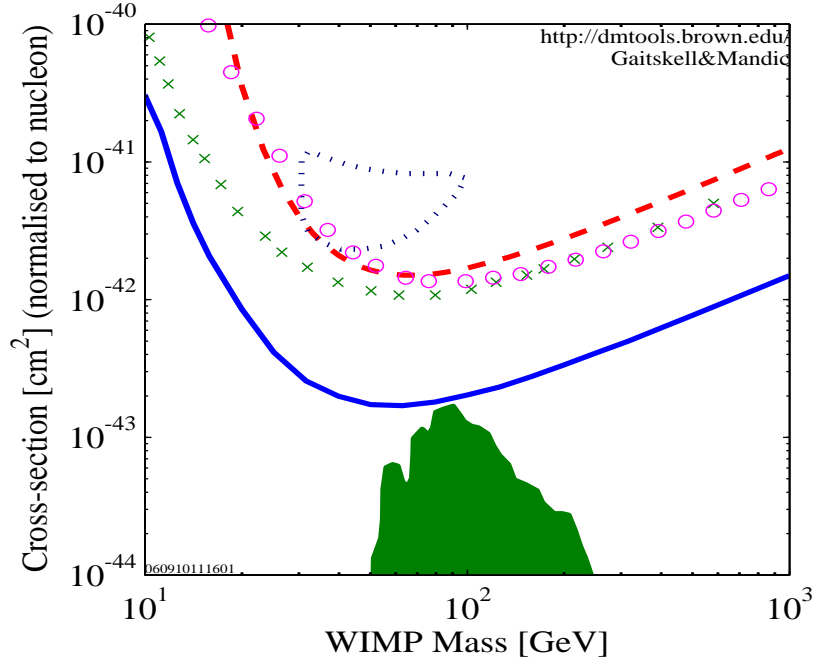}
\caption{Constraints upon direct detection of dark
matter~\cite{onlinedm,Akerib:2004fq,Angloher:2004tr,Bernabei:2000qi,Alner:2005pa}. The red dashed curve is from CRESST, purple
circles are from Edelweiss 1, green crosses are from ZEPLIN 1 and the solid
blue curve is from CDMS.  The region enclosed by dashes is the claimed
detection from the DAMA collaboration which does not seem to be confirmed by the other experiments. The dark green filled region
are a set of example SUSY models from Ref.~\cite{baltzgondolo}.}
\label{directdm}
\end{center}
\end{figure}

The coupling of the dark matter to SM particles is constrained
experimentally in two basic ways.  The first is by direct
experiments, which are looking for momentum exchange between the
dark matter and sensitive detectors such as CDMS, CRESST, DAMA and ZEPLIN
\cite{Akerib:2004fq,Angloher:2004tr,Bernabei:2000qi,Alner:2005pa}. Observations of rotation curves
of the Milky Way suggest that the density of dark matter near the
earth is approximately 0.3 GeV cm$^{-3}$ and the velocity
corresponding to the gravitational potential of the Milky Way is
around 200 km s$^{-1}$.  These numbers are based upon the assumption
of a smooth dark-matter halo and could turn out to be (locally)
wrong by more than an order of magnitude.  Using these numbers, the
experimental status is outlined in Fig.~\ref{directdm} and the
low-energy cross section between a 100 GeV dark-matter particle and
a nucleon is constrained to be less than $\sim 10^{-43}$ cm$^2$.
A heavy particle produced at the LHC undergoing Standard Model
strong or electromagnetic interactions cannot therefore be a dark-matter
candidate.

The annihilation of dark-matter particles into Standard Model
particles determines the rate of interaction of dark matter with the
rest of the plasma and hence determines if the dark matter is in
thermal equilibrium with the rest of the universe or not
\cite{bergstromgoobar}.  When the dark matter goes
out of thermal equilibrium, it essentially stops interacting and its
number density becomes constant (more precisely the number density
per co-moving volume - the universe is expanding, so the actual
number density will decrease due to dilution).

The analysis of the three-year WMAP data tells us that the density
of dark matter is $\Omega_{DM} h^2=0.102\pm0.009$ where
$\Omega_{DM}$ is $\rho_{DM}/\rho_{crit}$, $\rho_{crit}$ is the
density corresponding to a flat universe \cite{bergstromgoobar} and
$h$ is the Hubble constant in units of 100 km s$^{-1}$ Mpc$^{-1}$.
In order to obtain this relic density, the thermally averaged cross
section for dark-matter self-annihilation into Standard Model
particles should be \cite{jungmanprep}
\begin{equation}
\Omega_{DM} h^2\sim \frac{3\times 10^{-27}\rm
cm^3s^{-1}}{\langle\sigma v\rangle} \label{abundance}
\end{equation}
so that $\langle\sigma v\rangle\sim 3\times 10^{-26}\rm cm^3s^{-1}$
is favoured.  A cold dark-matter candidate produced at the LHC
should therefore have this annihilation cross section.  This
quantity leads us to the second method of measuring the coupling of
dark matter to Standard Model particles, namely through the search
for the annihilation or decay products of dark matter coming from
high-density regions such as the centre of galaxies \cite{larsberg}.
Since the WMAP results give us rather good information about
$\langle\sigma v\rangle$, the uncertainties in this approach lie in
the lack of knowledge of the exact density of dark matter in dense
regions such as the centre of galaxies and in separating the signal
from dark-matter annihilation from possible background signals.

\subsubsection{Dark matter at the LHC}
In Section~\ref{sec:scen} we have learnt about a number of dark-matter candidates.  The most common SUSY candidates in the
literature are neutralinos.  These are WIMPS, which are deemed
attractive candidates for dark matter since there is considerable
parameter space in SUSY models where these particles have
annihilation cross sections rather similar to what is required to
obtain the abundance suggested by WMAP, as described in
Eq.~\ref{abundance} \cite{jungmanprep,darksusy}.

Universal extra dimensions also provide a natural dark-matter
candidate if radiative corrections make the lowest excitation of the
hypercharge gauge boson $B$ the lightest KK mode \cite{Servant:2002aq}.  KK modes of other species will decay into the lightest mode by emitting zero mode excitations, in other words normal SM particles. The lightest mode is stable and if it is aslso weakly interacting it can be a good WIMP candidate.
There are tight constraints upon the radius and the number of
compact dimensions if this kind of UED KK dark matter is to provide
us with the correct abundance, in particular a KK mass around a TeV
is favoured, which is encouraging for LHC studies.  If the lightest
KK mode turned out to be a particle with EM or colour charge rather
than the photon then it would be difficult to see how UED could
remain a good candidate for dark matter.

There is a large amount of literature on WIMPS (see e.g.
\cite{jungmanprep,larsberg,darksusy,hoopersilk} and references
therein) and since they are only weakly charged we will not
concentrate on them here.  They do however form candidates for the
LSP in SUSY models when the NLSP is a charged particle, which then
decays into neutralinos.

\subsection{Cosmological constraints on quasi-stable sleptons\label{slepton}}

As described in Section~\ref{sec:scen} if the LSP is the gravitino
and the NLSP is the stau, i.e. the scalar super-partner of the tau
lepton, then the lifetime for the stau decay into tau lepton plus
gravitino will be relatively large due to the Planck-suppressed
couplings to the gravitino \cite{buchmuller0402179}.

The gravitino itself is a perfectly good cold dark-matter candidate,
although it would be impossible to observe through its coupling to
the Standard Model fields. There is however a potential problem with these dark matter scenarios related to nucleosynthesis.

The epoch of nucleosynthesis in the early universe occurred when the
temperature was low enough for photodisintegration of nuclei to have
ceased but high enough for nuclear fusion reactions to occur.
Protons and neutrons interchange regularly due to beta and inverse
beta decay driven by the background thermal bath of neutrinos and
electrons. Eventually the plasma becomes cool enough for various
light elements and isotopes such as deuterium, helium-3 and 4, and
lithium-6 and 7, to form.

Nucleosynthesis gives very specific predictions for the ratio
between the light elements H, He, D, Li etc. that one should observe
when one makes observations of the abundance of these
elements in regions of the universe where they have not been
processed in stars.

The strongest constraint upon the quasi-stable stau comes from this
observed light element abundance.  If the decay of the slepton into lepton plus
gravitino takes place during or after nucleosynthesis then energetic
particles will be injected into the plasma, leading to various
effects such as the dissociation of these light elements.

In particular, the injection of energetic protons into the plasma
due to the decay of unstable relic particles will lead to the
reaction $\rm p+^4\hspace{-0.1cm}He\rightarrow
^3\hspace{-0.1cm}He+D$ which increases the abundance of deuterium
\cite{jedamzik,arvanitaki}.  This deuterium can also subsequently
interact with helium to increase the abundance of $\rm ^6Li$.

The $\rm^3 He/D$ ratio is a strong function of the decay
of the slepton into gravitino \cite{Buchmuller0605164,Steffen:2006hw} and the observed value of  $\rm^3 He/D$ restricts the stau mass $m_{\tilde{\tau}}$ to be
greater than about $5 m_{\tilde{G}}$ for $m_{\tilde{\tau}}\sim 500$ GeV or $20 m_{\tilde{G}}$ for $m_{\tilde{\tau}}\sim 100$ GeV.  This cuts into the interesting
region for collider searches.  For instance, one can in principle
reconstruct the gravitino mass by simply measuring the mass of the
stau and the energy of the emitted tau \cite{buchmuller0402179}
\begin{equation}
m_{\tilde{G}}^2=m_{\tilde{\tau}}^2+m_\tau^2-2m_{\tilde{\tau}}E_{\tau}
\end{equation}
but $m_{\tilde{G}}$ can only be measured in this way if it is not too
small.  Since the constraints from light element abundance restrict
the gravitino mass for a given stau mass, much of the interesting
parameter space is ruled out.  One way of avoiding this is for there
to be entropy production at some temperature below the freeze-out of
staus from the plasma and the era of nucleosynthesis.  Such entropy
production will increase the number of relativistic particles in the
plasma and effectively dilute the sleptons, so that when they do
decay, their effect upon the light element abundances is diminished.
We will mention this more in Section~\ref{round}.

 One can also consider situations where the superpartner of the axion, the axino is the LSP.  This gives rise to analagous dark matter physics, although different constraints from nucleosynthesis \cite{Covi:2001nw}.

\subsection{Stable and quasi-stable gluino}
In this section we present the cosmological constraints upon the
stable and quasi-stable gluino.

\subsubsection{Stable gluino}

The case of the stable gluino is presented in detail in
Ref.~\cite{Baer:1998pg}.  If the gluino is the LSP then the relic
abundance is calculated by obtaining the thermally averaged
annihilation rate and comparing it to the expansion of the universe,
giving the approximate result
\begin{equation}
\Omega_{\tilde{g}} h^2\sim\left(\frac{m_{\tilde{g}}}{\rm 10
TeV}\right)^2. \label{gluest}
\end{equation}
 The stable gluino could not be the dark matter as its coupling
with Standard Model fields is greatly in excess of cross sections
that are already ruled out (see fig. \ref{directdm}).  We therefore
require $\Omega_{\tilde{g}} h^2\ll 0.1$, giving a maximum
 mass for the gluino of a few TeV.  These gluinos will be captured by nuclei,
leading to the possible detection of (highly) unusual isotopes
of nuclei \cite{heavyhydrogen}. Non-observance of these
isotopes restrict values of $m_{\tilde{g}}\ge$ 100 GeV although
non-perturbative effects could change the values predicted by equation (\ref{gluest}).

\subsubsection{Quasi-stable gluino}
\begin{figure}
\begin{center}
\includegraphics[height=8cm,width=14cm]{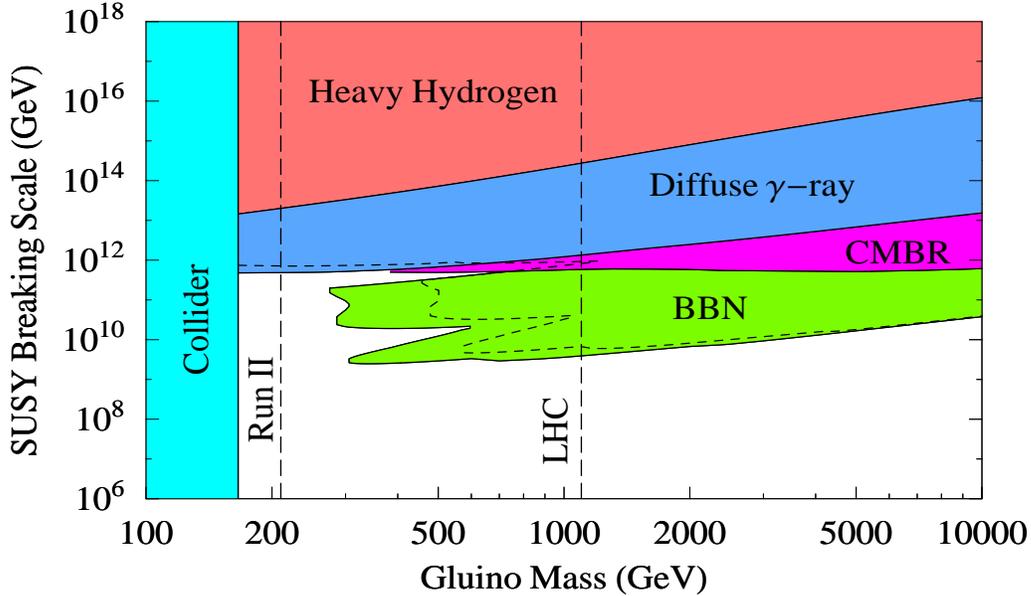}
\caption{Constraints upon the gluino mass and SUSY breaking
scale parameters in models of split supersymmetry. Shaded regions
are ruled out by different cosmological observations.  The tightest
constraints come from the diffuse gamma ray background for
$m_{\tilde{g}}<300$ GeV and from Big Bang nucleosynthesis (BBN) for
$m_{\tilde{g}}>300$ GeV. The plot is taken from
Ref.~\cite{arvanitaki}.} \label{gluino}
\end{center}
\end{figure}

The constraints upon the quasi-stable gluino come from the same
places as the constraints upon the slepton described in the previous
section.  We have seen in some versions of split supersymmetry
\cite{Arkani-Hamed:2004fb} that gluinos can be produced which then decay outside
the detector.  The lifetime of these gluinos is given by \cite{Gambino:2005eh}
\begin{equation}
\tau\sim 4\rm s\left(\frac{m_S}{10^9
GeV}\right)^4\left(\frac{1 TeV}{m_{\tilde{g}}}\right)^5
\label{gluinoeq}
\end{equation}
where $m_S$ is the scale of supersymmetry breaking, which is
typically very large in these models, $\gg 10^{10}$ GeV. The gluino lifetime as function of $m_S$ is displayed in Fig.~\ref{fig:gluinolife}.  As for
the stable gluino, constraints from the non-observance of anomalous
isotopes become stronger as one raises the energy of $m_S$, since
the anomalous isotopes will be increasingly stable for cosmological
time-scales.  Fig.~\ref{gluino} shows how different cosmological
constraints rule out values of $m_S$ which depend upon
$m_{\tilde{g}}$, for example, $m_S$ must be less than $10^{13}$ GeV
for a gluino mass $m_{\tilde{g}}\sim 200$ GeV or less than $10^{16}$
GeV for $m_{\tilde{g}}\sim 10$ TeV \cite{arvanitaki}.

As one lowers the value of $m_S$, the lifetime of the gluino falls,
and the annihilation products become progressively more troublesome
for cosmology.  As one would expect, as the lifetime drops, the time
at which the strongest constraints emerge relates to progressively
earlier epochs in cosmology.  For example, at values of $m_S$, lower
than those ruled out by searches for heavy isotopes, the parameter
space is ruled out through constraints upon the diffuse gamma-ray
background today made by the EGRET gamma ray observatory
\cite{kribs,egret}.

Values of $m_S$ greater than about $10^{12}$ GeV are ruled out since
the decay of such gluinos would lead to distortions in the
thermalisation of the CMB before recombination. The CMB is the radiation
 remaining from the time when the temperature of the universe decreased such that the mean free path for photons approaches infinity.  At
 this epoch, a few hundred thousand years after the Big Bang, the
 universe becomes transparent. The CMB is the most
perfect blackbody ever observed, so distortion of the spectrum by
the injection of high-energy photons into the plasma leads to strong
constraints \cite{husilk}.

Finally gluinos with very short lifetimes will, like the unstable
stau, cause problems for light-element abundance.  The decaying
gluinos will photo-dissociate the light elements left over after
nucleosynthesis although for $m_{\tilde{g}}< 300$ GeV the gluino
lifetime is long enough to avoid the nucleosynthesis constraint \cite{arvanitaki}.

Very roughly, the combined constraints upon the gluino are as
follows - if the gluino mass $m_{\tilde{g}}<300$ GeV then the SUSY
scale $m_S<10^{12}$ GeV whereas if $m_{\tilde{g}}>$ 300 GeV then
$m_S<10^{10}$ GeV.

\subsection{Consequences of the observation of a charged stable object at the LHC\label{round}}

\subsubsection{Electrically charged particles}

If a charged massive stable object is discovered at the LHC then it
may be very interesting for cosmology. If the object were truly
stable, then it would be difficult to understand why it is not
observed today.  We have already discussed that such a charged
object could not be the dark matter.  On the other hand, if the
object were found to possess a lifetime such that one would expect
its decay to be problematic for nucleosynthesis or the CMB, or that
it would overproduce gamma rays, again it would be difficult to
understand.

\begin{figure}
\begin{center}
\includegraphics[height=9cm,width=12cm]{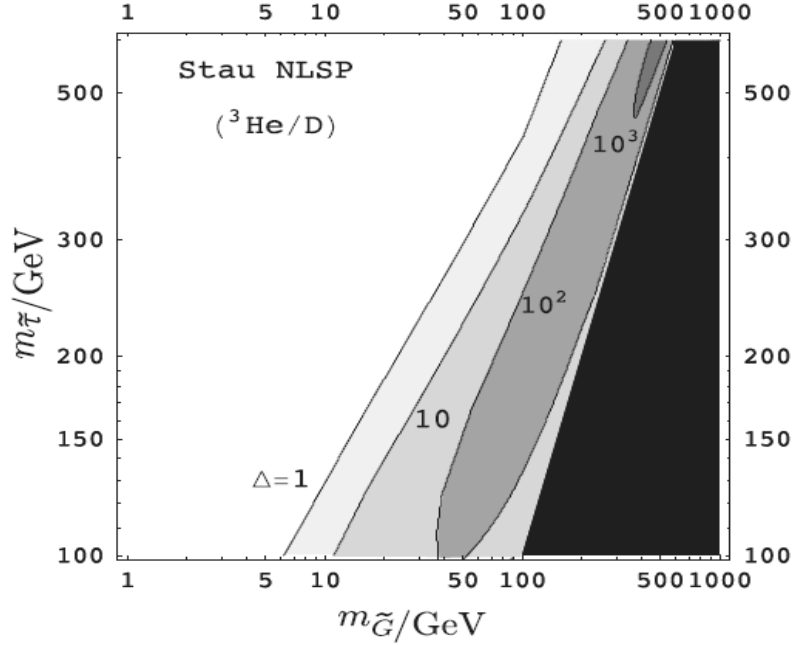}
\caption{Amount of entropy which needs to be produced in order for
the unstable stau to be consistent with nucleosynthesis constraints,
see Eq.~\ref{entropy}.  Plot taken from
\cite{Buchmuller0605164}.} \label{stauent}
\end{center}
\end{figure}

If such a situation were to arise, it would be necessary to invoke a
mechanism, which would dilute the density of the relic.  Let us call
our problematic charged stable particle $X$.  If there were another
species of particle $Y$, which decayed in between the time at which
the stable relic $X$ froze out of the plasma and the time at which
it ($X$) causes problems (by decaying or by not being observed), then
it might be possible to dilute the particle.  The decay of the
particle $Y$ may produce more light particles such as photons,
thereby reducing the number density per photon of the dangerous
relic $X$.  Another way of looking at this is to say that the decay
of particle $Y$ into light particles temporarily stops the
temperature in the radiation bath from dropping as the universe
expands.  At the same time, the number density of particle $X$ {\it
would} drop as it is no longer in equilibrium with the radiation
bath but has already frozen out.  In each particular scenario, the dangerous relic particles would have to be diluted by a factor $\Delta$ in order for their decay not to be dangerous for light element abundance. The
dilution of the dangerous $X$ particle (in this case the stau) is written
\begin{equation}
\frac{n_{Xbefore}}{s}=\frac{1}{\Delta}\frac{n_{Xafter}}{s}
\label{entropy}
\end{equation}
where $n_X$ is the number density of the $X$ particle and $s$ is the
entropy density, which is roughly equal to the number of relativistic
particles in the plasma per unit volume.    As an example,
Fig.~\ref{stauent} shows the amount of entropy, written as the
required value of $\Delta$, which should be produced after
freeze-out in order for the unstable stau described in Section~\ref{slepton} to be compatible with constraints from light element
abundance \cite{Buchmuller0605164}.

Another effectively equivalent way of diluting the density of a
stable relic is via inflation and reheating.  In principle there
could be a period of inflation at an energy scale lower than that of
the LHC. However, if this period of inflation were driven by an
inflaton it would have to be rather strongly coupled to the Standard
Model particles in order to achieve thermalisation before the beginning of
nucleosynthesis (see however
\cite{lowscaleinflation,Lyth:1995ka,Lyth:1995hj}).

\subsubsection{Magnetic Monopoles}

Theories which contain the possibility of spontaneous breaking of
some gauge symmetry to a true vacuum with a non-trivial topology
contain solitonic objects, which are topologically stable.  These
objects are simply classical field configurations of the equations
of motion.  The simplest example of this is when there are two
discrete degenerate vacua; this is the familiar case of a massive
real scalar field with self-interaction and a negative mass-squared
term.  In separate, causally disconnected regions in the early
universe, the field will take a random decision as to which vacuum
it will inhabit when the temperature drops below the critical
temperature corresponding to the phase transition.  Later, as the
horizon grows and different parts of the universe come into causal
contact with each other, regions are found where the field is forced
to interpolate (smoothly due to the equations of motion) between one
vacuum and another vacuum.  This region corresponds to a domain
wall, a wall trapped in the symmetric state of the early universe.
For different vacuum structures, different classes of objects can be
produced such as strings for a U(1) vacuum.  It is in this way that
monopoles are formed cosmologically, when the vacuum state is a
continuum of degenerate vacua with a spherical topology in field
space.  For more details on different topological objects, see
\cite{kibble,vilenkinprep}.

The idea that there should be on average one monopole created per
causally disconnected region in the early universe was first
intoduced by Kibble \cite{kibble}. Any theory of particle physics
where there is too much non-relativistic matter created in the early
universe gives problems for cosmology since no monopoles have been
detected.  One of the initial motivations for the theory of
inflation was to provide a dilution of the density of monopoles
created in the early universe \cite{guth}.

A very tight constraint upon the density of monopoles comes from the
observation by Parker that the presence of magnetic monopoles would
limit the ability of astronomical objects to build up magnetic
fields \cite{parker}.  The constraint for interstellar space is that
there should be less than one monopole per $10^{28}$ nucleons today.
This means any symmetry-breaking scale larger than $10^{10}$ GeV
would lead to an 
unacceptably large number of magnetic monopoles if
inflation does not occur.

\section{Modelling the production of SMPs at colliders}
In this section the techniques used to model the production of heavy
particles at colliders are discussed. The aim of this section is not
to provide an overdetailed description of the phenomenology of
proposed, undiscovered particles, but instead to show how techniques
and models developed for SM particles can profitably be used to
predict the gross features of SMP production at colliders. We focus
mainly on SMPs which do not possess magnetic charge in view of the
inapplicability of perturbative field theory to monopole processes.
A description of monopole production mechanisms which have been
considered in collider searches, and the approximations which have
been used to calculate their rates is given at the end in Section
\ref{monocal}.

\subsection{Production rates}\label{prodrate}

Exotic stable or long-lived new particles are usually thought to be
 pair-produced at a collider:
\begin{equation}
ab \to X_c X_d,
\end{equation}
where $a$ and $b$ are normal constituents of the incoming beams, and
$X_c$ and $X_d$ belong to the same new theory. While pair production
may occur at a large rate, the decay of particles $X_c$ and $X_d$
may be suppressed by the existence of a new (almost) preserved
quantum number, possibly resulting in long-lived particles which
could interact in the detector as an SMP. Besides direct production,
SMPs could be produced via the decay of a heavier particle, for
example if a state $X_i$ would decay rapidly into a lighter state,
which in turn cannot (rapidly) decay any further.

If the production of a particle $X$ is allowed singly
\begin{equation}
ab \to X,
\end{equation}
then  the decay $X \to ab$ may also occur. The production cross section is
directly proportional to the decay width, \mbox{$\sigma(ab \to X) \propto \Gamma(X \to ab)$}, where
the constant of proportionality involves mass, spin and colour factors, as well as
the $ab$ partonic flux, but no model-specific details. A singly-produced particle which
is sufficiently long-lived so as to penetrate through a detector, has, therefore,
a production rate which
will be negligible, as can be seen from the following example. Consider a colour-singlet
scalar state with mass 200 GeV, coupling only to gluons. A lifetime of $c\tau = 1$~mm will
translate into a small production rate of 0.0002 events per year, assuming full
LHC luminosity, 100 fb$^{-1}$. The larger the lifetime, the smaller is the production rate.
If the state is a colour octet, there is a factor eight enhancement, or if the particle has spin one,
another factor three must be added. However the
production rates will still be negligible. A potential exception could be off-shell decays, such as $X \to W^+ W^-$ for $m_X \ll 2 m_W$, where
$\Gamma$ may be strongly kinematically suppressed, while $\sigma$ would be
less affected since the incoming $W$'s, emitted off the beam constituent quarks, are spacelike. In practice, taking
into account current experimental limits discussed in Section~\ref{ss:search} as well as
the impact of loop-induced decays, such prospects are excluded.

\begin{figure}[h]
\begin{center}
\epsfig{file=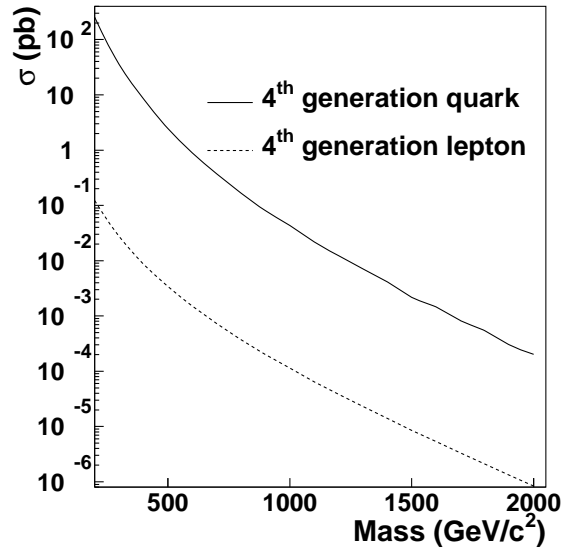,height=8cm,width=8cm} \caption{{\sc Pythia}
predictions of the pair-production cross section for
fourth-generation quarks of charge $\pm \frac{2}{3}e$, and leptons
of charge $\pm e$ at the LHC. \label{fig:newpart}}
\end{center}
\end{figure}

Given the Lagrangian of the theory, and the values of relevant
couplings and masses, the parton-level differential cross section
for $\mathrm{d} \hat{\sigma}(ab \to X_c X_d)$ is readily obtained.
Convolution with the $a$ and $b$ parton densities gives all relevant
differential distributions, and integration over phase space
provides the total cross section. The pair production cross section of coloured
particles is of \Oaa, while that of colour-singlet ones is of
$O(\alpha_{ew}^2)$, where $\alpha_{ew}$ represents electroweak
couplings. Thus the production rate for colour-singlet particles
would be a factor $(\alpha_{ew}/\alpha_{s})^2$ times smaller.
Additionally, the latter cannot couple directly to the gluonic
content of the beams, so typical production rates for colour-singlet
particles are about 2 to 4 orders of magnitude lower in rate. To
illustrate this, Fig.~\ref{fig:newpart} shows {\sc Pythia}
predictions for the pair-production cross section at the LHC of
exotic fourth-generation quarks with charge $\pm\frac{2}{3}e$ and
fourth-generation leptons with charge $\pm e$, as a function of the
mass of exotic particles. The lepton cross section is clearly
dwarfed by that of the quarks. However, this argument only concerns
the direct production mechanisms. Weakly interacting particles can
be produced at a large rate in the cascade decays of heavier,
strongly interacting particles.

Next-to-leading-order (NLO) calculations of event rates have been
performed for many exotic scenarios, such as
SUSY~\cite{Beenakker:1996ch,Beenakker:1999xh,Freitas:2003yp},
but not for all. Experience shows that
$K = \sigma_{\mathrm{NLO}} / \sigma_{\mathrm{LO}}$ typically range
between 1 and 2\cite{Spira:1998sx}.
In this article $K = 1$ will be assumed throughout,
which will provide conservative estimates of the
experimental feasibility of discovery.

\subsection{Event topologies}

\begin{figure}[t!]
\begin{center}
\epsfig{file=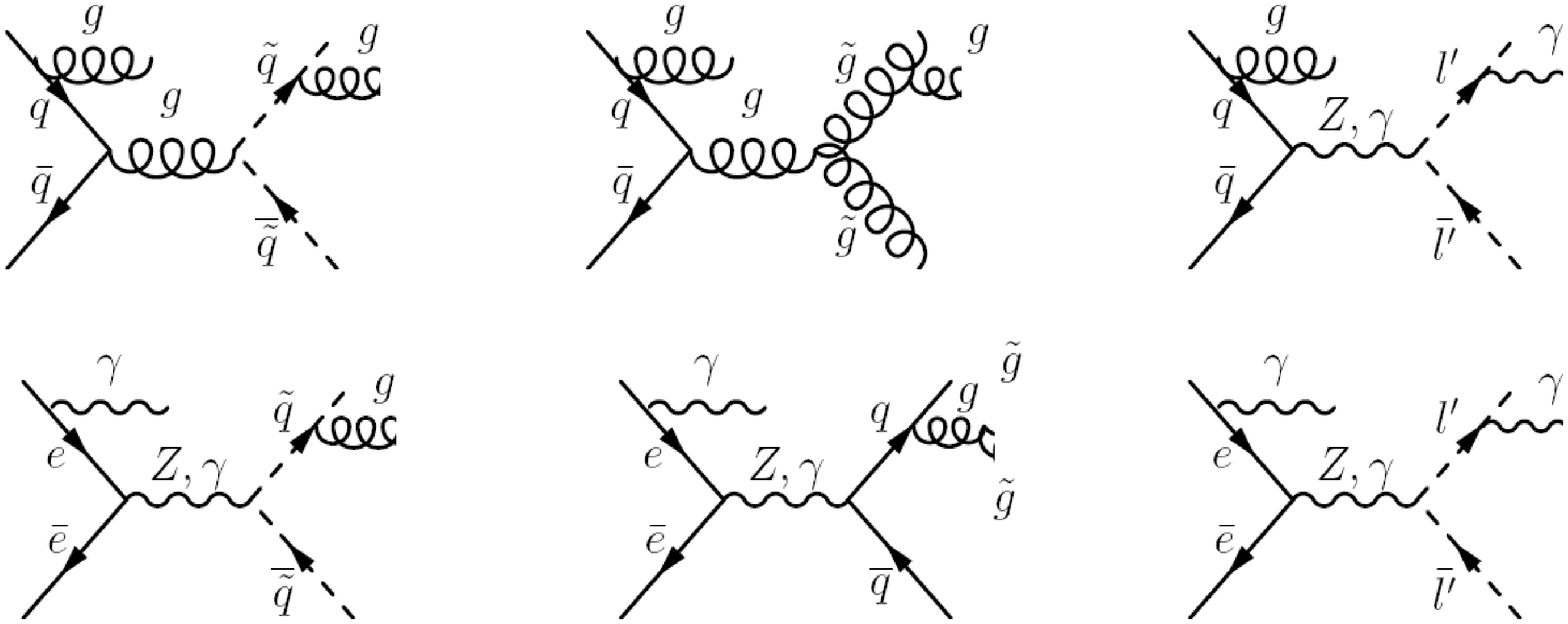,height=4cm,width=12cm}
\end{center}
\caption{Top: examples of production of squarks (top left), gluinos (top center) and heavy leptons (top right) in hadron-hadron colliders. Bottom: examples of production of squarks (bottom left), gluinos (bottom center) and heavy leptons (bottom right) in  $e^+e^-$-colliders. Examples of ISR and FSR are displayed.\label{fig:prod}}
\end{figure}

Higher-order perturbative calculations
involve emissions of further
partons. This can be understood as \ $ab \to X_c X_d$ generalising to
$ab \to X_c X_d \, e$ and $ab \to X_c X_d \, e f$, etc., where $e$ and $f$
will mainly be gluons for incoming hadron beams, but also could be
quarks, leptons or photons, Fig.~\ref{fig:prod}. Such emissions modify event shapes and
thereby the experimental signatures.

These processes can be described either by higher-order matrix elements
or by parton showers applied to lower-order matrix elements. There are
 relative advantages and limitations for both approaches. The former are more accurate
for a few well-separated emissions, and of course contain all the
model-specific details, but they diverge in the soft/collinear
limits and are not meaningful there. The latter are based on
model-independent approximations that work especially well in the
soft/col\-linear regions, in which dampening by Sudakov factors and
resummation of multiple emissions ensures a physically meaningful
behaviour. In practice, showers have turned out to be reasonable
approximations up to the scale of the lowest-order process, i.e.\
typically for the additional-jet transverse momenta below the mass
scale of the produced $X$ particles \cite{Plehn:2005cq}. For
exploratory studies the shower picture alone is therefore
sufficient, while, for precision studies, matrix-element information
also has to be supplied. Various methods to combine the two
approaches, so that the matrix-element behaviour is reproduced at
large separation scales and the parton-shower one at small scales,
are under active
study~\cite{Catani:2001cc,Lonnblad:2001iq,Krauss:2002up,Mrenna:2003if,Hoche:2006ph}.
Following a discovery, both higher-order matrix-element calculations
and corresponding matchings would have to be carried out.

In the shower approach, one distinguishes between initial-state
radiation (ISR) from the incoming $a$ and $b$ partons and final-state
radiation (FSR) from the outgoing $X_c$ and $X_d$, with interference
effects neglected. In the current case, since the $X_i$ particles
are heavy, radiation from them is strongly suppressed~\cite{deadcone} in comparison to light particles. Thus ISR
dominates.
The main experimental consequences of ISR is that the $X_c X_d$ pair
is produced in association with a number of further jets, and that
$X_c$ and $X_d$ do not have opposite and compensating transverse
momenta, as they do in the LO picture. This $p_{\perp}$ imbalance is
in itself a revealing observable, especially when the $X_i$ are not
directly detectable. On the other hand, the activity of the
additional jets, and of the underlying event, may be a nuisance for
some studies. Jets from FSR could be important, as will be discussed in the following section.

The kinematics of heavy particles produced in collisions can be
studied with the help of several programs. The two general-purpose
generators \textsc{Pythia} \cite{Sjostrand:2006za} and
\textsc{Herwig} \cite{Corcella:2000bw,Moretti:2002eu} are
traditionally used to study physics within and beyond the Standard
Model. They are mainly based on leading-order matrix elements, but
in a few cases also NLO matching is available. Their strength is
that they provide complete event topologies, such as they could be
observed in a detector, including descriptions of parton showers,
underlying events, and hadronisation. Many of these aspects are
handled in different fashions in the two programs, e.g. with respect
to the choice of shower evolution variables or hadronisation
schemes, but they tend to give similar results. They have both been
supplemented with routines to handle hadronisation into SMPs, as
will be described in the next subsection. Other general-purpose
generators, but currently without hadronisation into SMPs, are
\textsc{Isajet} \cite{Paige:2003mg} and \textsc{Sherpa}
\cite{Gleisberg:2003xi}.

Many other programs provide matrix elements to LO or NLO. A comprehensive
survey is given in the ``Les Houches Guidebook'' \cite{Dobbs:2004qw},
while dedicated BSM tools descriptions and an online repository
can be found in \cite{Skands:2006sk,Skands:2005vi}.
Examples of programs that can calculate arbitrary LO processes, once
the Feynman rules have been encoded, are \textsc{MadGraph/SMadGraph}
\cite{Stelzer:1994ta,Maltoni:2002qb,Cho:2006sx} and
\textsc{CompHep/CalcHep} \cite{Boos:2004kh,Pukhov:2004ca}.
Examples of programs that contain
NLO matrix elements are \textsc{Prospino} \cite{Beenakker:1996ed}
for SUSY particle production and \textsc{SDecay}
\cite{Muhlleitner:2003vg} for SUSY decays. In UED, production cross sections
for gluon and quark KK excitations at hadron
colliders have been calculated \cite{Macesanu:2002db} and are
available as an extension \cite{uedpythia} of the \textsc{Pythia} generator.
These programs do \textit{not} provide complete events, however, and
therefore have to be interfaced with general-purpose generators to
be fully useful. This process is rather well automated:
the original Les Houches Accord (LHA) \cite{Boos:2001cv,Alwall:2006yp} provides
a standard format to transfer simple parton-level configurations,
including information on flavours, colours and momenta, while the
SUSY Les Houches Accord (SLHA) \cite{Skands:2003cj} allows
standard exchange of SUSY parameters, couplings, masses, and branching
ratios.
\subsection{Hadronisation}\label{hadro}

For colourless new particles the story ends here. Once produced, the
$X_i$'s will sail out towards the detector. A coloured particle,
however, cannot escape the confinement forces. Therefore it will
pass through a hadronisation stage, during which it picks up
light-quark or gluon degrees of freedom to form a colour-singlet
``hadron''. For a colour-triplet, denoted $C_3$ to highlight its
coloured nature, this may either be a ``meson'' $C_3 \overline{q}$
or a ``baryon'' $C_3 q_1 q_2$. For a colour-octet, denoted $C_8$,
the alternatives are a ``meson'' $C_8 q_1 \overline{q}_2$, a
``baryon'' $C_8 q_1 q_2 q_3$ (or corresponding antibaryon), or a
``glueball'' $C_8 g$. Collectively, such states will be referred to
as $R$-hadrons, irrespective of the physics scenario which gives
rise to them. The name is borrowed from SUSY, where it refers to the
nontrivial $R$-parity possessed by such hadrons. Specific states
will be denoted by an $R$ with the flavour content
as lower indices and the charge as upper one%
\footnote{This is the convention adopted in the 2006 {\it Review of
Particle Physics}~\cite{rn}.}, e.g. $R^+_{\tilde{g}u\overline{d}}$
for a $\tilde{g}u\overline{d}$ state. In order to avoid too unwieldy
a notation, spin information is suppressed.

The hadronisation process does not appreciably slow down the
$R$-hadron relative to the original $C_{3,8}$, as can be seen from
the following. Consider a colour triplet $C_3$ with energy $E_{C_3}$
and longitudinal momentum $p_{\parallel C_3}$ (here longitudinal is
defined in the direction of the $C_3$), which hadronises into an
$R$-hadron with energy $E_R$ and longitudinal momentum $p_{\parallel
R}$, plus a set of ``normal'' hadrons that (approximately) take the
remaining energy--momentum. These hadrons are produced in the colour
field pulled out behind the $C_3$, and therefore ought to be limited
to have smaller velocities than that of the endpoint $C_3$ itself
or, more precisely, to have smaller rapidities $y$ defined with
respect to the $C_3$ direction. Now recall that $E + p_{\parallel} =
m_{\perp} e^y$, where $m_{\perp}$ is the transverse mass of a
particle. So if a normal hadron $h$, with a $\langle m_{\perp h}
\rangle \approx 1~\gev$, could at most reach the same rapidity as
the $C_3$, then it follows from the above that the ratio between the
$E + p_{\parallel}$ value of the normal hadron and that of the $C_3$
could at most be $\langle m_{\perp h} \rangle / m_{C_3} \approx
1~\gev / m_{C_3}$. Finally, assume that the normal hadrons are
produced behind the $C_3$, located at some rapidity $y_0$, with a
typical rapidity separation of $\langle \Delta y \rangle \approx 0.7
\approx \ln 2$, as in ordinary jets, i.e. at $y_0-\ln 2$, $y_0-2\ln
2$, $y_0-3\ln 2$, \ldots. The above upper bound for $E +
p_{\parallel}$ of a single hadron then is replaced by an average for
the summed effect of the normal hadrons produced in association with
the $C_3$: $m_{\perp} \exp(y_0) (\exp(-\ln 2) + \exp(-2\ln 2) +
\exp(-3\ln 2) + \cdots) = m_{\perp} \exp(y_0)$, where each term
expresses the $E + p_{\parallel}$ value of the next hadron. The
$R$-hadron will retain all the $C_3$ energy not taken away by these
normal hadrons. One therefore arrives at the expectation
\cite{Bjorken:1977md} that the fragmentation parameter $z$ typically
is
\begin{equation}
\langle z \rangle = \langle
\frac{E_R + p_{\parallel R}}{E_{C_3} + p_{\parallel C_3}} \rangle
\approx \langle \frac{E_R}{E_{C_3}} \rangle
\approx 1 - \frac{\langle m_{\perp h} \rangle}{m_{C_3}}
\approx 1 - \frac{1~\gev}{m_{C_3}} ~.
\label{zBjorken}
\end{equation}

That is, if $C_3$ has a large mass, $\langle z \rangle$ is close to
unity, and all other normal hadrons in the jet take a very small amount
of energy, $\sim 1~\gev \cdot \gamma_{C_3}$, where
$\gamma_{C_3} = E_{C_3} / m_{C_3}$ is not large for a massive
state $C_3$. For a colour octet the energy loss from the $R$-hadron
to the rest of the jet would be  about twice as big as for a colour
triplet ($C_A/C_F=9/4$), but otherwise the argument for colour triplets
applies in the same way.

\begin{figure}[h!]
\begin{center}
\epsfig{file=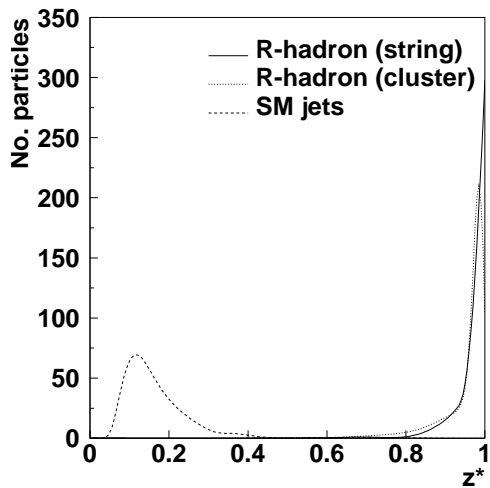,height=7cm,width=7cm} \caption{The predicted
distribution of the fragmentation variable $z^* = E_R /
E_{\mathrm{jet}}$ for gluino $R$-hadrons of mass 300 GeV produced at
the LHC. The expectations of the string and cluster hadronisation
schemes, implemented in \textsc{Pythia} and \textsc{Herwig},
respectively, are shown. Also shown is the prediction from
\textsc{Pythia} of the $z^*$ distribution of leading particles
within SM jets.}
  \label{fig:fragfunction}
\end{center}
\end{figure}

The above $z$ value is not accessible experimentally since, in a
busy hadronic environment, it is not possible to know which
particles come directly from the $C_{3,8}$ hadronisation. Instead,
with an $R$-hadron being part of a jet, the natural observable is
the fraction $z^*$ of the jet energy, $E_{\mathrm{jet}}$, which is
carried by the $R$-hadron: $z^*= E_R / E_{\mathrm{jet}}$.
Fig.~\ref{fig:fragfunction} shows the expected distribution of $z^*$
for gluino $R$-hadrons of mass 300 GeV produced at the LHC. The jets
were reconstructed with the CDF Run I cone algorithm with $R =
0.7$~\cite{Abe:1991ui}. Predictions are shown using the string and
cluster models of \textsc{Pythia} and \textsc{Herwig}, respectively.
The details of these models are given later in this section. Also
shown are \textsc{Pythia} predictions for the distribution of $z^*$
of the leading particles contained in jets produced in Standard
Model 2-to-2 QCD processes (labelled SM-jets).

Note that the $R$-hadron $\langle z^* \rangle$ is significantly
smaller than the prediction of Eq.~\ref{zBjorken},
$\langle z \rangle \approx 1 - 2/300 \approx 0.993$,
including a factor of two for the colour-octet nature of gluinos,
but not taking into account the possible loss of particles outside
the cone. Instead several other physics components contribute to the
final curve, by depositing further particles inside the jet cone.\\
$\bullet$ Final-state radiation from the gluino, although small,
is still larger than the nonperturbative effect discussed above. Since
collinear FSR is strongly suppressed \cite{deadcone}, the additional
FSR jets will not be correlated with the gluino direction, and so
may or may not end up inside the $R$-jet cone. At hadron colliders
the FSR jets tend to drown among the more numerous ISR jets. They
could stand out at a lepton collider, at which there is no QCD ISR.\\
$\bullet$ Initial-state radiation from the incoming quarks and gluons
is not suppressed by any mass effects. The ISR jets may appear anywhere
in phase space, by chance also inside the $R$-jet cone, and may then
deposit a significant amount of extra energy.\\
$\bullet$ The underlying events (``beam remnants'' and
``multiple interactions'') tend to deposit particles inside
any jet cone, in proportion to the size of the cone. Normally these
particles would be rather soft, but upwards fluctuations can occur.

For the actual hadronisation of the coloured heavy object there are
two main models available. In \textsc{Pythia} the Lund string
fragmentation model \cite{Andersson:1983ia} is used, wherein an
assumed linear confinement potential is approximated by a string
with a constant tension of $\kappa \approx 1$ GeV/fm. A \emph{colour
triplet} $C_3$ (or $C_{\overline{3}}$) is at the endpoint of such a
string. When the $C_3$  moves away from its production vertex it
pulls out the string behind itself, to which it loses energy. This
string may then break by the production of a light quark-antiquark
or diquark-antidiquark pair, where the $\overline{q}$ or $q_1 q_2$
is in a colour-antitriplet state which can combine with the $C_3$ to
form a singlet. Further breaks of the string may occur, which causes
the formation of a jet of hadrons. The composition of the different
light flavours is assumed universal and thus constrained e.g. by LEP
data\cite{Abreu:1996na,Alexander:1995bk} ($u : d : s \approx 1 : 1 :
0.3$, with diquarks further suppressed).  A mixture of different
charge states is finally produced. The energy-momentum fraction $z$
retained by the $R$-hadron is described by an appropriate
fragmentation function, with parameters that fit e.g.\ $B$ meson
spectra\cite{Heister:2001jg,Abe:2002iq}. These functions have a mass
dependence consistent with Eq.~\ref{zBjorken} for the
extrapolation from $b$ to $C_3$ hadronisation.

A \emph{colour octet} $C_8$ in the Lund string model is viewed as
the incoherent sum of a colour and an anticolour charge (the planar
or $N_C \to \infty$ approximation~\cite{tHooft:1973jz}), such that
an octet $C_8$ is attached to two string pieces instead of one only.
Each of these pieces is allowed to break as above. One gives a quark
or an antidiquark, the other an antiquark or diquark, and these
combine to form an $R$-meson or $R$-baryon. Diquark-antidiquark
``hadrons'' are rejected. In addition, a new mechanism is
introduced: a $g g$ pair may be produced, such that an $R$-glueball
is formed and the leftover gluon attaches to the two string pieces.
Its relative importance is not known. The equivalent mechanism for
normal gluons would lead to the production of glueballs. There are a
few indications that this may
occur,~\cite{Minkowski:2000qp,Chua:2002wp}, but it cannot be at a
significant rate. Based on the absence of real evidence for normal
glueballs\cite{Klempt:2006uy}, the default value for the
$R$-glueball fraction in \textsc{Pythia} is 10\%, and can be
changed. Clearly a scenario in which glueballs are predominantly
formed in the hadronisation step will have a large impact on
experimental searches which rely on the reconstruction of tracks.
However, as is outlined in Section~\ref{rprop}, the behaviour of an
$R$-glueball when it interacts in material is expected to be similar
to that of a neutral $R$-meson and it may therefore convert into a
charged $R$-hadron which would leave behind a track.

Predictions from \textsc{Pythia} of the fractions of different
$R$-hadron species formed in the hadronisation of a gluino and a
stop are shown in Tab.~\ref{tab:flavcomp}. The predicted species of
an $R$-hadron arising from an antistop are almost exactly the charge
conjugates of a stop $R$-hadron. The key observation is that, when
neglecting the {\it a priori} fraction of $R$-glueballs of 10\%,
roughly 50\% of the produced states are charged. The gluino
$R$-baryon numbers are somewhat lower than expected. This is an
artifact of approximations used in hadronisation model. However,
these should anyway  contribute less than 10\% of the $R$-hadrons
produced. It is also interesting to note that, as discussed in
Section~\ref{sec:hadscat}, although $R$-hadrons will largely start
out as mesons, nuclear scattering in calorimeters will ensure that
they mostly end up as baryons as they leave the detector.

\textsc{Herwig} is based on cluster fragmentation. In this picture
all gluons are nonperturbatively split into quark-antiquark pairs
at the end of the perturbative cascade. Partons from adjacent such
breaks, and from original (anti)triplets, are then grouped into
colour-singlet clusters. Large-mass clusters are broken into smaller
ones, along the ``string'' direction. These clusters then decay to
two hadrons, using phase-space weights to pick between allowed
flavours. A colour octet $C_8$ is combined with a $q_1$ and a
$\overline{q}_2$ from two adjacent nonperturbative splittings to
form a cluster \cite{Kilian:2005kr}. The cluster decay can be
either of $C_8 q_1 \overline{q}_2 \to %
C_8 q_3 \overline{q}_2 + q_1 \overline{q}_3$, $\to C_8 q_1
\overline{q}_3 + q_3 \overline{q}_2$, or $\to C_8 g + q_1
\overline{q}_2$. The two former possibilities are handled as usual
based on phase space weights, while the latter $R$-glueball
possibility is added with a free normalisation. By default it is set
to zero. The resulting gluino $R$-hadron composition is shown in
Tab.~\ref{tab:flavcomp}.

Thus, although the technical details in \textsc{Pythia} and
\textsc{Herwig} are quite different the basic principles are
similar. In both models, $R$-hadron production occurs predominantly
by the $C_{3,8}$ picking up $u$ and $d$ quarks, while $s$ is more
rare, and $c$ or $b$ can only be produced in the shower, at an even
lower rate. Apart from the unknown fraction of R-glueballs, the
$u$-$d$ charge difference ensures that approximately half of the
produced $R$-hadrons will be charged. The two generators should
therefore provide similar phenomenology for the flavours and momenta
of produced $R$-hadron events. Any differences would reflect true
uncertainties in our current understanding.

\begin{table}[t]
\begin{center}
\hspace*{-0.1cm}\begin{tabular}{|c|r|r|@{\protect\rule{0mm}{0.9\baselineskip}}}
\hline
$R$-hadron & \textsc{Pythia} & \textsc{Herwig}\\
           & Fraction & Fraction\\
           & (\%) & (\%) \\
\hline
$R^+_{\tilde{g}u\overline{d}}$, $R^-_{\tilde{g}d\overline{u}}$ & 34.2 & 28.2\\
$R^0_{\tilde{g}u\overline{u}}$, $R^0_{\tilde{g}d\overline{d}}$ & 34.2 & 28.2\\
$R^+_{\tilde{g}u\overline{s}}$, $R^-_{\tilde{g}s\overline{u}}$ & 9.7 & 17.5\\
$R^0_{\tilde{g}d\overline{s}}$, $R^0_{\tilde{g}s\overline{d}}$,
$R^0_{\tilde{g}s\overline{s}}$  & 10.4 & 26.1 \\
$R^0_{\tilde{g}g}$ & 9.9 & --- \\
$R^{++}_{\tilde{g}}$, $R^{--}_{\tilde{g}}$ (anti)baryons & 0.1 & --- \\
$R^+_{\tilde{g}}$, $R^-_{\tilde{g}}$ (anti)baryons & 0.8 & --- \\
$R^0_{\tilde{g}}$ (anti)baryons  & 0.7 & --- \\
\hline
\end{tabular}%
\hspace{10mm}%
\begin{tabular}{|c|r|@{\protect\rule{0mm}{0.9\baselineskip}}}
\hline
$R$-hadron & Fraction\\
           & (\%) \\
\hline
$R^+_{\tilde{t}\overline{d}}$ & 39.6 \\
$R^0_{\tilde{t}\overline{u}}$ & 39.6 \\
$R^+_{\tilde{t}\overline{s}}$ & 11.8 \\
$R^{++}_{\tilde{t}}$ baryons & 0.8 \\
$R^+_{\tilde{t}}$ baryons & 6.7 \\
$R^0_{\tilde{t}}$ baryons & 1.5 \\
\hline
\end{tabular}
\end{center}
\vspace{0.45cm} \caption{Predictions from \textsc{Pythia} of the
fractions of different species of $R$-hadrons following the
hadronisation of a gluino (left) and a stop (right) of mass 500 GeV
produced at the LHC. The \textsc{Herwig} gluino predictions
\cite{Kilian:2005kr} are for a 2000 GeV mass, but almost identical
for 50 GeV. \label{tab:flavcomp}}
\end{table}

\subsection{$R$-hadron properties}\label{rprop}

The mass splittings of $R$-hadrons are of critical importance when
designing a search strategy. If one state would be significantly
lighter than another, one would expect this
state to be dominantly present in the detector. If the
lightest state would be neutral, a completely different
(and far more experimentally challenging) signature is expected
than if the lightest state would be charged.

The masses of the produced $R$-hadrons are best understood from
the mass formula for the lowest-level (i.e. no radial or orbital
excitation) hadrons \cite{DeRujula:1975ge, Kraan:2004tz}
\begin{equation}\label{masss}
m_{\mathrm{hadron}} \approx \sum_i m_i - k \sum_{i \neq j}
\frac{(\mathbf{F}_i \cdot \mathbf{F}_j) \, (\mathbf{S}_i \cdot
\mathbf{S}_j)}{m_i \, m_j}
\end{equation}
where $m_i$ are the constituent masses, $\mathbf{F}_i$ are the
colour \textbf{SU(3)} matrices, $\mathbf{S}_i$ the spin
\textbf{SU(2)} ones, and $k$ a parameter related to the wave
function at the origin. The $C_{3,8}$ is so heavy that it provides
an almost static colour field in the rest frame of the $R$-hadron,
and therefore its spin, if any, is decoupled (cf. ``Heavy Quark
Effective Theory'' \cite{Isgur:1989vq}). The heavy $C_{3,8}$ has
thus a strongly localised wave function, while the light degrees of
freedom are spread over normal hadronic distance scales. The
relative localisation of the wave functions largely accounts for the
expected energy loss and scattering behaviour of $R$-hadrons in
matter, as discussed in detail in Section \ref{sec:hadscat}. It also
ensures that the  mass splittings, given by the second term in
Eq.~\ref{masss}, are determined by the light degrees of freedom. It
must be noted that the
 ``experimentally observable'' constituent masses
used below are different from the renormalisation-scheme-dependent
running masses found in the Lagrangian of a theory although it is
possible to provide an approximate translation for any specific
case. The mass splittings discussed below agree with those obtained
earlier with a similar bag model
approach~\cite{Chanowitz:1983ci,Buccella:1985cs}, and with lattice
calculations~\cite{Foster:1998wu}.

For a colour-triplet $C_3$ the hadron mass is
easily obtained from standard quark and diquark constituent masses.
For $C_3$
 mesons (${C}_{\bar{3}}{{q}}$, ${C}_3\bar{{{q}}}$), no significant mass splitting is expected to occur\cite{Kraan:2004tz}, in analogy with the absence of mass splittings of B-hadrons. For $C_3$ baryons such as ${C}_3{q}{q}$ states (and, by symmetry for the ${C}_{\bar{3}}\bar{{q}}\bar{{q}}$ baryons)
the mass spectrum would be (all units in GeV)
\begin{eqnarray*}
M_{{C}_3{q}{q}}& \approx & M_{{C}_3}+0.3+0.3-0.026\times\frac{(-\frac{2}{3}\times-\frac{3}{4})}{0.3\times 0.3}\hspace{2cm} {s_{{q}{q}}=0}\\
&\approx& M_{{C}_3}+0.46\\
M_{{C}_3{q}{q}}& \approx& M_{{C}_3}+0.3+0.3-0.026\times\frac{(-\frac{2}{3}\times+\frac{1}{4})}{0.3\times 0.3}\hspace{2cm} {s_{{q}{q}}=1}\\
& \approx &  M_{{C}_3}+0.65
\end{eqnarray*}
where $s_{qq}$ denotes the total spin of the $qq$ system.

In the case of a $C_8$ state, $R$-mesons, $R$-baryons and
$R$-glueballs could arise from hadronisation. The $R$-mesons $C_8
q_1 \overline{q}_2$ give a colour factor $\mathbf{F}_1 \cdot
\mathbf{F}_2 = 1/6$ for the light quarks, to be compared with $-4/3$
in a normal $q_1 \overline{q}_2$ meson. For a ${C}_8 q\bar{q}$
state, the mass spectrum is thus given by:
\begin{eqnarray*}
M_{{C}_8{q}\bar{{q}}}& \approx & M_{{C}_8}+0.3+0.3-0.043\times\frac{(\frac{1}{6}\times-\frac{3}{4})}{0.3\times 0.3}\hspace{2cm} {s_{{q}\bar{{q}}}=0}\\
&\approx& M_{{C}_8}+0.66\\
M_{{C}_8{q}\bar{{q}}}& \approx& M_{{C}_8}+0.3+0.3-0.043\times\frac{(\frac{1}{6}\times+\frac{1}{4})}{0.3\times 0.3}\hspace{2cm} {s_{{q}\bar{{q}}}=1}\\
& \approx &  M_{{C}_8}+0.58
\end{eqnarray*}
The $\rho$-$\pi$ mass difference therefore flips sign and is
considerably reduced in size for the corresponding $R$-mesons.
Further, a gluon constituent mass is about twice a
light-quark\cite{Cornwall:1982zn,Hou:2001ig}, so the $C_8 g$ state
is almost mass degenerate with the $C_8 u \overline{u}$ and $C_8 d
\overline{d}$ ones, and will have similar properties. Assuming that
the constituent mass of a gluon is approximately 700
MeV~\cite{Cornwall:1982zn,Hou:2001ig}, the mass of a $C_8g$ state is
then
\begin{equation}
M_{ C_8 g}\approx M_{ C_8}+0.7.
\end{equation}
 The expressions for the masses of the
 $R$-baryons ($C_8qqq$) are somewhat more
cumbersome, since there are more colour and spin combinations possible,
but do not offer any peculiar features. Splittings are expected to be less
than the order of a 100 MeV~\cite{Kraan:2004tz}, and the mass
of the ${C}_8{q}{q}{q}$ states is
thus approximately
\begin{equation}
M_{C_8 qqq}=M_{C_8}+0.9.
\end{equation}

Furthermore, the orbital and radial excited states would follow patterns
not too dissimilar from normal hadronic spectra. However, before a
discovery of $R$-hadrons it is not useful to attempt to calculate
the full hadron spectrum including excited states. In case such an
excited state is produced it would probably decay electromagnetically or
strongly at such a short time scales that no secondary vertex is
resolved. Owing to the small mass splittings and the small boost factor of
a massive $R$-hadron, the additional photon or pion produced will also be
of low momentum, and so drown in the general hadronic environment. For the
current studies it is sufficient to simulate the
production of the lowest-lying state of each allowed flavour combination.


Although this paper is only concerned with stable $R$-hadrons, it is
worthwhile to briefly consider the possibility of $R$-hadron decays.
If unstable over nanosecond time scales, $R$-hadrons could decay
inside the detector. This would take place as an almost free decay
of the $C_{3,8}$, with the rest of the hadron acting as spectators,
and could therefore be described by the standard perturbative
picture. Showering and hadronisation can be added, the latter also
involving the spectators. These decays may well violate baryon or
lepton number, via such processes as $\tilde{g} \to u d s$, and give
rise to unusual hadronisation scenarios~\cite{Sjostrand:2002ip,
richardson}.

A further property of $R$-hadrons about which little is known is
their oscillatory behaviour. Neutral $R$-mesons will be able to
convert into their anti-particles without violating any known
conservation laws. Since the oscillation length depends on couplings
and masses of particles which have yet to be discovered (if ever)
then either minimal or maximal oscillations are
conceivable\cite{Sarid:1999zx,Culbertson:2000am}. As discussed in
Section~\ref{sec:hadscat} this may have implications for the ability
of experiments to both discover and quantify a heavy squark which
may be produced at a collider.

\subsection{Sources of uncertainties}\label{uncert}

The understanding of the production of non-magnetically charged SMPs
is not complete, but should be fully adequate for the purposes of a
search-and-discovery mission. Further sophistication would be added
once the first signals for new physics would point the way to a more
specific scenario, rather than the generic ones considered here. In
any search relying on QCD models, the following areas of uncertainty
may be relevant:
\begin{itemize}
\item Production rates: are very much model-dependent, and contain
some uncertainties from uncalculated higher-order contributions.
These can be parameterised in the form of a $K$-factor.
\item Event topologies: contributions from further parton emissions
are treated by the parton-shower approximation, which should be
sufficient for the bulk of the cross section, but not necessarily
for topologies with well-separated further jets.
\item Hadronisation, momenta: the fragmentation function is guaranteed
to give most of the momentum to the $R$-hadron, so uncertainties are
restricted to the minijet of a few further normal hadrons produced
in the same general direction.  The typical minijet energy could
easily be uncertain by a factor of two.
\item Hadronisation, flavours: some uncertainty, especially in the
glueball and baryon sectors. The rate of the former is a completely
free parameter. It should not affect the bulk of the production, mesonic
states with $u$ and $d$ quarks only, which also will guarantee a rather
even mixture of charged and uncharged states.
\item $R$-hadron properties: the general pattern of states and masses
appears to be well understood, especially in the dominant meson sector.
Should, against all expectations, decays like
$C_8 u \overline{d} \to C_8 g + \pi^+$ be kinematically allowed,
they would have consequences for the charged/neutral ratio.

\end{itemize}

\subsection{Production mechanisms of for magnetic monopoles}\label{monocal}
The interactions of magnetic monopoles at high energies are
difficult to treat theoretically since perturbation theory is
inapplicable due to the size of the electromagnetic coupling
constant for monopoles ($\alpha_m=g^2/\hbar c = 137/4 n^2$ for a
Dirac monopole). Nevertheless, the scattering cross sections for
monopoles incident on charged particles have been computed using
non-perturbative techniques (see ref. \cite{Milton:2006cp} for a
review of monopole scattering calculations). This has allowed the
calculation of the stopping power of monopoles in
material\cite{Ahlen:1978jy,Ahlen:1982mx,Drell:1982zy,Derkaoui:1998gf}.
In addition the binding energy of monopoles to nuclear magnetic
dipoles has been calculated \cite{Bracci:1983fe,Gamberg:1999tv}.
Since these calculations concern the interactions of monopoles in a
detector, they are discussed in \ref{dedxMM} and \ref{stopmm},
respectively.
\begin{figure}[h!]
\begin{center}
\epsfig{file=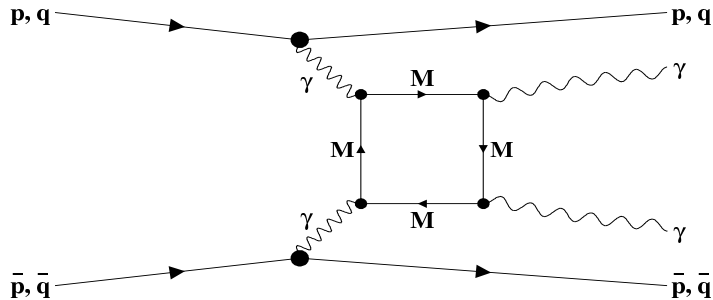,height=5cm,width=10cm} \caption{A Feynman
diagram of a multi-photon process mediated by an internal monopole
line in $p\bar{p}$ collisions.} \label{fig:inloop}
\end{center}
\end{figure}

Even though a perturbative treatment of monopoles is impossible,
experiments nevertheless must assume certain production processes in
order to estimate the detector acceptance. The production mechanisms
which have been assumed by experiments are typically those which
would correspond to leading order pair production in a perturbative
picture. A Drell--Yan-like mechanism to produce a
monopole-antimonopole pair is a commonly considered scenario in
hadron-hadron collisions\cite{Abulencia:2005hb}. Searches at
$e^+e^-$ experiments have hypothesised monopole production through
the annihilation reaction $e^+ + e^- \longrightarrow Z^0/{\gamma}
\longrightarrow m\bar{m}$\cite{Pinfold:1993mq}. In lepton-hadron
collisions, pair production via photon-photon fusion has been
envisaged\cite{h1mon}. To obtain calculations of cross-sections and of the 
kinematic distributions of produced monopoles, the formalism for 
the electroweak production of $\mu^+\mu^-$ is used,
with appropriate phase space modifications to account for the larger
monopole mass. However, it is important to emphasise that
little is known about processes in which monopoles could be directly
produced, and that the above reactions are 'best guesses'. Thus, some
experiments also use random phase-space models to calculate acceptance.
Examples of the methods used by experiments to model monopole
pair production are discussed in Section~\ref{monsearch}.

It has also been proposed in Refs.\cite{DeRujula:1994nf} and
\cite{Ginzburg:1999ej}, respectively, that virtual monopoles
mediating $e^+e^-$ and hadron-hadron collisions would be manifest
via photonic radiation, as illustrated in Fig.\ref{fig:inloop} for
$p\bar{p}$ collisions. Calculations of cross sections of such
process assume that perturbation theory can be used for monopoles.
Limits\cite{Acciarri:1994gb,Abbott:1998mw} obtained using these
calculations, which are discussed later in Section~\ref{mmin} have
been criticised\cite{Gamberg:1998xf}.

\section{Interactions of SMPs}\label{sig}

This section describes the expected interactions of SMPs in particle
detectors. The dominant types of interactions are electromagnetic
and strong interactions. An outline is given of the theory of
electromagnetic energy loss for electrically and magnetically
charged particles. Compared to the electromagnetic case, nuclear
interactions of SMPs containing heavy coloured objects are poorly
understood. A review is given here of various models which have been
proposed to describe these nuclear interactions and their
uncertainties are also pointed out.

\subsection{Ionisation energy loss}
The most commonly used observable in SMP searches is the measurement
of the continuous ionisation energy loss $\frac{dE}{dx}$. Both
electrically and magnetically charged SMPs lose energy principally
through ionisation energy loss as they propagate through matter and
for both types of particle the theory of electromagnetic energy loss
is well established.

\subsubsection{Ionisation energy loss by electrically charged
particles in material} \label{inte}

As an electrically charged particle moves through a material it
loses energy either by interactions with atomic electrons or by
collisions with atomic nuclei in the material. The first of these
results in the liberation of electrons from the atoms in the
material (i.e. ionisation) while the second results in the
displacement of atoms from the lattice.
The energy loss due to the second process is called the
Non-Ionising Energy Loss (NIEL).  The differential energy loss
(stopping power), $(dE/dx)$, due to the ionisation energy is much
larger than the NIEL \cite{Chilingarov:1999vi} in practical particle
detectors.


For fast particles of charge $Z_1$ in a medium of atomic number
$Z_2$ the mean ionisation energy loss is given by the
Bethe-Bloch formula \cite{rn}.
\begin{equation}
\frac{dE}{dx} = \frac{4 \pi e^4 Z_1^2}{m_e c^2 \beta^2} n \bigg(\frac{1}{2}
\ln\big(\frac{2 m_e c^2 \beta^2 \gamma^2 T_{max}}{I_e^2}\big)- \beta^2 -
\frac{\delta}{2}\bigg)
\label{dedxE}
\end{equation}
where $e$ and $m_e$ are the charge and mass of the electron, $n$ is
the number of electrons per unit volume in the material, $\beta$ is
the relativistic velocity of the incident particle, $\gamma =
1/\sqrt{1-\beta^2}$ and $I_e$ is the mean ionisation potential of
the material.  The latter can be parameterised by \cite{lewin};
\newline
$I_e(Z_2) = (12 Z_2 + 7)$ for $Z_2 \leq 13$
or $(9.76 Z_2 + 58.8 Z_2^{-0.19})$ eV for $Z_2 > 13$.
\newline
The quantity $T_{max}$ is the maximum kinetic energy which can be imparted
to a free electron in a single collision and is given, for a particle of
mass $M$, by
\begin{equation}
T_{max} = \frac{2 m_e c^2 \beta^2 \gamma^2}{1+2 \gamma m_e/M + (m_e/M)^2}.
\end{equation}\label{eq:tmax}
The term $\delta$ represents the density effect which limits the
relativistic rise at high energy and it has been calculated by
Sternheimer {\it et al.}~\cite{Sternheimer:1983mb}. This term is
only relevant for particles with $\beta \gamma  \gg 3$ and for
massive particles $T_{max} \approx 2 m_e c^2 \beta^2 \gamma^2$.
Slight differences occur for positive and negative particles moving
with low velocity~\cite{barkas2}.

For low energies when the velocity of the incident particle is
comparable or less than the velocity of the electrons in the atom,
the so-called Lindhard region,  this formula is no longer valid. The
energy loss is then proportional to the particle velocity $\beta$~\cite{lindhard}:
\begin{equation}
 \frac{dE}{dx} = N \xi_e 8 \pi e^2 a_o \frac{Z_1
 Z_2}{Z}  \frac{\beta}{\beta_o},
\label{Lind}
\end{equation}
where $N$ is the number of atoms per unit volume, $\xi_e \approx  Z_1^{1/6}$,
  $a_o$ is the Bohr Radius of the hydrogen atom and
$Z^{\frac{2}{3}}\>= Z_1^{\frac{2}{3}}+Z_2^{\frac{2}{3}}$.
This formula holds for $\beta < Z_1^{\frac{2}{3}}  \beta_o$ where
$\beta_o = \frac{e^2}{2 \epsilon_0 h c}$ ($\approx .0073$)
is the electron velocity in the classical lowest Bohr orbit of the
hydrogen atom.

The intermediate region, in which neither the Bethe--Bloch formula
(Eq.~\ref{dedxE}) nor the Lindhard formula (Eq.~\ref{Lind}) are
valid, is defined by the velocity range $\beta_1\leq\beta\leq\beta_2
$, where $\beta_1 = \max [\alpha Z_1^{1/3}/(1+\alpha Z_1^{1/3}), (2
Z_2^{0.5}+1)/400]$, $\beta_2=\alpha Z_1/(1+\alpha Z_1)$ and $\alpha$
is the fine structure constant.  This region is described by
Anderson and Barkas \cite{Anderson}. In this region a polynomial can
also be used to join up the two regions \cite{lewin} of the form;
\begin{equation}
 \frac{dE}{dx} = A \beta^{3} + B \beta^{2} + C \beta + D
 \label{middle}
\end{equation}
where $A$, $B$, $C$, $D$ are derived from the four simultaneous
equations obtained by equating $ \big(\frac{dE}{dx}\big)_{el}$ and $
\frac {d \big(\frac{dE}{dx}\big)_{el}} {d\beta}$ at $\beta_1$ and
$\beta_2$ for the Eqs.~\ref{dedxE}, \ref{Lind} and \ref{middle}. This gives:
\medskip \\
$ A=\frac {1} {\triangle \beta^{2}} \big(\frac {y_1}{\beta_1} + k -
\frac {2 \triangle y}{\triangle \beta} \big)\\
B=\frac {1} {\triangle \beta^{2}} \big( 3 (\beta_1 + \beta_2)
\frac{\triangle y}{\triangle \beta} - (\beta_1 + 2 \beta_2)
\frac {y_1}{\beta_1} -
\triangle y^{2} k\big)\\
C=\frac {1} {\triangle \beta^{2}} \big( \beta_1 (\beta_1 +2 \beta_2) k +
\frac {\beta_2 (2 \beta_1 + \beta_2) y_1}{\beta_1} -
\frac { 6 \beta_1 \beta_2 \triangle y}{\triangle \beta} \big)\\
D=\frac {1} {\triangle \beta^{2}} \big[ \frac {1}{\triangle \beta}
\big( \beta_1^{2} y_2 (3 \beta_2 - \beta_1) + \beta_2^{2} y_1 (
\beta_2 - 3 \beta_1) \big) - \beta_1 \beta_2 \big( \beta_1 k + \beta_2
\frac {y_1}{\beta_1} \big) \big] $
\newline
where $\Delta\beta=\beta_2-\beta_1$, $y_1$,$y_2$ are the
$\frac{dE}{dx}$ values computed from Eqs.~\ref{dedxE} and~\ref{Lind}
at velocities $\beta_1$ and $\beta_2$,
respectively,$\Delta y=y_1-y_2$,
$k = \frac {1}{\beta_2^{3}} \big[ 2  \frac{4\pi e^4 Z_1^2}
{m_e c^2} n
( \gamma_2^{2} - \ln\big({\frac{2 m_e c^2  \beta_2^2 \gamma_2^2}
 {I(1-\beta_2^2)
 }}\big)\big)] $
and $\gamma_2=1/\sqrt{1-\beta_2^2}$.

\begin{figure}
\begin{center}
\hspace{-7mm} \epsfig{file=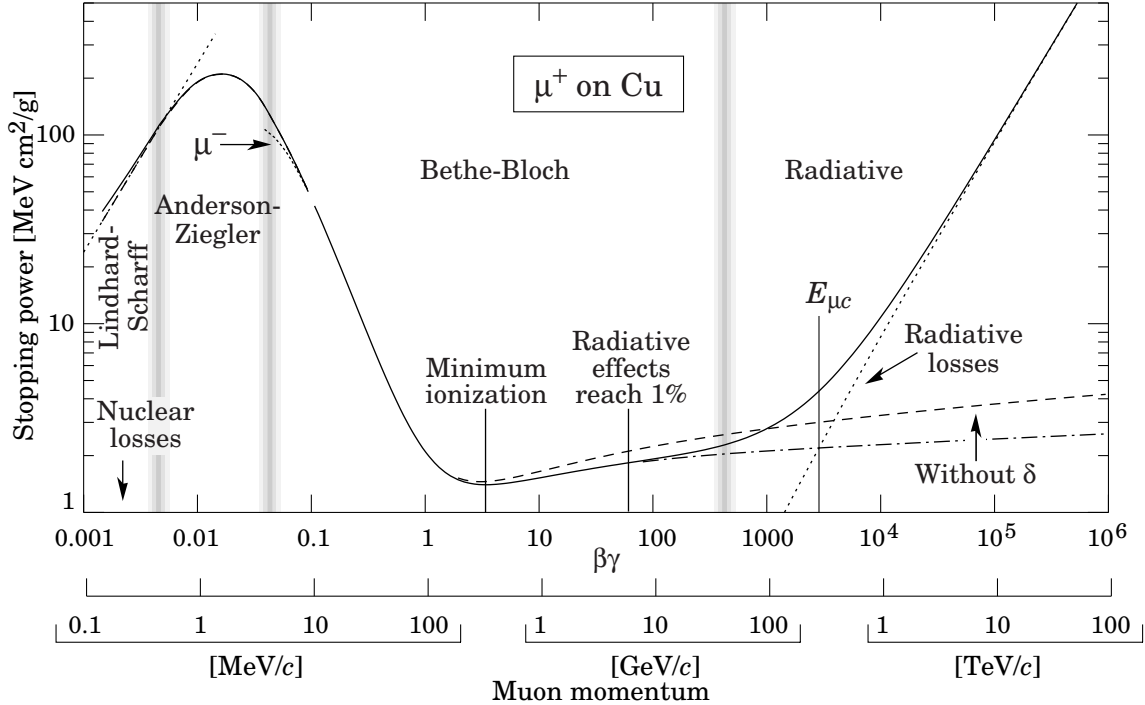}\caption{ Stopping
power ($\frac{dE}{dx}$) for positively charged muons in copper as a
function of $\beta \gamma = p/M$ taken from the Particle Data Group
\cite{rn}. The solid curve indicates the total stopping power. The
radiative effects apparent at very high energies are much less
relevant for particles heavier than muons. The different regions
indicated by the vertical bands are described in Ref.~\cite{rn} as
are the small difference between positive and negative charges at
low values of $\beta\gamma$ (the Barkas effect~\cite{barkas}) shown
by the short dotted lines labelled $\mu^-$.} \label{dedx1}
\end{center}
\end{figure}

\begin{figure}
\begin{center}
\hspace{-7mm} \epsfig{file=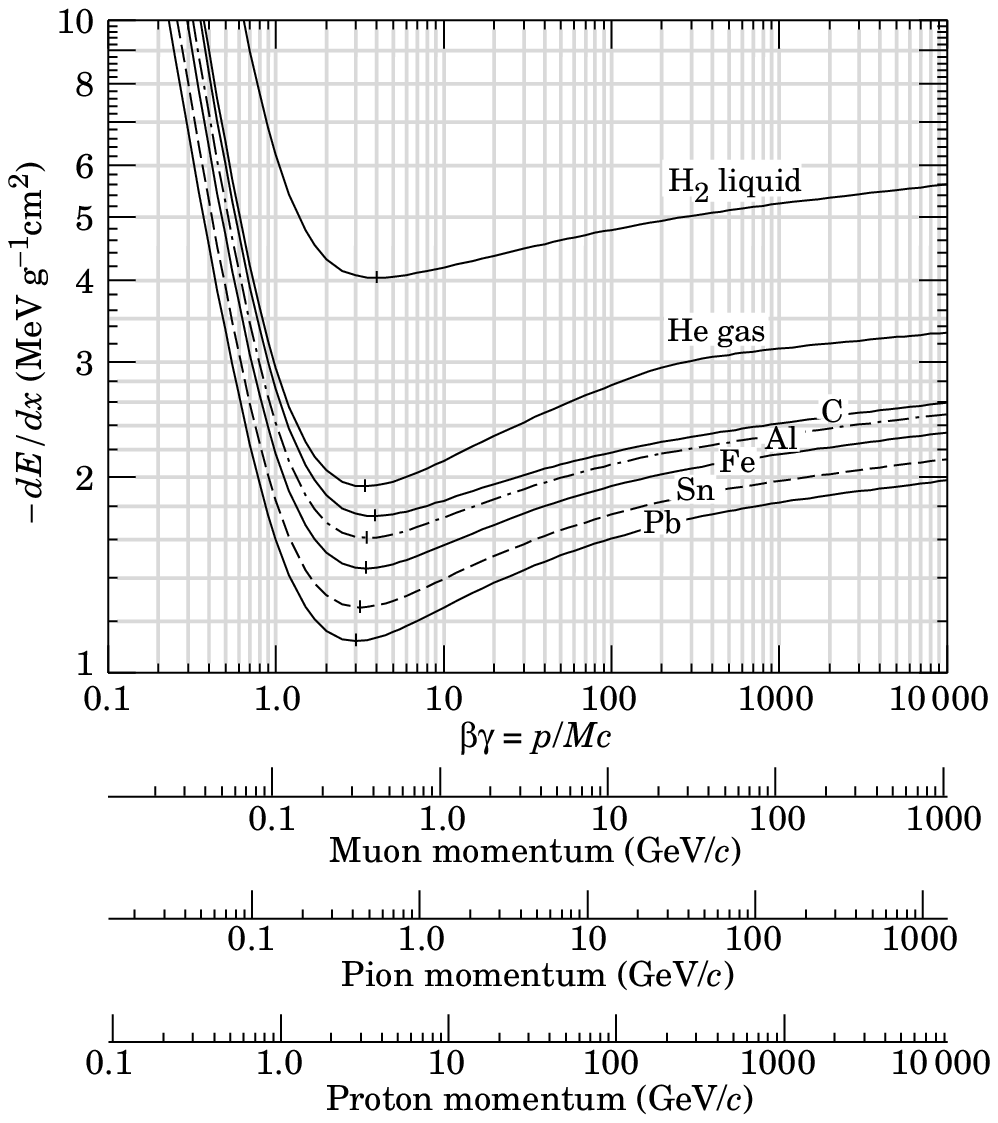,height=9cm} \caption{
The stopping power ($\frac{dE}{dx}$) for particles in different
materials (omitting radiative effects)~\cite{rn}.} \label{dedx2}
\end{center}
\end{figure}

The variation of $\frac{dE}{dx}$ with $\beta \gamma$ is illustrated
in Fig.~\ref{dedx1} by a  calculation made for $\mu^+$ in copper
\cite{rn}. The rise with $\beta \gamma$ in the Lindhard region from
Eq.~\ref{Lind} turns into a fall as $\beta \gamma$ increases, as
expected from Eq.~\ref{dedxE}. The accuracy of the rather arbitrary
polynomial procedure between the two regions can be assessed from
extrapolating the rise at low $\beta \gamma$ and the fall at higher
$\beta \gamma$ into the worst possible case of a discontinuous join.
The variation of $\frac{dE}{dx}$ with material is shown in
Fig.~\ref{dedx2}~\cite{rn}.

\subsubsection{Ionisation energy loss by magnetically charged
particles in material} \label{dedxMM}

Since a Dirac monopole has a magnetic charge equivalent to an
electric charge of $137e/2$, such a particle would be expected to
suffer ionisation energy loss at a rate several thousand times
greater than that of a particle with electric charge
$e$\cite{Ahlen:1978jy,Ahlen:1982mx}. A Dirac monopole would thus be
expected to leave a striking ionisation signature.
Ionisation energy loss could also lead to a monopole becoming
stopped in detector material before reaching a tracking chamber.

\begin{figure}
\begin{center}
\epsfig{file=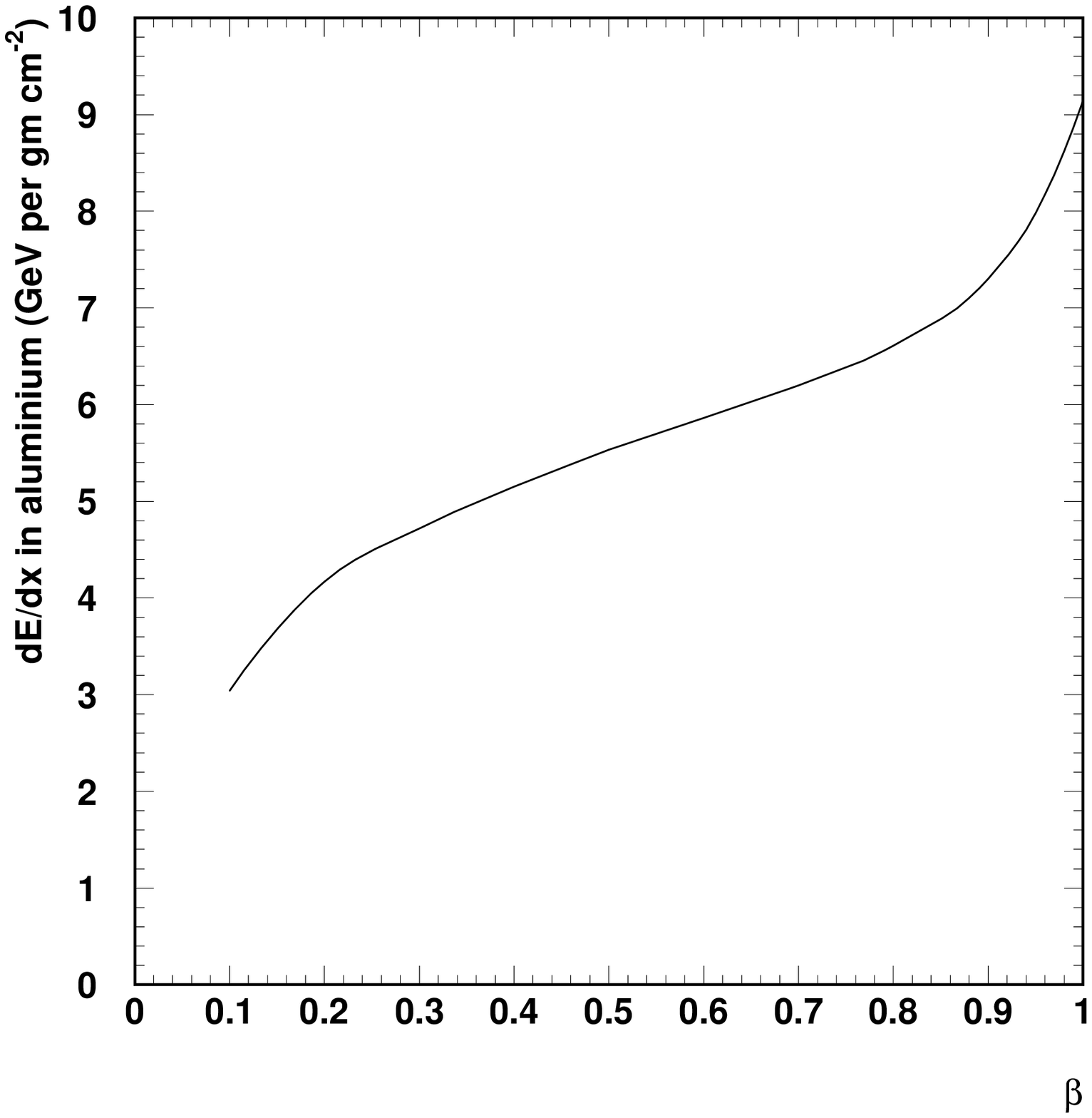,height=7cm}\epsfig{file=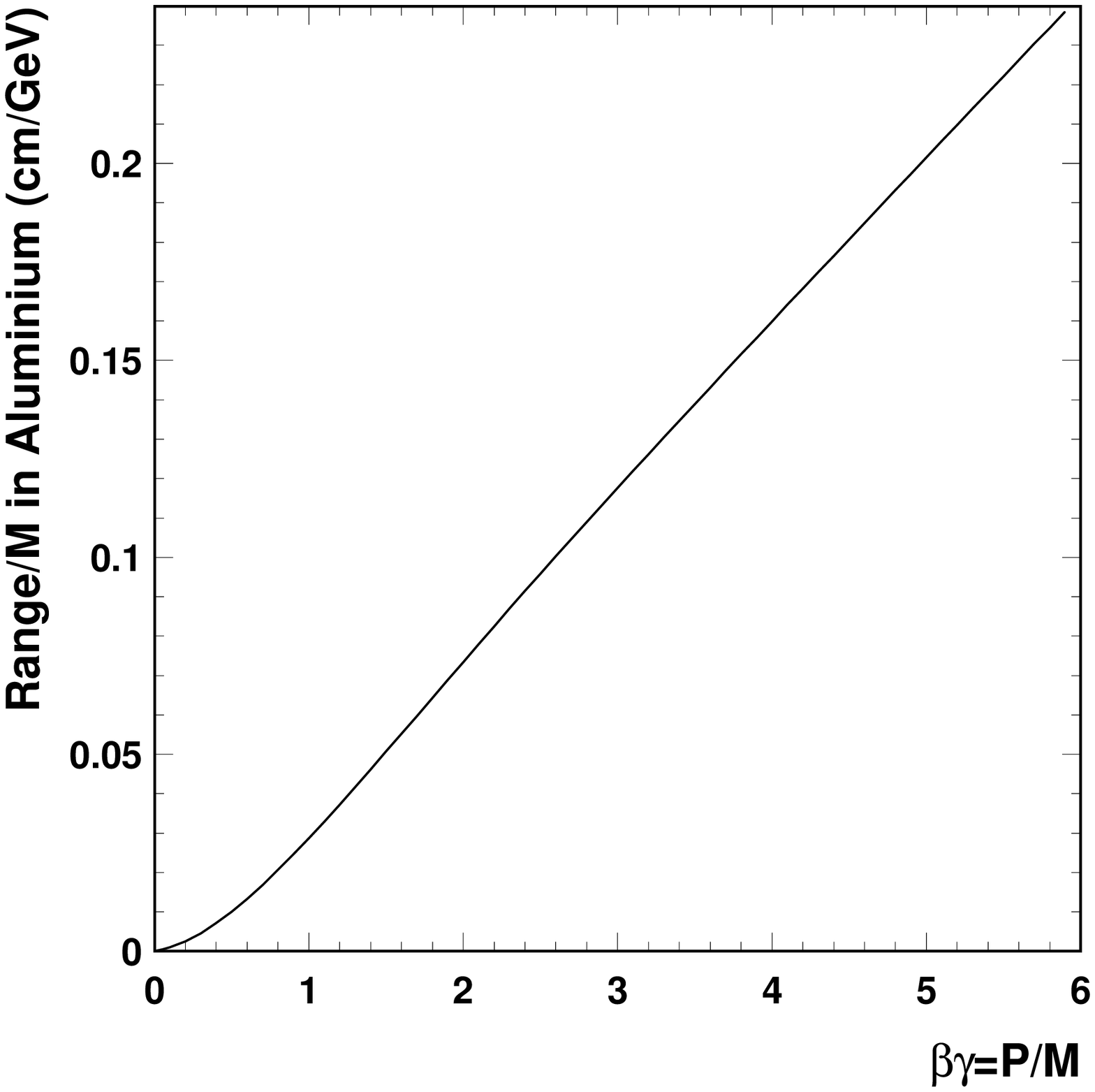,height=7cm}
\caption{Left: the $\frac{dE}{dx}$ for a Dirac monopole in aluminium
as a function of the velocity of the monopole taken
from~\cite{Ahlen:1978jy} and adjusted for the electron density in
aluminium. Right: the ratio of range to mass for a Dirac monopole in
aluminium versus $\beta \gamma$, calculated from the stopping power,
$\frac{dE}{dx}$. } \label{mono_dedx}
\end{center}
\end{figure}

The energy loss in the form of ionisation of magnetic monopoles
passing through material due to interactions with atomic electrons
has been shown to follow a form similar to the Bethe-Bloch equation
(Eq.~\ref{dedxE}) but without the multiplicative factor $1/\beta^2$
\cite{Ahlen:1978jy,Ahlen:1982mx}. The velocity-dependent Lorenz
force is responsible for the interaction between a moving magnetic
monopole
and an atomic electron in the material. In contrast, for a moving
electrically charged particle the velocity-independent Coulomb force
is responsible for the interactions. The velocity dependence of the
Lorenz force causes the cancellation of the $1/\beta^2$ factor in
the formula for $\frac{dE}{dx}$ for magnetic monopoles (compare
Eqs.~\ref{dedxE} and \ref{dedxM}). The detailed formula for the
stopping power of a magnetic monopole of strength $g$ is given in
\cite{Ahlen:1978jy} as
\begin{equation}
\frac{dE}{dx} = \frac{4 \pi e^2 g^2}{m_e c^2} n \bigg(
\frac{1}{2}\ln\big(\frac{2 m_e c^2 \beta^2 \gamma^2
T_{max}}{I_m^2}\big)- \frac{1}{2} - \frac{\delta}{2}
+\frac{K(|g|)}{2} -B(|g|)\bigg) \label{dedxM}
\end{equation}
and the modifications at very low velocity in \cite{Ahlen:1982mx}.
Here $I_m$ is the mean ionisation potential for magnetic monopoles
which is close in value to $I_e$ (Eq.~\ref{dedxE}). The relationship
between $I_m$ and $I_e$ can be expressed as $I_m=I_e \exp{-D/2}$.
Sternheimer \cite{stern} has shown, for several solids, that
$D({\textrm{Li}})=0.34$, $D({\textrm{C}})=0.22$, $D({\textrm{Al}})=0.056$, $D({\textrm{Fe}})=0.14$, $D({\textrm{Cu}})=0.13$,
and $D({\textrm{W}})=0.07$. The correction terms $K(|g|)=0.406, 0.346$ and
$B(|g|)=0.248, 0.672$ for $g_D=1,2$ Dirac Monopole strengths,
respectively \cite{Ahlen:1978jy}. Fig.~\ref{mono_dedx} (left) shows
the stopping power for a unit Dirac magnetic monopole in aluminium
as a function of the velocity of the monopole. Inspection of
Eqs.~\ref{dedxE} and \ref{dedxM} shows that the ratio of the
stopping power for a unit Dirac monopole and a unit electric charge
moving with velocity $\beta$ is $\sim 4700 \beta^2$. It can be seen
from Fig.~\ref{mono_dedx} that, as a monopole slows down, the
ionisation becomes less dense, in contrast to the case of
electrically charged particles for which the opposite is true. This
adds to the striking nature of a track left by a monopole.

The large differential ionisation energy loss of monopoles makes it
relevant to discuss the range. The range $R$, of monopoles in
aluminium is computed \cite{h1mon} by integrating the stopping power
shown in fig \ref{mono_dedx}. So that
\begin{equation}
R= \int_0^E \frac{dE}{dE/dx} = M \int_0^\gamma
\frac{d\gamma}{dE/dx(\beta \gamma)}
\end{equation}
Fig.~\ref{mono_dedx} (right) shows the computed range (normalised to
mass), for a Dirac monopole versus $P/M=\beta \gamma$ where $P$ and
$M$ are the momentum and mass of the monopole, respectively.


The theory of energy loss described above is implemented as part of
a GEANT package to describe monopole interactions in a
detector\cite{Bauer:2005xc}.

\subsection{Nuclear interactions of SMPs}\label{sec:hadscat}
As heavy (charged or neutral) hadrons propagate through a medium,
they may undergo scattering from the nuclei of the material of the
apparatus. For $R$ hadrons the interaction cross sections are
expected to be of the order of those for pion scattering (see
below).  As is shown in this section it is expected that the energy
loss of an $R$-hadron through such scattering can be sufficiently
small as to allow it to penetrate through to an outer muon chamber
and be reconstructed as a slow moving exotic particle. However, in
extreme cases such interactions can have a large impact on
experimental searches. For example, these interactions can lead to
$R$-hadrons being 'stopped' inside the detector material, because
they come to the end of their ranges due to electromagnetic and
hadronic energy losses. Also, R-hadrons could undergo charge
exchange reactions in a hadron-absorbing material such as a
calorimeter, in which case the initial charge at the production
vertex is not necessarily the same after traversing the material. An
understanding of such effects is crucial to quantify a discovery or
assess a discovery potential. This section provides an overview of
the theory of $R$-hadron scattering processes and a description of
the phenomenological approaches which have been used to describe
them. Before discussing these different approaches, we summarise
which general observations can be made, independently of any
specific model.


One important feature of $R$-hadron scattering common to all
approaches is the passive nature of the exotic heavy coloured
object. The probability that the parton ${C}_i$ of colour state $i$
will interact perturbatively with the quarks in the target nucleon
is small, since such interactions are suppressed by the squared
inverse mass of the parton. As a consequence, the heavy hadron can
be seen as consisting of an essentially non-interacting heavy state
${C}_i$ acting as spectator, accompanied by a coloured hadronic
cloud of light constituents, responsible for the interaction. Hence
the interaction cross section will be typical of that of a meson. In
addition, the effective interaction energy of the heavy object is
small. As an example, consider a ${C}_8{q}\bar{{q}}$ state with a
total energy $E$=450 GeV and a mass $m$ of the ${C}_8$ parton of 300
GeV, the Lorentz factor will be $\gamma$=1.5. Although the kinetic
energy of the $R$-hadron is 150 GeV, the kinetic energy of the
interacting ${q}\bar{{q}}$ system is only $(\gamma-1)
m_{{q}\bar{{q}}}\approx$ 0.3 GeV, (if the quark system consists of
up and down quarks). For $R$-hadrons produced at the Tevatron or LHC
with masses above $100$ GeV, the centre-of-mass energy of the system
of quarks and a stationary nucleon can thus be at most around a few
GeV. Thus, the energy scales relevant for heavy hadron scattering
processes from nucleons are low and comparable with low-energy
hadron-hadron scattering for which Regge theory is often applied.
The heavy state ${C}_i$ serves only as a reservoir of kinetic
energy.

\begin{figure}[t!]
\begin{center}
\begin{tabular}{cccc}
\epsfig{file=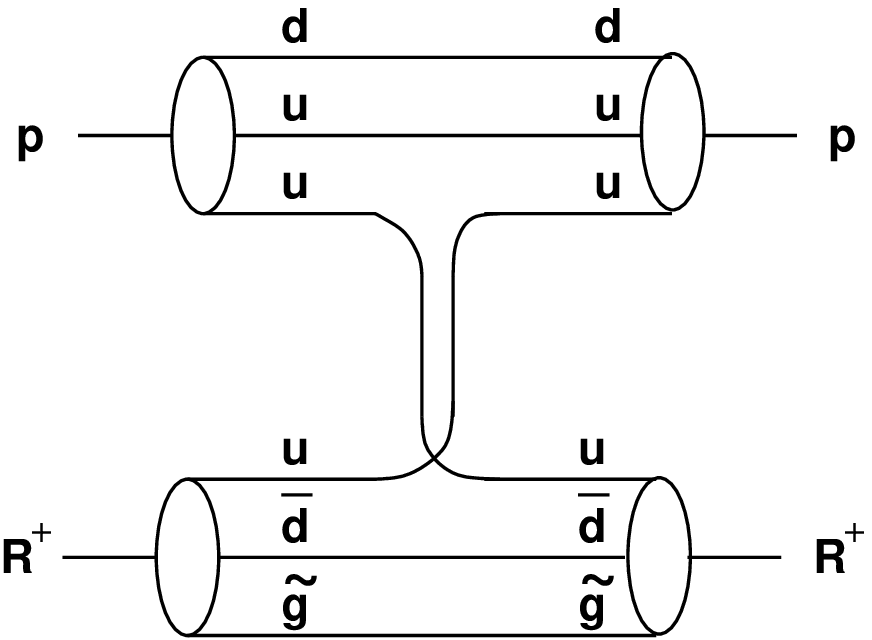,height=2.5cm,width=2.75cm}&\hspace{0.4cm}\epsfig{file=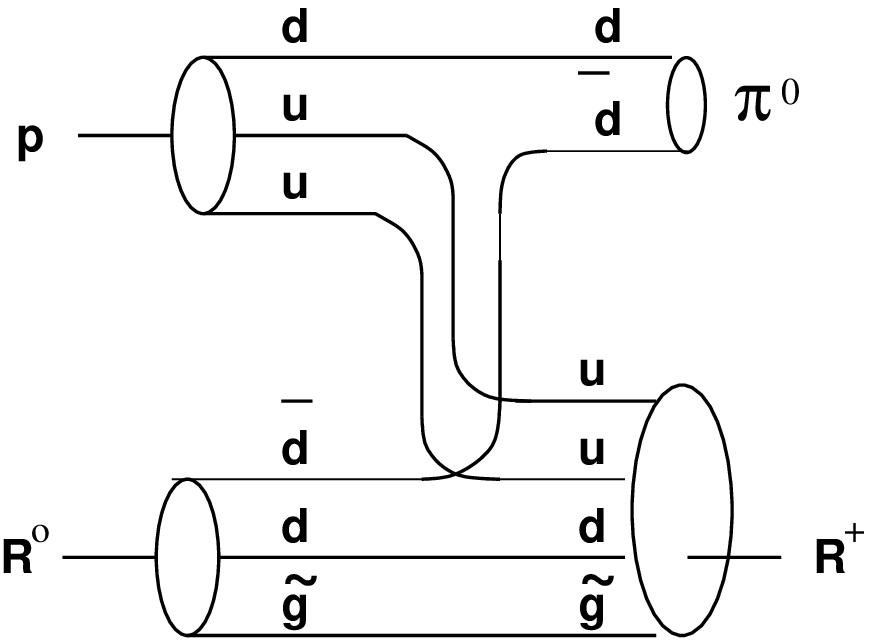,height=2.5cm,width=2.75cm}
&\hspace{0.4cm}\epsfig{file=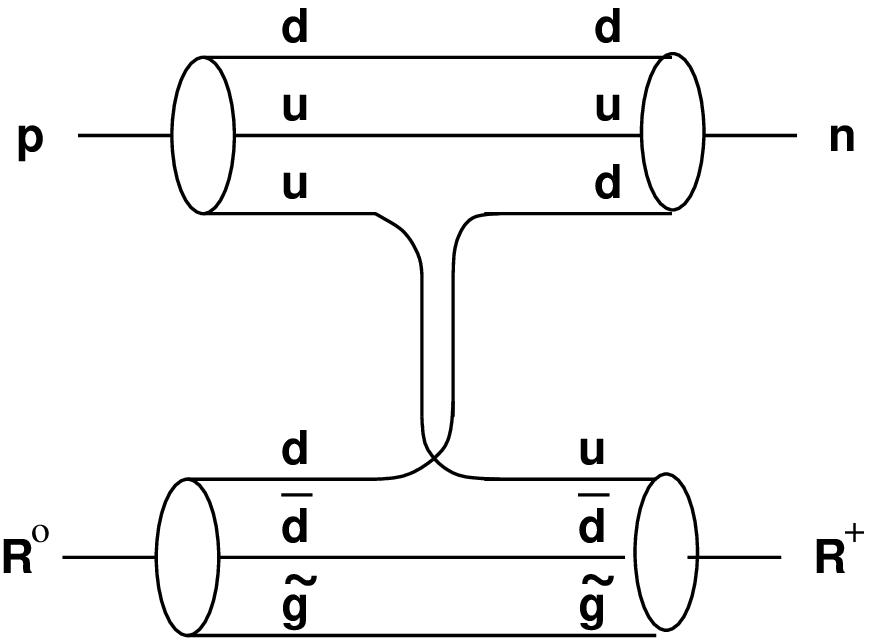,height=2.5cm,width=2.75cm}&\hspace{0.4cm}\epsfig{file=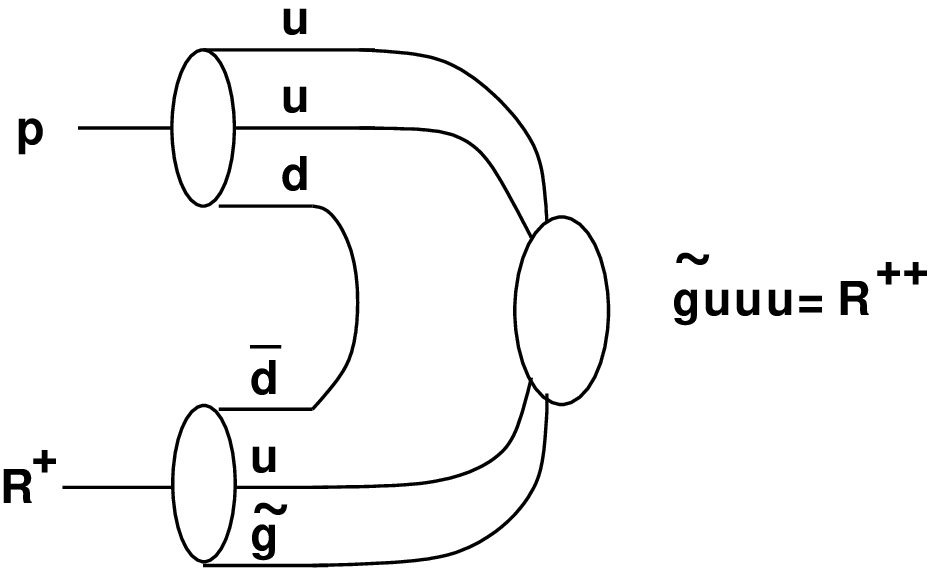,height=2.5cm,width=2.75cm}\\
(a) & (b) & (c) & (d)
\end{tabular}
\end{center}
\caption{$R$-hadron-proton scattering processes. (a) Elastic
scattering, (b) Inelastic scattering leading to baryon and charge
exchange, (c) Inelastic scattering leading to charge exchange, (d)
Resonance formation. \label{fig:pro}}
\end{figure}

Although $R$-hadrons may scatter elastically or inelastically the
energy absorbed in an elastic scattering process, such as that
illustrated in Fig.~\ref{fig:pro}  (a), is expected to be
small\cite{drees}, since the high-mass $R$-hadron scatters on a
lower mass target nucleus, and inelastic collisions are expected to
be largely responsible for the energy loss of an $R$-hadron. These
inelastic collisions may cause the conversion of one species of
$R$-hadron to another in two ways: baryon exchange, which was
overlooked until recently~\cite{Kraan:2004tz}, and charge exchange,
as shown in Fig.~\ref{fig:pro} (b) and (c), respectively. In the
first process, an exothermic inelastic $R$-meson-nucleon interaction
results in the release of a pion. The reverse reaction is suppressed
by phase space and because of the relative absence of pions in the
nuclear environment. Thus, most $R$-mesons will convert early in the
scattering chain, in passing through hadron absorbing material, e.g.
a calorimeter, to baryons and remain as baryons. This is important,
since baryons have larger scattering cross sections. Baryon
formation offers one opportunity for a charge exchange process to
take place. Charge exchange may arise in any meson-to-meson,
meson-to-baryon, or baryon-to-baryon process. Although exact
predictions of individual processes are difficult to make, the low
energies involved in $R$-hadron scattering imply that reggeon and
not pomeron-exchange will dominate, and thus charge exchange
reactions may well form a substantial contribution to all
interactions. This may lead to striking topologies of segments of
tracks of charged particles with opposite signs of charge on passage
through hadron absorbers or calorimeter material. It is also
interesting to note that such a configuration can also arise if a
neutral $R$-meson, formed as an intermediate state during
scattering, oscillates into its own anti-particle and then
subsequently interacts to become a charged
$R$-hadron\cite{Sarid:1999zx,Culbertson:2000am}.

\begin{figure}[b!]
\begin{center}
\epsfig{file=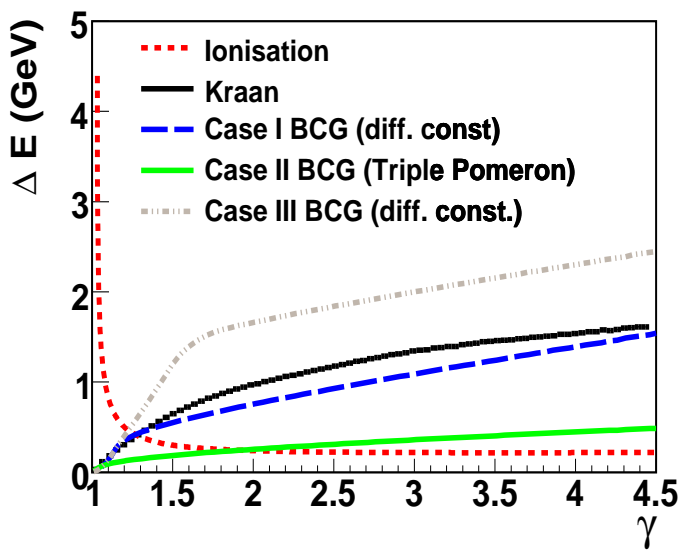,height=8cm,width=10cm}
\caption{Predictions from four phenomenological models of expected
hadronic energy loss per interaction as a function of the Lorentz
factor $\gamma$. Also shown is the ionisation energy loss corresponding to the
passage of an $R$-hadron with charge $\pm e$ through 18 cm of iron.\label{fig:fig_et} }
\end{center}
\end{figure}

Several phenomenological approaches have been developed
\cite{gunion,mafi,Kraan:2004tz} to describe $R$-hadron nuclear
scattering which are described later in Sections~\ref{sec:bcg} and
\ref{sec:aafke}. Although these differ in the phenomenology used,
they are largely based on the generic picture of $R$-hadron
 scattering described above, much of which was first introduced
  in\cite{drees}. Low-energy hadron-hadron data are typically
   used to estimate scattering
cross sections and several of these models are based on modified
Regge paramaterisations of the data.
Uncertainties in the models arise from several sources. Most models
assume $R$-meson
 scattering processes to be dominant despite it being likely that a meson will convert into
 a baryon and then stay baryonic as it propagates through matter. A
further theoretical uncertainty arises from resonance production.
The formation of resonant $R$-hadron states, as shown in
Fig.~\ref{fig:pro} (d), is expected to take place during hadronic
interactions. To date, no explicit modelling of $R$-hadron
resonances has yet been attempted. However, it has been argued
that the the minimum centre-of-mass energy required for a
scattering process is above the expected mass of the main
resonances\cite{Kraan:2004tz}. Nevertheless, since little is known about such resonances 
it is important to note that they may play an important role in accounting for $R$-hadron energy loss 
and conversions.

Below, we will discuss for each of the phenomenological approaches the interaction cross section or interaction length, scattering processes included, and energy losses.
\subsubsection{Models based on Regge theory}\label{sec:bcg}

Consider a gluino $R$-hadron scattering proces $RN\rightarrow R'X$, where $R$ is the initial R-hadron before scattering, $N$ is a single nucleon (proton or neutron), $R'$ is the R-hadron after scattering, and $X$ is the remaining system produced in the scattering, the latter is expected to be a nucleon and possibly a number of other light hadrons. Baur, Cheung and Gunion (BCG)\cite{gunion} propose three different ansatzes for the functional form of the cross section for gluino $R$-hadron scattering processes:

\begin{enumerate}
\item  $\frac{d\sigma}{d|t|dm_X}\propto$ 1 for $|t|\le$ 1 GeV$^2$ and 0
for $|t|>1$ GeV$^2$
\item  $\frac{d\sigma}{d|t|dm_X^2}$ given by a triple-pomeron form used to describe single inclusive
particle production in pion-nucleon scattering\\
$\frac{d\sigma}{d|t|dm_X^2} \propto
\frac{1}{m_X^2}\beta^2(|t|)(\frac{s}{m_X^2})^{2(\alpha_P(|t|)-1)}(m_X^2)^{\alpha_P(0)-1}$
where $\alpha_P(|t|)=1-0.3|t|$ \\ and
$\beta(|t|)=\frac{1}{(1+|t|/0.5)^2}$.
\item  $\frac{d\sigma}{d|t|dm_X} \propto
1$ for $|t|\le$ 4 GeV$^2$ and 0 for $|t|>$ 4,
\end{enumerate}
where $t$ is the four-momentum transferred from $R$ to $R'$, and $m_X$ is the mass of the remaining system of final state particles produced.
The cross section $\sigma_T(RN)$ for gluino $R$-hadron scattering on a nucleon is derived from the cross section $\sigma_T(\pi N)$ for pion nucleon scattering. The normalisation of the cross section is determined by constraints
on the collision length $\lambda_T(R)$ of an $R$-hadron. The
evaluation of $\lambda_T(R)$ is made by correcting the pion
collision length $\lambda_T(\pi)$.
\begin{equation}
\frac{\lambda_T(R)}{\lambda_T(\pi)}\equiv\frac{\sigma_T(\pi
N)}{\sigma_T(RN)}  =(\frac{C_F}{C_A})\frac{\langle r^2_\pi
\rangle}{\langle r^2_R \rangle}
\end{equation}
The colour factors $C_F=4/3$ and $C_A=3$ arise due to the low-mass
colour-octet constituent of the $R$-hadron. The term $\langle r^2
\rangle$ is the squared transverse size of the particle. In the case
of a pion and an $R$-hadron $r^2$ is given by $\langle {r^2_R}
\rangle \propto 1/m_g^2$, and $\langle {r^2_\pi} \rangle \propto
4/m_q^2$. Here, $m_q$ and $m_g$ are the constituent masses of light
quarks and gluon constituent masses, respectively, which were
assumed equal in this approach. This leads to a collision in length
in iron of $R$-mesons of around $19$ cm. However, the uncertainties
in the constituent masses of the partons would affect the collision
length, and a value of $38$ cm has also been considered within the
BCG approach.

 As mentioned above,
this model does not take into account $R$-baryons or conversion from
mesons into baryons. Charge exchange reactions are possible since
the $R$-hadron is considered to be stripped of its system of quarks
following a nuclear scattering. A refragmentation process, governed
by a probability to fragment into a charged or neutral state then
allows the formation of a new type of $R$-hadron.

The energy loss of hadron $R$ scattering on nucleon $N$ in the
process $ RN\rightarrow R'X$ is given by
\begin{equation}
\Delta E=\frac{m_X^2-m_N^2+|t|}{2m_N}.
\end{equation}
Here $\Delta E$ is then
evaluated according to the parameterisation chosen for the
differential cross section $\frac{d\sigma}{d|t|dm_X^2}$. The
distribution of mean energy loss per collision as a function of
$\gamma$ for the three different scenarios of this approach is shown
in Fig.~\ref{fig:fig_et}. All three approaches show a rising energy
loss with $\gamma$. However, there are large differences between the
various scenarios. For comparison, the ionisation energy loss corresponding
to the passage of an $R$-hadron with charge $\pm e$ through 18 cm of iron (1 interaction length for R-mesons in the approach by Kraan discussed below in Section~\ref{sec:aafke}) is also shown.

This approach has been extended by Mafi and Raby\cite{mafi}(MR),
where again a single particle inclusive scattering process $RN\rightarrow R'X$ involving only $R$-mesons was considered. However, MR consider two Regge
trajectories: an isosinglet pomeron, and an isovector reggeon $\rho$
trajectory. Using these, triple pomeron and reggeon-reggeon-pomeron
cross-section forms were extracted. The presence of reggeon exchange
incorporates charge exchange processes naturally without relying on
the BCG assumption that an $R$-hadron is stripped of its soft
partonic system in an interaction and then forced to fragment.
Two values of the
collision length $\lambda_T=19$ cm and $\lambda_T=38$ cm are used in
their analysis to estimate $R$-hadron stopping in different
scenarios and the scattering cross section functional form was
rescaled to achieve these values. The relative proportion of reggeon
and pomeron were also allowed to vary between the extreme cases of
100\% reggeon and 100\% pomeron exchange. As mentioned above, charge
exchange processes are naturally included in this approach. Baryon
exchange processes are however omitted. The typical energy loss per collision is of the order of several GeV
and is comparable with the BCG approaches.

\subsubsection{Model based on geometrical cross sections}\label{sec:aafke}

Since the behaviour of the hadron-hadron scattering cross sections
at values of the centre-of-mass energy below several GeV is specific
to the type of hadrons interacting, it is not necessarily a reliable
general guide to the scattering of $R$-hadrons. Therefore, in a
complementary approach to the Regge-based models of MR and BCG the
constant geometrical cross section is used by
Kraan\cite{Kraan:2004tz} over the full scattering energy regime.

The total nucleon interaction cross section is approximated by the
asymptotic values for the cross sections for normal hadrons
scattering off nucleons. The model assumes that only $u$ and $d$
quarks are present in $R$-hadrons and that each quark which can
interact represents a contribution of $12$ mb to the total
scattering cross section. Thus, the scattering cross sections of a
gluino $R$-meson and $R$-baryon are $24$ mb and $36$ mb,
respectively.
 A gluino-gluon state can be assumed to have the same cross section
as a gluino $R$-meson, since the geometrical cross section is
approximated by the high-energy hadron cross section, where gluon
exchange would dominate. The gluon-gluon coupling is a factor 9/4
larger than the quark-gluon coupling, but a meson has two quarks,
resulting in a cross section of a gluino-gluon state which is
(9/4)/(1+1)$\approx$1 times the cross section for a gluino R--meson.
Translating this into interaction lengths, for $R$-baryons the
average nuclear interaction length (i.e. amount of material where on
average one interaction takes place) is 12 cm in iron, 31 cm in
carbon, and 660 cm in hydrogen. For $R$-mesons these numbers are 3/2
larger.

\begin{figure}[b!]
\begin{center}
\hspace{-7mm} \epsfig{file=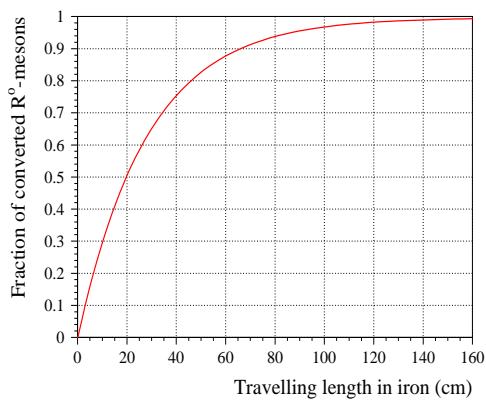,
height=6.0cm,width=7.0cm,angle=0} \vspace*{-3mm} \caption{Fraction
of $R$-mesons converted into $R$-baryons as predicted in the model
of Kraan~\cite{Kraan:2004tz}. } \label{fig:mesbar}
\end{center}
\end{figure}
This model includes predictions for all possible $2 \to 2$ and $2
\to 3$ processes. The relative rates of $2\to 2$ and $2 \to 3$
processes are, in the asymptotic region, set to 15\% and 85\%,
respectively, as suggested by hadron-hadron scattering data. A phase space
factor is used to determine the proportion of $2\to 2$ and $2 \to 3$
processes as the scattering centre-of-mass energy is reduced. Within
the sets of $2 \to 2$ or $2 \to 3$ interactions each allowed process
is assigned the same matrix element and the different rates of
processes is determined by phase space. Included in these processes
is charge exchange, and, for the first time, baryon exchange.
Fig~\ref{fig:mesbar} shows the fraction of $R$-mesons which convert
into $R$-baryons when travelling in iron. As was already mentioned
in Section~\ref{sec:hadscat}, conversion from baryons back to mesons
is highly suppressed.

Energy losses are determined by implementation of
the model into the GEANT framework (GEANT-3~\cite{Kraan:2004tz}
and GEANT-4~\cite{rasmus}), which allows a more sophisticated
treatment of energy loss in a nuclear reaction than was possible
 in the analytical approaches of BCG and MR. Issues related
  to nucleus scattering like Fermi motion, binding energy
   of the nucleons inside the nucleus, evaporation energy
    and instability of a nucleus are thus included.
    Figure~\ref{fig:fig_et} displays the nuclear energy loss per interaction
     for an $R$-meson (the curve labelled Kraan). The energy losses from this model
      are found to lie between the different BCG approaches.
       Ionisation losses are also shown and they dominate energy losses only at
        low beta values ($\beta<0.75$).
\section{Search techniques}\label{searchtechniques}

Following the previous section on the interactions of SMPs in
materials, this section describes how these interactions can be used
to search for SMPs in colliders. Search techniques include the use
of ionisation energy loss, Cherenkov radiation and time of flight.
Finally, we summarise search techniques which are only applicable to
magnetically charged SMPs.

\subsection{Methods based on ionisation energy losses}
The most commonly used observable in SMP searches is the measurement
of the continuous ionisation energy loss $\frac{dE}{dx}$.

\subsubsection{Measurements of ionisation energy losses in tracking systems}

Ionisation energy loss can be used to identify heavy charged
particles with tracking chambers. The measurement of $\frac{dE}{dx}$
is part of the routine program of calibrating charged particle
tracking chambers. However, a number of systematic studies must be
performed in order to optimise the detector calibration.  Examples
of the application of ionisation loss measurements in different
collider environments can be found in Section ~\ref{sec:esearch},
and the accompanying table~\ref{directsearches}.

\begin{figure}[htb]
\begin{center}
\hspace{-7mm} \epsfig{file=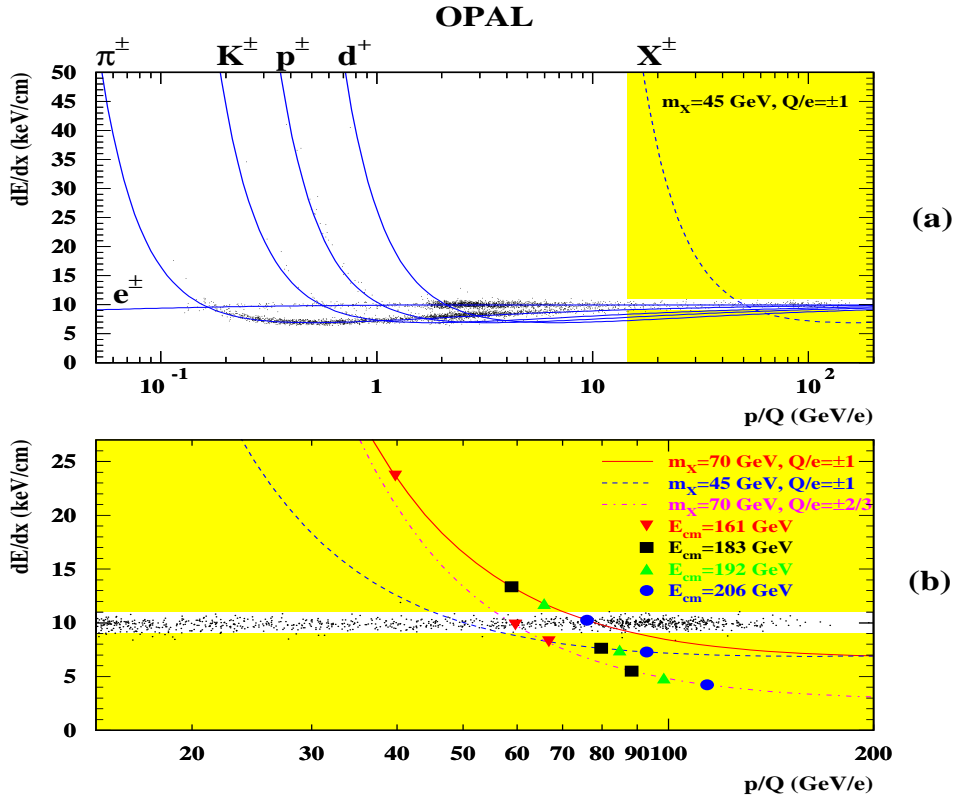,
bburx=582,bbury=600,bbllx=12,bblly=50,
height=10.0cm,width=13.0cm,angle=0} \vspace*{18mm}
\caption{Measurements of $\frac{dE}{dx}$ vs $\frac{p}{Q}$ as
measured by the OPAL  experiment~\cite{chargedopal1}, where $Q$ is the charge of the particle. The smooth
curves show the expected values of the mean values of
$\frac{dE}{dx}$ for the known particles, while the dotted points are
the measured values. The regions in which unknown massive particles
are sought are shaded. Also shown are lines
which illustrate the expected $\frac{dE}{dx}$ values of SMPs with
specific masses and charges. } \label{alephdedx}
\end{center}
\end{figure}

It can be seen from Eq.~\ref{dedxE} that the value of
$\frac{dE}{dx}$ is dependent on the velocity factors $\beta,\gamma$.
The particle's momentum ($p=M \beta \gamma$) is measured
independently, usually from the curvature of the track in the
magnetic field. Hence, comparing the measured value of
$\frac{dE}{dx}$ with an independent measurement of track momentum,
the particle mass can be determined. Fig.~\ref{alephdedx} shows an
example of what mass separation was achieved in the OPAL experiment
when simultaneously measuring $\frac{dE}{dx}$ and momentum. The
regions in which SMPs possessing various values of charge and mass
could be manifest are shaded.

Several issues must be taken into account in the identification of
charged particles by $\frac{dE}{dx}$ measurements. First of all,
considerable fluctuations occur in single measurements of
$\frac{dE}{dx}$. These are not Gaussian but are asymmetric with a
high-energy tail due to the emission of energetic secondary
electrons ($\delta$ rays), giving rise to the Landau distribution
\cite{Rossi}. Examples of methods adopted to avoid complications due
to these high-energy tails include disregarding abnormally large
single measurements of $\frac{dE}{dx}$ and taking the mean of the
remaining measurements~\cite{Atwood:1991bp} or suppressing tail
distributions~\cite{Abt:1996xv}. Another method is to perform a
maximum likelihood fit of a Landau distribution to the sample of
measurements on a track \cite{h1highion} to determine the mean value
of $\frac{dE}{dx}$ for the track. The particle masses determined
from the $\frac{dE}{dx}$ at a fixed momentum had smaller high-mass
tails using this method than with the others \cite{h1highion}. Hence
this is likely to be the best technique to employ in searches for
unknown particles, since there will be less background by smearing
from the lower mass region. The large numbers of measurements needed
for reliable $\frac{dE}{dx}$ determination means that silicon vertex
detectors with small numbers of detector layers are likely to be of
limited usefulness in searches for unknown SMPs.



Further problems in measuring $\frac{dE}{dx}$ for unknown massive
particles could arise due to saturation of the electronics used in
the track detectors. For such searches the electronics must have a
wide enough dynamic range otherwise large values of  $\frac{dE}{dx}$
become unmeasurable due to saturation. This is particularly true for
magnetic monopoles, for which $\frac{dE}{dx}$ is significantly
 larger than that for electrically charged particles, but may already play a role
  for slow electrically charged ones. Although saturated hits have a tracking
   resolution much worse than that of typical unsaturated data, the presence
    of many saturated channels is itself a distinctive feature of the presence of a high-mass charged particle. A characteristic pattern of a charged particle helix
    can generally easily be recognised from the spatial distribution of saturated hits. The technique of searching for saturated hits has been applied in Ref.~\cite{chargedaleph2}. Saturation of electronics
 as a consequence of highly ionising particles may in particular have a
  non-negligible effect in the future LHC collider experiments where the resulting dead time as a result of electronics saturation may be of the order of the bunch crossing time.
 The effect of highly ionising particles on the CMS silicon strip
tracker has been studied in
Refs.~\cite{Adam:2005pz,Bainbridge:2002wt}.

There are several kinds of backgrounds producing highly ionising
signals.  An important background source to such studies is
positively charged nuclei. These typically arise as spallation
products from secondary interactions of particles produced in the
primary interaction. Such interactions take place with nuclei in the
material of the apparatus. This accounts for stronger limits for
negatively charged SMPs using this approach, since spallation
products are positively charged. Another background arises from the
finite resolution of a tracking chamber, causing, for example, two
overlapping tracks to be measured as one; this results in a highly
ionizing signal. These backgrounds are mainly caused by photon
conversions. Finally, interactions of the colliding beams with
residual gas atoms in the beam pipe can also produce highly ionising
spallation products. However, most background events can usually be
effectively removed by the application of a lower cut on the
momentum~\cite{chargedaleph3}.


\subsubsection{Methods using ionisation damage in plastic foils}
\begin{figure}[htb]
\begin{center}
\hspace{7mm}
\epsfig{file=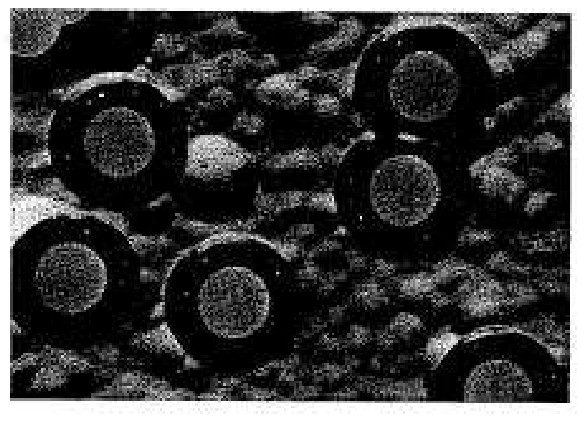,bburx=582,bbury=725,bbllx=12,bblly=164,
height=20.0cm,width=20.0cm,angle=0} \vspace*{-16cm}
\caption{Microphotograph of a heavily etched CR39 sheet (type EN3)
exposed to sulphur ions of 200 GeV/nucleon \cite{Bertani:1990tq}.
The sheet thickness was reduced from 1.4 mm to 0.2 mm by the heavy
etching. Note the holes produced by the sulphur nuclei. }
\label{lexan} \vspace{0.8cm}
\end{center}
\end{figure}

Next to $\frac{dE}{dx}$ measurements to identify heavy particles in
tracking chambers, heavily ionising single particles (magnetic or
electrically charged) can be detected by the damage they cause to
certain plastic foils eg. foils made from plastics such as lexan,
CR39 and Makrofol \cite{Cecchini:2005ix}. Such damage is caused by
both the NIEL and the ionisation energy loss. The effect of the
damage is made visible by chemically soaking the foil in a
concentrated alkali solution which etches the region around the
damage centre into small holes (pits) which are visible under a
microscope~\cite{Fleischer:1975ya}. The value of $\frac{dE}{dx}$ of
a heavily ionising track can be determined from the etch rate i.e.
the ratio of the rate of increase in depth of the pit to the
decrease in overall thickness of the foil due to the chemical
action. Fig. \ref{lexan} shows the pits produced by sulphur ions in
a heavily etched CR39 foil (type EN3) as reported in
\cite{Bertani:1990tq}. The etching process only makes the tracks
visible for ionisation levels above a threshold value so that the
lightly ionising tracks remain invisible and only the heavily
ionising tracks are seen. For heavy ions of velocity $\beta$ and
atomic number $Z$, the experiment of Ref.~\cite{Bertani:1990tq}
found that the threshold corresponded to ions with $Z/\beta \sim 8$
for CR39 (type EN3). In collider
experiments~\cite{Price:1987py,Price:1990in,Bertani:1990tq,Pinfold:1991ix,
Kinoshita:1992wd, Pinfold:1993mq}, the layers of foils surround the
beam pipe at an interaction point and are left to be exposed to the
products of the beam interactions. After the exposure the foils are
soaked in the appropriate chemical to etch out the damage centres.
Usually, the inner layer foil is scanned under a microscope to
search for etched pits. The outer layers are then scanned for pits
which align with those found in the inner layer to search for
heavily ionising continuous tracks.

\subsection{Ring Imaging Cherenkov detection}\label{rich}
Another way to exploit the electromagnetic interactions of SMPs to identify them is via the Cherenkov
effect\cite{Cerenkov:1937vh,rn}. A particle, travelling with
velocity $\beta$ which is faster than the speed of light in a medium
of refractive index $n$, emits light waves by the Cherenkov effect
in a cone at an angle of $\cos \theta = 1/n \beta$ relative to the
track direction. The number of Cherenkov photons per unit path
length is proportional to $\sin^2 \theta$. Particles with velocity
less than the speed of light in the medium do not emit Cherenkov
light. Unlike the other LEP experiments, the DELPHI detector
contained Ring Imaging Cherenkov (RICH)
systems\cite{Anassontzis:1992jq,Adam:1994yp,Adam:1996gb,Albrecht:1999ri}
comprising two Cherenkov radiators with different refractive
indices: one in the liquid and one in the gaseous phase. To select
low-velocity massive particles it was required that neither detector
should give a detected Cherenkov light signal for high-momentum
tracks. The DELPHI experiment was able to use this technique for a
number of SMP
searches\cite{chargeddelphi1,chargeddelphi3,Abdallah:2002rd}, the
results of which are described in Section~\ref{sec:esearch}.

\subsection{Methods based on time of flight}\label{sec:tof}

Another technique which can be used for the detection of SMPs is
the method of time of flight. Although less widely used than the
$\frac{dE}{dx}$ method, this method has been applied in several
collider environments, examples of which can be found in
Section~\ref{ss:search}. Massive particles are produced with a
smaller velocity than light particles, and thus would have a larger
time of flight. For electrically charged particles the mass of the
particle can be determined by correlating the flight time with the
track momentum, $p$, measured from another source e.g. from the
track curvature in a magnetic field. Timing can be obtained from the
detection of the tracks in scintillation counters. Since $M=p/\beta
\gamma$, particles are identified by plotting a graph of $1/\beta^2$
against $1/p^2$ which is linear with slope $1/M^2$ for particles of
mass $M$. The accuracy of determination of the mass due to the
momentum and time of flight uncertainties, which are assumed to be
uncorrelated, is
\begin{equation}
\bigg(\frac{\Delta M}{M}\bigg)^2 = \bigg(\frac{\Delta p}{p}\bigg)^2 + \bigg(\gamma^2\frac{\Delta \beta}{\beta} \bigg)^2
\end{equation}
where the uncertainty in the velocity $\Delta \beta/\beta = \Delta t
/t$ with $t$ the time of flight. For a typical timing resolution of
$\Delta t \sim 1$ ns and performing time of flight over a distance
of $\sim 3$ m~\cite{Aarnio:1990vx}, the mass resolution $\Delta M/M$
varies between $2\%$ to $23\%$ for particles in the range of $0.2 <
\beta < 0.8$, neglecting the momentum resolution. Hence the
technique is comparable to the $\frac{dE}{dx}$ technique in the
precision of the mass measurement~\cite{h1highion}.

Backgrounds in time of flight measurements can arise from
 instrumental effects such as mismeasured times
 or random hits in the scintillation counters. Cosmic
 rays may form a substantial background, as may $K$-decays in flight. In
 addition,
  at colliders with high-frequency bunch
 crossings and a large number of interactions per bunch, such
 as the LHC, the particles from
one bunch may become confused with particles from another bunch. In
this situation, it may not be possible to make a unique
determination of the time of flight.

A slightly different technique which has not been used so far but
could be used to detect heavy charged particles is to measure the
track velocity with a tracking system based on wires and gas. The
detection time of the signal of a traversing charged SMP is in simplified form given
by
\begin{equation}
t=t_0+t_{TOF}+t_{drift}+t_{electronics}
\end{equation}
where $t_0$ is bunch crossing time, $t_{TOF}$ is the time of flight
from the beampipe to the wire, $t_{drift}$ is the drift time of the
signal inside the gas to the wire, and $t_{electronics}$ is the time
for the transmission of the signal. For a slow particle, $t_{TOF}$
is large. Since default track reconstruction programs are based on
expectations for $t_{TOF}$ for light particles, a misalignment
occurs for slow moving particles. If the reconstruction of the track
is successful, this pattern is very distinctive for heavy slow
particles. It could in addition be used to support a heavily
ionizing signal. This method has not yet been applied in data, but
has been studied in Ref.~\cite{thesisaafke} with the ATLAS muon
reconstruction software~\cite{Shank:2003yx}.

\subsection{Specific techniques for magnetic monopoles}
In addition to the techniques described above, a number of
approaches have been used purely to search for particles with
magnetic charge. Examples of applications of the techniques discussed below are given in Section~\ref{monsearch}.

\subsubsection{Parabolic tracks}\label{sec:para}

Monopoles experience a
force $g{\bf B}$ in a magnetic field ${\bf B}$ which causes them to accelerate. With the field aligned along the $z$ axis
they assume a parabolic trajectory with
\begin{equation}
z(r)-z_v = \frac{g|{\bf B}| r^2}{2 e P_T \beta_T~10^9} +
\frac{r}{\tan \theta_0} = \frac {g_D 20.54 |{\bf B}| r^2}{2 P_T
\beta_T} + \frac{r}{\tan \theta_0}
\end{equation}
where $z_v$ is the $z$ coordinate of the vertex and $z(r)$ is the
coordinate of a point on the trajectory at distance $r$ from the
proton beam with lengths in metres and $|{\bf B}|$ in Tesla. The
transverse momentum and transverse velocity of the monopole are
$P_T$ GeV/c and $\beta_T$, respectively, and $g_D$ is the pole
strength in units of the Dirac Monopole. The initial angle of the
monopole to the magnetic field direction is $\theta_0$ and $e$ is
the unit of electric charge.  In this equation $g$ is the magnetic
pole strength which is negative (positive) for South (North) poles
which decelerate (accelerate) in the $+z$ direction in the magnetic
field.

\begin{figure}[htb]
\begin{center}
\hspace{-7mm} \epsfig{file=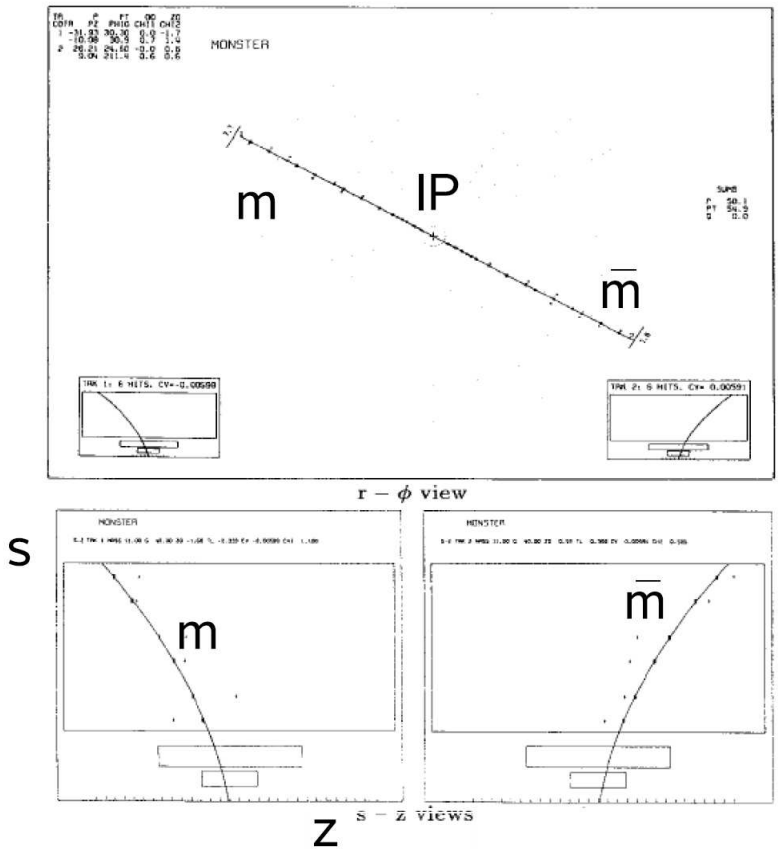,
height=12.0cm,width=12.0cm,angle=0} \vspace*{-3mm}
\caption{Simulated monopole, anti-monopole production at the TASSO
detector~\cite{Braunschweig:1988uc}. The top plot shows the paths of
the particles (produced at the centre) in the $r/\phi$ view. The
lower left (right) plot shows the path of the monopole (anti-monopole)
as it is accelerated (decelerated) in the magnetic field, in the $s-z$ view, where $s$ is the total distance travelled. }
\label{tasso}
\end{center}
\end{figure}

A simulation of the passage of a monopole, anti-monopole
$(m\bar{m})$ pair at the TASSO experiment~\cite{Braunschweig:1988uc}
is shown in Fig.\ref{tasso}. Fig.\ref{tasso} (top) shows the
$r-\phi$ plane (transverse to the magnetic field direction) of the
paths of the two particles which are produced at the primary
interaction point (IP). However, as seen in the lower plots, the
paths of the particles are clearly disturbed by the magnetic field
in the $s-z$ view, where $z$ is a distance parallel to the beam axis
and $s$ is the total distance travelled by the monopole along its
path.

\subsubsection{Searching for stopped magnetic monopoles}\label{stopmm}

\begin{figure}[t!]
\begin{center}
\epsfig{file=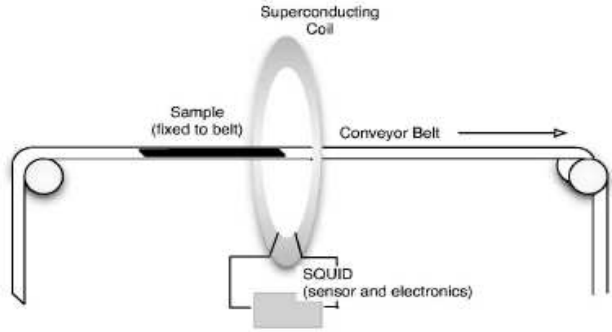,height=5cm,width=10cm} \caption{A
schematic diagram showing the SQUID apparatus used by the H1
experiment~\cite{h1mon}. The conveyer belt travelled in small steps
until the sample was passed completely through the coil. At each
step the current in the superconducting coil was read.}
\label{fig:h1mon}
\end{center}
\end{figure}

A further method to search for monopoles utilises the fact that the
$\frac{dE}{dx}$ of magnetic monopoles is so large that they tend to
stop in the beam pipe or apparatus surrounding the interaction
region. When a monopole stops, i.e. reaches a speed comparable to
that of an electron in a Bohr orbit of the atom ($\beta \sim 0.01$),
it is expected to become bound to the nuclei of the atoms of the
material. The binding energy is dependent on the magnetic dipole
moment of the nucleus
\cite{Bracci:1984zr,Olaussen:1984xb,Milton:2001qj} which depends on
the nuclear spin. Hence magnetic monopoles should remain bound in
materials such as aluminium (nuclear spin 5/2). Since the binding
energy of a magnetic monopole to a nucleus with zero magnetic moment
is expected to be small, the monopoles may diffuse out of materials
made of mainly even-A and even-Z nuclei with spin zero i.e. zero
magnetic moment, e.g. carbon. The strong divergence of the monopole
magnetic field, $B=\mu_o g/4 \pi r^2$, causes a persistent current
to flow if the monopole is passed through one or more
superconducting (sense) coils . In contrast, divergenceless magnetic
fields from ubiquitous magnetic dipoles and higher moments cause the
current to return to zero after complete passage through the coil.
If the material surrounding the beam is cut into small pieces which
are then passed through the superconducting coil a residual
persistent current signifies the presence of a monopole in the
sample. The current is measured using a Superconducting Quantum
Interference Device (SQUID) connected to the sense coils. This
method was invented to search for monopoles in lunar rocks
\cite{Alvarez:1963zp,Alvarez:1970zu,Alvarez:1971zt,Ross:1973it}. The
response of the superconducting
coils can be calibrated by the traversal of the coil by one end of a long,
thin solenoid, the magnetic field at the end of which approximates
to that of a magnetic monopole. Since the induced monopole current
is persistent, it is also possible to increase the sensitivity
of the apparatus by repeated traversals of the same sample of
material. This technique of gradually building up a signal is
especially useful when searching for particles with values of the
magnetic charge substantially less than the Dirac charge $g_D$.
Using this technique experiments have searched for magnetic charges
as low as
$\frac{1}{10}g_D$\cite{Alvarez:1963zp,h1mon}. The
apparatus used for the H1 search \cite{h1mon} is typical of such
searches and is illustrated in Fig.\ref{fig:h1mon}.

In addition to using a SQUID, it is also possible to use strong
external magnetic fields to look for stopped monopoles. In
Refs.~\cite{Carrigan:1973mw,Carrigan:1974un,Carrigan:1977ku}
materials in which monopoles could be stopped were placed in front of
a pulsed solenoid. A magnetic field of around 80 kG is expected to
be large enough to liberate a monopole and accelerate it towards
the detection systems comprising scintillators and plastic track
detectors. Searches have been sensitive to charges in the range
0.03-0.24$g_D$.
\section{Searches at colliders}\label{ss:search}
This section gives an overview of the various \emph{direct} searches
which have been performed for SMPs at colliders, i.e. searches based
on detecting SMPs by their passage through the detector. The results
of direct SMP searches broadly fall into three different
classifications.

\begin{itemize}

\item {\bf The first class} concerns general searches for SMPs,
which are made without assumptions on the properties of the SMP,
other than the values of the electric or magnetic charge, spin and
mass. Predicted cross sections are not available and
model-independent upper limits on production cross sections are
extracted within a specific mass acceptance region. Such searches are designed to be
sensitive to SMPs with a wide range of values of mass and charge, and thus cover many of the
SMPs predicted in Section~\ref{sec:scen}, as well as unexpected particles.

\item {\bf The second class} of searches assumes the existence of
certain types of SMPs with specific quantum numbers. These searches
are based on minimal theoretical scenarios in which the production
mechanisms and mass-dependent cross sections are assumed to be
known. The influence on the SMP production cross sections of any
other exotic particles which may be predicted within these
scenarios, eg via loop diagrams, is small. This allows lower limits
on SMP masses to be derived which depend dominantly on the
properties of the SMP and not on other model parameters.
\item {\bf The third class} concerns searches made within theoretical scenarios
for which the search results are quoted in terms of the complicated
model parameter space.
\end{itemize}
Searches for electrically charged particles have been made in all
three classes whereas, owing to the difficulties in calculating
short-distance monopole interactions, monopole searches are
typically made within the first class, with comparatively few
searches also quoting a mass limit. It should be noted that a given
study can present sets of results which fall into different classes.

 This section is organised as follows. In Section~\ref{sec:esearch} direct searches for electrically charged
particles are reviewed.
Searches at $e^+e^-$, hadron-hadron, and lepton-hadron facilities
are described. Searches in each collision environment are described
in the class order given above.
We then make a short summary of all
of the direct collider searches, pointing out the most stringent
limits which have been obtained. Finally, some \emph{indirect}
searches, i.e. searches for physics signals in which SMPs play an
indirect role (for example from their presence in loop diagrams) but
may not manifest themselves as final-state particles, are briefly
touched upon at the end of Section~\ref{sec:esearch}. The same
structure is adopted for the description of monopole searches in
Section~\ref{monsearch}, although in this case the vast majority of
search results fall within the first classification.

The intention in this section is not to provide an exhaustive
compendium of all results in this field, such as that which can be
found in Ref.~\cite{rn}. However, tables of selected results showing
the most stringent limits extracted for electrically charged
particle searches which belong to class 1 and to classes 2 and 3 are
summarised in Tabs.~\ref{class1searches},~\ref{directsearches},
respectively. Tab.~\ref{monosearches} summarises magnetic monopole searches.

\subsection{Searches for electrically charged SMPs}\label{sec:esearch} Searches for SMPs
have been performed at  $e^+e^-$, lepton-hadron, and hadron-hadron colliders. The majority of recent searches has been
performed at LEP
and comparatively few studies have been made at the Tevatron and
HERA.



\subsubsection{Searches at $e^+e^-$ experiments}\label{sec:eeSMPs}
The early studies at low-energy $e^+e^-$ facilities in the 1980s,
such as PETRA ($\sqrt{s}=27-35$ GeV), PEP ($\sqrt{s}=29$ GeV) and
Tristan ($\sqrt{s}=50-61$ GeV), comprised generic searches for SMPs with
unexpected mass and charge, and, in particular, searches for particles possessing
fractional charges, as inspired by the possibility of the existence of free
quarks\cite{rn}.
Although the notion of the existence of free quarks had fallen out of favour in the
1990's, the experiments at LEP ($\sqrt{s}=91.2, 130-209$ GeV)
 continued to search for
fractionally charged objects, and, in addition, made a
number of SMP searches within exotic scenarios, such as SUSY. Owing to
the high collision energies and large luminosity samples at LEP-1 (LEP-2), typically $\sim 100$ ($\sim 700$) pb$^{-1}$, the most stringent results were usually
extracted at LEP. Unless stated to the contrary, the LEP results quoted
here were based on luminosity samples of approximately these magnitudes.

The $e^+e^-$ searches for generic SMPs used simple topological cuts
to look for the exclusive or inclusive production of SMPs. The
principal experimental observable  used was the ionisation energy
loss measured in a tracking system. This was the approach adopted by
the low energy experiments~\cite{Bartel:1980zw,Weiss:1981fu,
Besset:1982xa,Guryn:1984qv,Aihara:1984px,Albrecht:1985zu,Bowcock:1989qj,Adachi:1990pg}
and by ALEPH~\cite{chargedaleph3,chargedaleph2},
OPAL~\cite{chargedopal4,chargedopal2,Ackerstaff:1998si,chargedopal1}
and L3~\cite{chargedl3_2,Acciarri:2000wy,Achard:2001qw}. For the
DELPHI studies, a RICH detector was used, in combination with
ionisation measurements from a
TPC~\cite{chargeddelphi2,chargeddelphi3,chargeddelphi1}.

The most stringent $e^+e^-$ limit, in terms of mass reach, for fractionally
charged particles, was obtained recently at LEP-2 by
OPAL~\cite{chargedopal1}. For this study, the collision
centre-of-mass energy spanned 130-209 GeV. An exclusive pair-production
mechanism $e^+e^- \rightarrow X \bar{X}$ was assumed, and model-independent
 upper limits on the production cross section for weakly interacting
scalar and spin $\frac{1}{2}$ particles with charge $\pm e$ of
between 0.005 and 0.03 pb were extracted at 95\% confidence level (CL), for
a sensitive mass region between 45 and 103 GeV.
This work also made a search for particles possessing fractional
charges $\pm\frac{2}{3}e$, $\pm \frac{4}{3}e$, $\pm \frac{5}{3}e$,
with the resultant cross-section limits lying between 0.005 and 0.02 pb at 95\% CL.
Using a substantially smaller luminosity sample ($\sim$ 90 pb$^{-1}$), and
for centre-of-mass energies of 130-183 GeV, DELPHI was able to place
limits on the
production cross-section of SMPs with charges
$\pm \frac{2}{3}e$~\cite{chargeddelphi3}, for masses
in the range 2-91 GeV.
In this work a free squark model was
assumed, and cross-section limits of between 0.04 and 0.6 pb were
obtained at 95\% CL. Using an exclusive slepton pair-production model,
DELPHI also extracted upper cross-section limits in the range 0.05-0.3 pb
for SMPs with charge $\pm e$ for masses up to 93
GeV~\cite{chargeddelphi3,chargeddelphi1}. Less stringent cross section
limits for SMPs with charge $\pm e$ were also extracted by
ALEPH~\cite{chargedaleph2}. The L3 experiment~\cite{Achard:2001qw,chargedl3_2,Acciarri:2000wy}
searched for signatures of charged heavy leptons at LEP2. However, the
results of these studies did not include any upper cross-section limits
which are relevant for generic SMPs.

\begin{figure}[t!]
\begin{center}
\epsfig{file=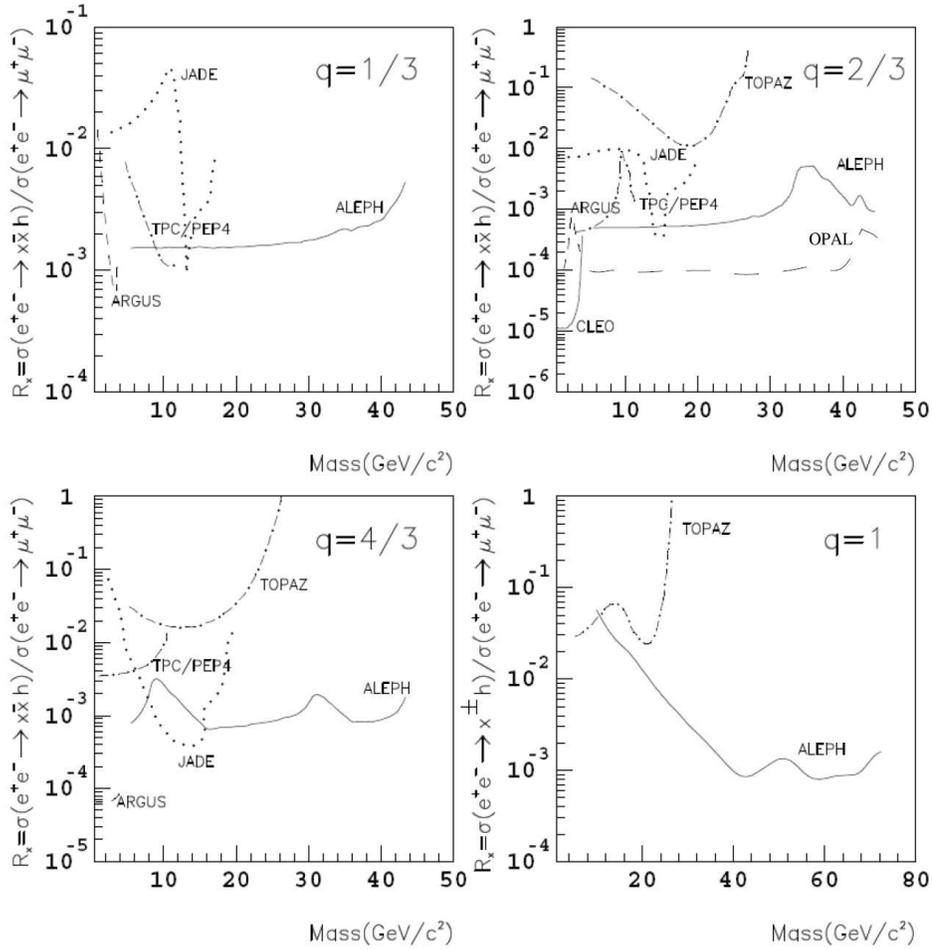,height=13cm,width=15cm} \caption{The upper
limits at 90\% CL on the ratio $R_X$ as extracted at LEP-1
(ALEPH~\cite{chargedaleph3} and OPAL~\cite{chargedopal2}) and
lower-energy experiments~\cite{Bartel:1980zw,Weiss:1981fu,
Besset:1982xa,Guryn:1984qv,Aihara:1984px,Albrecht:1985zu,Bowcock:1989qj,Adachi:1990pg}.
}\label{fig:smp-free}
\end{center}
\end{figure}

Searches at LEP-1 for fractionally charged objects
made by ALEPH~\cite{chargedaleph3,chargedopal2} and lower energy $e^+e^-$ colliders typically
expressed their results in terms of limits on $R_X$, the ratio of
the cross section for the single or pair production of SMPs to that of
exclusive $\mu^+\mu^-$ production. Fig.~\ref{fig:smp-free} shows
limits on $R_X$ (to 90\% CL), which were obtained from $e^+e^-$
experiments for fractionally charged SMPs with masses below 45
GeV~\cite{chargedaleph3,Bartel:1980zw,Weiss:1981fu,
Besset:1982xa,Guryn:1984qv,Aihara:1984px,Albrecht:1985zu,Bowcock:1989qj,Adachi:1990pg}
for the putative charges $\pm \frac{1}{3}e$, $\pm \frac{2}{3}e$,
 and $\pm \frac{4}{3}e$. Also shown is the ratio of the upper limit on the single inclusive production cross section
of an SMP with charge $\pm e$ to the exclusive dimuon cross section. The LEP-1 results from
ALEPH~\cite{chargedaleph3} and OPAL~\cite{chargedopal2} were based on luminosity samples
of 8 and 74 pb$^{-1}$, respectively and extend down to 5-10 GeV in mass. The
lower energy $e^+e^-$ limits exclude SMPs with masses as low as
$\sim 1$ GeV.
To obtain the results shown in Fig.~\ref{fig:smp-free} the experiments assumed that the SMPs
follow a momentum distribution suggested for a heavy particle:
$E\frac{d^3N}{dp^3}=$constant~\cite{Bjorken:1977md} (see Section~\ref{hadro}). The
extracted limits are very sensitive to the
form of the momentum distribution. ALEPH demonstrated that the limits can
change by more than a factor of five if the momentum dependence is derived
from a fit to the Feynman-$x$ spectra of inclusively produced
hadrons~\cite{chargedaleph3}.

In addition to the results described above, OPAL also
extracted mass-dependent cross-section
limits on the single inclusive production of SMPs with charges $\pm
e$,$\pm \frac{4}{3}e$, $\pm \frac{5}{3}e$, and $\pm 2e$. However,
the experiment was not sensitive to particles with charge $\pm
\frac{1}{3}e$. Furthermore, ALEPH and OPAL and some of the lower
energy experiments~\cite{Bartel:1980zw,Weiss:1981fu,Guryn:1984qv,Albrecht:1985zu}
also presented limits on the exclusive pair-production of SMPs with specific charges.
The obtained limits on SMPs with charges $\pm \frac{2}{3}e, \pm e, \pm \frac{4}{3}e, 2e$
were of similar values to the limits they quoted for the inclusive cases.

 The possibility of the production of SMPs with values
of charge substantially in excess of $e$, such as $Q$-balls, has
been investigated by experiments using plastic track detectors.
While these were primarily monopole searches, the experiments were
also sensitive to non-magnetically charged particles with unexpected
values of electric charge. These works are often overlooked in
studies which survey SMP searches although they are extremely
important in extending the range in charge to which collider
experiments are sensitive. Several such experiments took place at
LEP-1~\cite{Pinfold:1993mq,Kinoshita:1992wd}. In
Ref.~\cite{Kinoshita:1992wd} the MODAL passive detector was deployed
at the vacant I5 intersection point at LEP and was exposed to
$e^+e^-$ collisions corresponding to an integrated luminosity of
48 nb$^{-1}$. The experiment was sensitive to SMPs with electric
charges of up to approximately $|240e|$ at low masses, with decreasing charge
sensitivity as the mass increases. An upper limit on the production cross-section of
highly ionising particles of 70 pb was extracted, assuming that the efficiency was equal to
the maximum value of acceptance for a pair of highly ionising particles. A second
search~\cite{Pinfold:1993mq}, using passive detectors deployed at
the OPAL intersection point, was exposed to a far higher luminosity
(8.7 pb$^{-1}$). However, this experiment quoted its sensitivity only
for particles possessing various values of magnetic charge (see Section~\ref{sec:monee}).

In addition to the generic searches, model-dependent (class 2)
searches for electromagnetically charged and weakly interacting SMPs
have also been made. Using the $\frac{dE}{dx}$ technique heavy
lepton limits have been extracted at LEP~\cite{chargedopal1,
chargedl31} for a range of putative non-SUSY, heavy leptons, such as
so-called sequential and mirror leptons (see
Section~\ref{peter:fourth}). Using centre-of-mass energies up to
$209$ GeV, the L3 experiment~\cite{Achard:2001qw} was able to
exclude new leptons with masses below around $103$ GeV to 95\% CL.
Varying the exotic scenario changes the limit by less than $1\%$.

A number of searches have been made by the LEP experiments for
stable SMPs within SUSY scenarios. ALEPH~\cite{aafke2} has obtained
lower-mass limits for stable coloured sparticles through an analysis
of LEP-1 and LEP-2 data. In Ref.~\cite{aafke2}
 MSSM scenarios with a stable gluino or squark and R-parity conservation was
 assumed (see Sections~\ref{peter:mssm}-~\ref{peter:amsb}). For a stable
gluino hypothesis, the process $e^+e^- \rightarrow q\bar{q}\tilde{g}\tilde{g}$ with LEP-1 data was considered. In
this scenario a radiated gluon splits into two stable gluinos.
The hypotheses that $R$-hadrons would be formed as either as neutral or charged states were both considered. In
the former case, $R$-hadron jets were identified via a discrepancy between their measured hadronic energy
loss and the measurement of the total momentum of charged particles within the jet. A range of topological
event shape variables were also used to suppress background. In the latter case $\frac{dE}{dx}$
measurements, together with topological variables, were used to identify $R$-hadron events. A
lower-mass limit for gluinos of 27 GeV was evaluated at 95\% CL with this work. A similar analysis by
DELPHI~\cite{Abdallah:2002qi} gave a lower limit of 18 GeV. Differences in the limits arise due to
several factors. ALEPH used around 4 million hadronic $Z^0$ events compared
with 1.6 million events used by DELPHI. Furthermore, different topological variables were considered and
different analysis techniques used to suppress background. In addition, ALEPH employed a model of
$R$-hadron scattering in material which is based on the geometric cross-section, whilst DELPHI used the
first BCG ansatz (see Section~\ref{sec:hadscat}). The gluino mass limits of ALEPH and DELPHI were
extracted assuming
a direct production mechanism and are applicable to any MSSM scenario containing long-lived
gluinos. Thus, these works can be considered to belong to the second class of searches.

The ALEPH work also sought evidence of stable stops and sbottoms via the
reaction $e^+e^- \rightarrow \tilde{q}\bar{\tilde{q}}$, using
$\frac{dE}{dx}$ measurements.  Since the partners of the left and right handed states of the top (or
bottom) mix to form mass eigenstates, the lower mass limits depend on the choice of mixing angle~\cite{Martin:1997ns,Tata:1997uf}. In this
study, a stop (sbottom) mixing angle of $56^\circ$ ($68^\circ$) was used. At these values the stop and
sbottom couplings to the $Z$ vanish. This allowed the extraction of conservative  mass limits: 95 GeV
(stop) and 92 GeV (sbottom), at 95\% CL. As for the stable gluino work, the squark mass limits
are relevant for any MSSM scenario accommodating long-lived squarks.

Similarly, stable heavy slepton limits which are valid for all MSSM
scenarios in which the slepton is stable have also been evaluated.
Direct pair-production mechanisms $e^+ e^- \rightarrow
\tilde{l}_R^+ \tilde{l}_R^-$ and $e^+ e^- \rightarrow
\tilde{l}_L^+ \tilde{l}_L^-$ were assumed in a DELPHI
search~\cite{Abdallah:2002rd}. Within the MSSM, the cross section
for each of these processes depends only on the relevant slepton
mass. Lower-mass limits for the superpartners of the left (right) muon and tau
of 98 (97) GeV were obtained at 95\% CL. Mass limits of a similar
value were also obtained by ALEPH~\cite{chargedaleph1} and
OPAL~\cite{Abbiendi:2005gc}. Mass limits on selectrons are very model dependent owing to
an additional t-channel neutralino exchange process.

Searches for sleptons are usually interpreted within a slepton NLSP
or a sleptons co-NLSP GMSB scenario, as described in Section~\ref{peter:gmsb}. Such class 3 results exclude
regions in parameter space defined by quantities such as the slepton
masses, the effective SUSY-breaking scale, and the gravitino
mass~\cite{Abdallah:2002rd,chargedaleph1,Abbiendi:2005gc}. Since
stable slepton scenarios represent only a small part of the GMSB
parameter space,
we do not provide a detailed description of these results here.

Another type of class 3 SUSY search looks for NLSP charginos which
are nearly mass degenerate with the LSP
neutralino~\cite{Acciarri:2000wy,Heister:2002mn,Abreu:1999qr,
opalmassdeg} , as can occur in AMSB and gravity mediated
SUSY-breaking scenarios (see Section~\ref{peter:amsb}). For mass differences of less than
approximately the pion mass, the charginos are stable during their
passage through the detector. A search by
ALEPH~\cite{Heister:2002mn} provided the most stringent lower-mass
limit of 101 GeV for stable charginos at 95\% CL. However, as for
the GMSB searches, this result is dependent on the SUSY model
parameters. Refs.~\cite{Acciarri:2000wy,Heister:2002mn,Abreu:1999qr,
opalmassdeg} provide a detailed description of the model parameter
spaces excluded by collider studies.

\subsubsection{Searches at lepton-hadron experiments}\label{sec:lhad}

There have been few searches for electrically charged SMPs in
lepton-hadron scattering experiments. Those which have taken place
are generic (class 1) searches. A search for free quarks was
performed by the EMC experiment~\cite{Aubert:1983jy} in low-energy
deep-inelastic $\mu$-Beryllium scattering (DIS). This experiment was
sensitive to values of the photon virtuality, $Q^2$, and the
invariant mass of the entire hadronic final state, $W$, which
extended to around 100 GeV$^2$ and 20 GeV, respectively. Sets of scintillator
counters were used to gain sensitivity to the anomalously low ionisation energy loss
expected from free quarks. The ratio of the upper limit on the single inclusive production
cross section for SMPs to the total inelastic muon cross section was determined to be
around $10^{-6}$. The experiment was sensitive to SMPs with charges
$\pm \frac{1}{3}e$ and $\pm \frac{2}{3}e$ for masses up to $15$ GeV
and $9$ GeV, respectively. The limits on the production cross-section were smaller than
those obtained in neutrino-nucleon scattering~\cite{Basile:1980xt,Bergsma:1984yn,Allasia:1987ia}.

There has been one recent dedicated search for SMPs produced in
high-energy lepton-hadron scattering, which was made by the H1
experiment~\cite{h1highion}. This used a data sample corresponding
to processes with an average $W$ of 200 GeV, and in which the
exchanged boson was quasi-real (photoproduction). In photoproduction
processes, the exchanged photon can be ascribed a partonic
structure, and, in this picture, the H1 study is therefore
equivalent to the hadron-hadron searches described
in~\ref{sec:hadhad}. However, there are differences between final
states produced by photoproduction and hadron-hadron interactions,
for example, due to the contributions of so-called direct and
anomalous photoproduction processes~\cite{Abt:1994dn}. Thus, the two
environments are not identical, and this search is therefore
complementary to searches performed at a hadron-hadron facility.

 The H1 experiment used measurements of
$\frac{dE}{dx}$ in its drift chamber to search for SMPs depositing anomalously high
amounts of ionisation energy. A minimum bias sample corresponding to
$6$ pb$^{-1}$ of luminosity was used. At 95\% CL an upper
cross-section limit of 0.19 nb was extracted for the production of
SMPs. No production model was assumed for this study and the mass
range explored is determined by the capability of the H1 tracking
chamber. Since the apparatus was fully sensitive to SMPs with charge
$\pm e$ in the region of transverse momentum and mass
$0.2<p_T/M<0.7$, this implies a sensitivity in mass of up to around
100 GeV, assuming an SMP pair-production mechanism.

\subsubsection{Searches at hadron-hadron experiments}\label{sec:hadhad}

 Early class 1 searches for fractionally charged particles
produced in $pp$ interactions at the ISR ($\sqrt{s}=52$ GeV) were
able to set upper cross-section limits for particles possessing
fractional charges $\pm \frac{1}{3}e$,$\pm \frac{2}{3}e$, and $\pm
\frac{4}{3}e$, and masses less than around 20
GeV~\cite{Giacomelli:1979nu}. A host of lower energy hadron-hadron
experiments also performed such searches, usually for masses
less than around 10 GeV~\cite{rn}.

Searches for SMPs in $p\bar{p}$ collisions at the Tevatron have been
made by the CDF
collaboration~\cite{Abe:1989es,Abe:1992vr,chargedcdf1}. These were
based on relatively low-luminosity samples of
26 nb$^{-1}$~\cite{Abe:1989es}, 3.5 pb$^{-1}$~\cite{Abe:1992vr} and
90 pb$^{-1}$~\cite{chargedcdf1}. By the end of the Tevatron Run-II
data-taking period, the D0 and CDF experiments should each have
collected several fb$^{-1}$ of luminosity and we anticipate that
the above mentioned studies be updated. Here we consider the works
in Refs.~\cite{Abe:1992vr,chargedcdf1} which supersede the early,
very low-luminosity CDF study\cite{Abe:1989es}.

In Ref.\cite{Abe:1992vr} SMPs leaving a slow ($0.25<\beta<0.65$)
muon-like track were sought. The analysis also relied on
measurements of ionisation energy loss, muon-like tracks and
time of flight information from the calorimeter. Model-independent
upper limits on the production cross section (calculated to 95\%
CL) of between 120 and 5 pb were derived for the pair production
of fractionally charged fermions of masses between 50 and 500 GeV. The
electric charges considered were $\pm \frac{2}{3}e$,$\pm e$, and
$\pm \frac{5}{3}e$.

Using theoretical models, these data were also used to extract mass
limits under the assumption that the SMP possessed a specific colour
charge (singlet, octet or decuplet). These, and the remaining search
results described here belong to the second classification of
searches. The mass limits for an SMP with a given colour and
electric charge vary between 140 and 255 GeV at 95\% CL.

The third CDF work\cite{chargedcdf1} did not employ time of  flight
information, and relied on searches for particles losing anomalous
amounts of ionisation energy. This study considers the possibility of fourth-generation
quarks (see Section~\ref{peter:fourth}), and was able to set lower-mass limits at 95\% CL for
exotic quarks with charges $\pm \frac{2}{3}e$ and $\pm \frac{1}{3}e$ at 220
GeV and 190 GeV, respectively. Furthermore, this work considered the
production of fourth generation quarks without hadronisation and charge
exchange effects. The limits obtained (0.3-2 pb over a mass range
100-270 GeV) broadly correspond to a generic search for the pair
production of particles with charge $\pm e$, assuming a strong
production mechanism.

Stable gluinos have not been explicitly considered in the Tevatron
works. However, it has been argued~\cite{Hewett:2004nw} that
existing Tevatron limits on stable charged
particles\cite{chargedcdf1} and anomalous mono-jet
production~\cite{Acosta:2003tz} imply lower limits on the gluino
mass to be roughly 170-310 GeV; the lower (upper) value corresponds
to an extreme scenario in which gluino $R$-hadrons are produced
solely as neutral (charged) states.

A heavy slepton hypothesis which assumed a Drell-Yan like production
mechanism was also studied in~\cite{chargedcdf1} and production
cross-section limits were extracted. However, the expected
cross section in exotic scenarios such as GMSB is more than an order
of magnitude below this level of sensitivity. Contrary to the
situation for exotic quark searches, heavy lepton mass limits from
SMP searches have, so far, only been extracted by the LEP
experiments.





\subsubsection{Summary and discussion of direct searches}

A selection of class 1 searches is
 summarised in Tab.~\ref{class1searches}. Since these studies
focus on generic SMPs, for which the production mechanisms are {\it
a priori} unknown, it is important that searches were conducted in
each of the different collision environments. Evidence for particles
with fractional charges $\pm \frac{1}{3}e$, $\pm \frac{2}{3}e$, $\pm
e$, $\pm \frac{4}{3}e$, and $\pm \frac{5}{3}e$ was sought. Plastic track detectors
deployed at LEP extend the SMP charge sensitivity to around $|240e|$.
The mass sensitivity at the various accelerators for the pair production of generic SMPs with
charge $\pm e$ extended to $100$ (HERA),101 (LEP), and 500
GeV (Tevatron).
None of the above generic searches
were sensitive to particles with charge much below $\sim
|\frac{1}{3}e|$.  Dedicated searches for millicharged particles with
charges as low as $\sim 10^{-5}e$ have been made at accelerators and
elsewhere albeit typically for particles with masses less than $\sim
1$ GeV~\cite{Davidson:2000hf,Perl:2001xi}. These searches are thus
beyond the scope of this report. It is relevant for the massive
particles considered here to note that there exist no direct
experimental searches which are sensitive to particles with charge
 $\lapprox |\frac{1}{3}e|$ and mass $1 \lapprox M \lapprox 1000$ GeV, and that indirect
astrophysical constraints tend only to exclude such SMPs with
charges less than around $|10^{-6}e|$~\cite{Dubovsky:2003yn}.

\begin{table}[t]\scriptsize \begin{tabular}{cccccccc}
\hline $\sqrt{s}$ & Collision & Experiment  & Mass range& Charge (e) & Cross-section& CL (\%) &Ref. \vspace{-.12cm}\\
(GeV) & &  & (GeV)&     & limit (pb)   &   & \\ \hline

       1800 & $p\bar{p}$  & CDF  &  100-270   & $\pm 1$      & 0.3-2 &  95 &     \cite{chargedcdf1}\\
       1800 & $p\bar{p}$  & CDF  &  50-500 & $\pm \frac{2}{3}$ & 10-100  & 95 & \cite{Abe:1992vr}       \\
            &             &       &          & $\pm 1$            & 5-50      &                          \\
            &             &      &           & $\pm \frac{4}{3}$ & 5-70  &  & \\
       300 &  $ep$  & H1   &    $< 100$  & $\pm 1$  & 190  & 95   & \cite{h1highion}\\
       130-209 &  $e^+e^-$    & OPAL &  45-102   & $\pm \frac{2}{3}$ & 0.005-0.02 & 95 & \cite{Ackerstaff:1998si,chargedopal1} \\
       &              &      &  45-102   & $\pm           1$ &    0.005-0.03&  &  \\
           &              &      &  45-100   & $\pm \frac{4}{3}$ & 0.005-0.02 &            \\
           &              &      &  45-98    & $\pm \frac{5}{3}$ & 0.005-0.02 &            \\
     189 & $e^+e^-$ & DELPHI & 68-93 & $\pm 1$  & 0.02-0.04 & 95 & \cite{chargeddelphi1} \\
    130-183 & $e^+e^-$ & DELPHI & 2-91 & $\pm 1$  & 0.05-0.3 & 95 & \cite{chargeddelphi3} \\
    130-183 & $e^+e^-$ & DELPHI & 2-91 & $\pm \frac{2}{3}$  & 0.04-0.6 & 95 & \cite{chargeddelphi3} \\
    130-172 & $e^+e^-$ & ALEPH & 45-86 & $\pm 1$  & 0.2-0.5 & 95 & \cite{chargedaleph2} \\
       91.2& $e^+e^-$     & ALEPH & 5-45     & $\pm \frac{1}{3}$ &$3-10$& 90 &\cite{chargedaleph3}\\
           &              &       &          & $\pm \frac{2}{3}$ &$1-12$        &\\
           &              &       & 10-72    & $\pm 1^*$& 1.6 - 140   &   \\
           &              &       &          & $\pm \frac{4}{3}$& $1.4-4$  &\\
       91.2& $e^+e^-$     & OPAL  & 3-45     & $\pm \frac{2}{3}$ & 0.2-1.0  & 90 &\cite{chargedopal2}\\
           &              &       &          & $\pm \frac{2}{3}^*$ & 0.15-0.9    & 95 &                   \\
           &              &       &          & $\pm 1^*$  & 0.15-3.0 & & \\
           &              &       &          & $\pm \frac{4}{3}^*$ & 0.18-0.21  &  \\
           &              &       &          & $\pm 2^*$  &0.27-0.3 & & \\

       91.2& $e^+e^-$     & Kinoshita {\it et al.} & 1-45 & $\lapprox |240e|$ & 70
      & 95  &\cite{Kinoshita:1992wd} \\
 \hline
\end{tabular}
\vspace{0.15cm} \caption{A summary of selected direct searches for
electrically charged SMPs belonging to class 1. The searches are
categorised according to centre-of-mass energy, colliding particles,
and experiment. A range in cross-section limit is provided for an
SMP with a given charge. The mass range corresponds to the region
for which the upper cross-section limit is quoted. The limits were
derived under the assumption of a pair-production mechanism, with
the exception of those results marked with
the symbol $^*$. For these limits, a single SMP inclusive production
mechanism was assumed. The confidence level to which the limits were
extracted is also shown.}
\label{class1searches}
\end{table}





\begin{figure}[t!]
\begin{center}
\epsfig{file=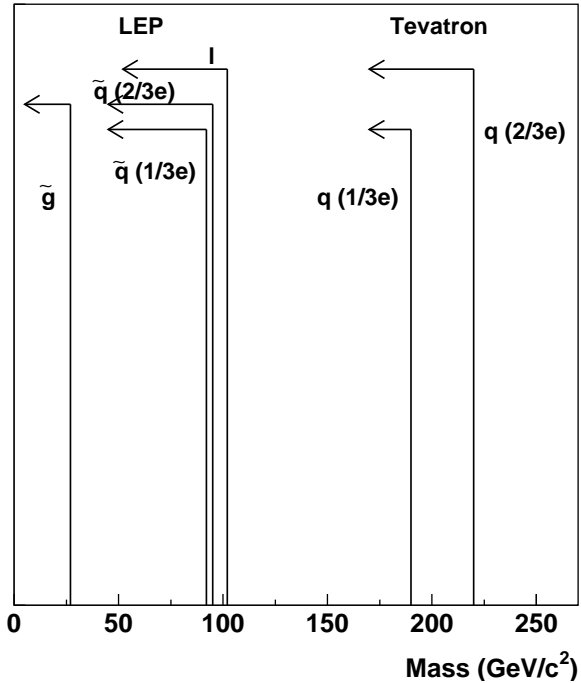,height=10cm,width=10cm} \caption{Lower
mass limits for stable gluinos~\cite{aafke2}, squarks~\cite{aafke2},
fourth-generation quarks~\cite{chargedcdf1}, and heavy
leptons\cite{Achard:2001qw}. The lengths of the vertical lines have
no significance.}\label{fig:smp-lim}
\end{center}
\end{figure}

For the class 2 searches, lower mass limits were obtained when
considering a specific type of SMP and an assumed production
cross section. A compilation of lower limits for masses of specific
types of particles is shown in Fig.~\ref{fig:smp-lim}.
The stable quark limits are taken from Ref.~\cite{chargedcdf1} which
assumed a fourth-generation quark model. 
The stable squark and gluino mass limits from LEP~\cite{aafke2} are
$\sim 90$ and $27$ GeV, respectively.
Stable heavy squark or gluino hypotheses have not yet been
considered by Tevatron experiments although it could naively be
expected that, in a minimal SUSY scenario with squark or gluino
LSPs, mass limits with Tevatron data would be similar to those
already obtained for stable quarks.
Stable stau and smuon mass limits of around 98 GeV have been
obtained at
LEP~\cite{chargedaleph1,Abbiendi:2005gc,Abdallah:2002rd}.
Non-SUSY heavy lepton mass limits
have also been extracted at
LEP\cite{chargeddelphi1,Abbiendi:1999sa,Achard:2001qw}. The mass limit
is largely insensitive to the type of new lepton, and we quote here the L3 mass
limit of 103 GeV\cite{Achard:2001qw} for mirror leptons. A summary of selected
searches made within classes 2 and 3 is given
in Tab.~\ref{directsearches}.

\vspace{0.0cm} {\tiny
\begin{table}[t]
\scriptsize
\begin{tabular}{p{1.5cm}p{1.1cm}p{1.1cm}p{2.7cm}p{2.7cm}p{2.2cm}}
 \hline
$\sqrt{s}$  (GeV) &  Collisions & Experiment  & Particle & Mass limit (GeV) & Ref.\\
\hline
1.8 TeV &  $p\bar{p}$  & CDF  &  $4^{th}$ generation quark  & $m_{t'}>$220, $M_{b'}>$190 & \cite{chargedcdf1}\\
130-209  &  $e^+e^-$  & OPAL & Heavy leptons &  $m_{l'}> 102$  &  \cite{chargedopal1} \\
133-208  & $e^+e^-$   & L3   & Heavy leptons & $m_{l'}>103$  & \cite{chargedl31}\\
91.2-209  & $e^+e^-$  & ALEPH &  Squark & $m_{\tilde{t}}>95$, $m_{\tilde{b}}>92$ & \cite{aafke2}\\
130-183  & $e^+e^-$   & DELPHI&  Squark & $m_{\tilde{t}}>80$, $m_{\tilde{b}}>40$ & \cite{chargeddelphi3}\\
130-183  &      $e^+e^-$     & DELPHI      &  Free squark & $m_{\tilde{t}}>84$ & \cite{chargeddelphi3}  \\
91.2  & $e^+e^-$  & ALEPH &  Gluino & $m_{\tilde{g}}>27$ & \cite{aafke2}\\
91.2  & $e^+e^-$  & DELPHI &  Gluino & $m_{\tilde{g}}>18$ & \cite{Abdallah:2002qi} \\
189-209  & $e^+e^-$   & ALEPH &  Slepton &  $m_{\tilde{\tau}}>97$, $m_{\tilde{\mu}}>97$       &
~\cite{chargedaleph1}\\
130-208  & $e^+e^-$   & DELPHI&  Slepton & $m_{\tilde{\tau}}>98$, $m_{\tilde{\mu}}>98$
&~\cite{Abdallah:2002rd}\\
189-209  &  $e^+e^-$  & OPAL &   Slepton  &  $m_{\tilde{\tau}}>98$,$m_{\tilde{\mu}}>98$ &
\cite{Abbiendi:2005gc} \\
189-209  & $e^+e^-$  & ALEPH &  Chargino & $m_{\chi^{\pm}}>101 $& \cite{Heister:2002mn}\\
130-189  & $e^+e^-$  & DELPHI & Chargino & $m_{\chi^{\pm}}>93$ & \cite{Abreu:1999qr}\\
189       & $e^+e^-$   & L3   &  Chargino  & $m_{\chi^{\pm}}>94$ & \cite{Acciarri:2000wy} \\
\hline

\end{tabular}
\vspace{0.15cm} \caption{A summary of selected classes 2 and 3
searches which have been made for electrically charged SMPs. The
searches are described according to the collision centre-of-mass
energy, colliding particles, experiment, the type of SMP, and the
resultant mass limit. The squark, gluino, and slepton searches are
valid for any MSSM scenario in which these particles are long-lived.
The squark and slepton limits correspond to values of mixing angles
which predict the lowest possible cross-sections. The chargino
limits are model-dependent and were extracted in AMSB and AMSB-like
scenarios. All limits were extracted at 95\%
CL.\label{directsearches}}
\end{table}}
\normalsize
\subsubsection{Indirect collider searches}

Indirect mass limits of up to around $\frac{M_Z}{2}$ for fourth generation quarks and leptons,
squarks (assuming a non-vanishing coupling to the $Z^0$), sleptons, charginos, and exotic particles with
electric charges as low as $\sim 0.2e$ can be inferred from measurements
of the $Z^0$ invisible width~\cite{rn,Dubovsky:2003yn}. Each limit is
generally applicable for a scenario in which SMPs are pair produced
through the decay of a $Z^0$, and not recorded in a detector. Whilst the
majority of the aforementioned particles would certainly be detected in
the direct searches, it is, however, conceivable that fractionally charged
particles may have been missed. LEP-1 limits for exclusive pair
production processes were not made for SMPs with charges below
$\pm \frac{2}{3}e$.  Furthermore, in a free quark picture, these particles
may possess a large interaction cross-section, and rapidly become stopped
in detector material. In these scenarios, the indirect limits
provide a valuable complement to the direct searches.

Similarly, indirect mass limits can be used to support the stable gluino limits from direct searches,
which are dependent on the phenomenological models used to describe $R$-hadron energy loss. These models
possess theoretical uncertainties which are not trivial to estimate. Indirect lower limits on the masses
of colour octet particles of around 6 GeV can be set from the partial width of any purely
hadronic exotic contribution (negative or positive) to Z-decays~\cite{janot}.

Mass constraints on gluinos have also been set by studying the QCD colour factors
from 4-jet  angular correlations and the differential 2-jet rate in
$Z$-decays~\cite{Barate:1997ha}. In addition, the modification in the
running of $\alpha_s$ as a consequence of the existence of stable
coloured particles has been used to set limits on the mass of these
particles~\cite{Csikor:1996vz}. These lower mass limits also extend to around 6 GeV.

\subsection{Searches for magnetically charged particles}\label{monsearch}
A body of searches has been performed for magnetic monopoles produced at
accelerators, in cosmic rays, and bound in matter~\cite{rn}.
Although no reproducible evidence for the existence of monopoles has
been found, several apparent observations have intermittently caused
great excitement in the field. In 1973 a particle consistent with
being a monopole with charge around $2g_D$ was recorded at a cosmic
ray experiment~\cite{Price:1975zt}. While the event remains to be
fully understood, the authors later established that a monopole
could not be responsible for the observation~\cite{Price:1976mr}. In
the 1980's isolated observations~\cite{Cabrera:1982gz,Caplin:1986kw}
of cosmic ray monopole candidates were made using SQUID detectors.
However, instrumental effects could not be ruled as being
responsible for the signals and the observations are inconsistent
with monopole flux limits set at other cosmic ray
facilities~\cite{Ambrosio:2002qq}.

During the 1990's monopole searches remained a routine part of the
experimental programs of collider experiments. This section
discusses searches for monopoles directly produced at
electron-positron annihilation, hadron-hadron and lepton-hadron
facilities. Direct searches include, for example, searches for
monopoles leaving tracks in plastic track detectors, and searches
for monopoles trapped in detector material. As will be shown, most
searches are concerned with the possibility of monopoles possessing
the Dirac charge. However, even in the absence of theoretical
motivation, it is prudent to also consider scenarios with particles
possessing as wide a charge-range as is experimentally possible. We
therefore describe the regions in magnetic charge and mass which
have been explored. The discussion will be followed by a brief
discussion of indirect monopole searches.

\subsubsection{Searches at $e^+e^-$ experiments}\label{sec:monee}
A variety of methods have been used to search for monopoles at
$e^+e^-$ experiments. The most commonly used approach, the
deployment of plastic track detectors around the beam interaction
point, has been used in searches at the PETRA~\cite{Musset:1983ii},
PEP~\cite{Kinoshita:1982mv,Fryberger:1983fa}, {\sc
KEK}~\cite{Kinoshita:1989cb,Kinoshita:1988cn} and
LEP-1~\cite{Kinoshita:1992wd,Pinfold:1993mq} facilities.
The most competitive exclusion limit for Dirac monopoles was
extracted with passive detectors deployed at the OPAL intersection
point at LEP-1~\cite{Pinfold:1993mq}. No monopoles were observed and
an upper cross-section limit for the production of Dirac monopoles
of 3 nb was computed at 95\% CL. Since the acceptance depends on the
monopole mass and charge, the maximum value of acceptance was used
in this extraction. It was also assumed that monopoles would be
exclusively produced in pairs and distributed isotropically in phase
space. The experiment was sensitive to monopoles with magnetic
charge between $0.9g_D$ and $3.6g_D$ and for masses up to $45$ GeV.

An earlier experiment at LEP by the MODAL\cite{modal} group used
CR-39 plastic sheet detectors at the I5 LEP interaction
point\cite{Kinoshita:1992wd}, as mentioned in
Section~\ref{sec:eeSMPs}. As in the case of \cite{Pinfold:1993mq}
this experiment was sensitive to monopole masses up to around $45$
GeV albeit with poorer sensitivity to the monopole production
cross section ($\sigma <$ 0.7$\mu$b at 95\% CL for Dirac
monopoles). However, the experiment was sensitive to lower values of
magnetic charge ($0.1g_D$). Lower-energy plastic track experiments
at $e^+e^-$ facilities have also been made albeit with a lower mass
sensitivity
~\cite{Kinoshita:1982mv,Fryberger:1983fa,Musset:1983ii,Kinoshita:1988cn,Kinoshita:1989cb}.

Searches for monopoles leaving parabolic tracks have been
made\cite{Gentile:1986sf,Braunschweig:1988uc} (see
Section~\ref{sec:para}). These assumed an exclusive pair-production
mechanism. Together, these works gave sensitivity to values of the
magnetic charge between $0.03-0.1g_D$ for monopole masses below $17$
GeV.

\subsubsection{Searches in lepton-hadron collisions}
There has been only one search for monopole production in
lepton-hadron scattering. A search was made for monopoles which
could have been stopped in the aluminium beampipe of the H1
experiment at HERA\cite{h1mon}. The beampipe was exposed to an
integrated luminosity of 62 pb$^{-1}$. The SQUID apparatus, which
was used in this experiment is shown in Fig.~\ref{fig:h1mon}.
To calculate the acceptance in this study it was assumed that
monopoles were produced in pairs via photon-photon fusion processes.
Limits were extracted for two different assumed reactions: $ep
\rightarrow eM\bar{M}p$ and $ep \rightarrow eM\bar{M}X$. In the
former (latter) case the monopoles were treated as spin 0
($\frac{1}{2}$) objects. No evidence was found for the presence of trapped magnetic
monopoles. This work set mass-dependent upper limits on the
production cross section of monopoles with charges $g_D,2g_D,3g_D$,
and $6g_D$ for masses up to around $140$ GeV. The cross-section limits (calculated at
95\% CL) vary between around 0.2 and 100 pb depending on the
charge, mass and assumed monopole spin.

\subsubsection{Searches at hadron-hadron experiments}\label{hhmon}
A recent study at CDF~\cite{Abulencia:2005hb} obtained upper
cross-section limits for monopoles possessing the Dirac charge for masses up to 900 GeV.
This work used 35.7 pb$^{-1}$ of integrated luminosity and employed
a specially built highly ionising particle trigger requiring large
light pulses at both ends of a TOF scintillator bar. In addition,
track reconstruction software was optimised to search for
characteristic parabolic trajectories. A Drell-Yan-like
pair-production mechanism, as described in Section~\ref{monocal},
was used to simulate the kinematic properties of monopoles. Using
this model, cross sections of greater than 0.2 pb (at 95\% CL) were
ruled out for masses between 200 and 700 GeV. Using the same
Drell-Yan formalism as a cross-section prediction, this can be
interpreted as a lower mass limit of $360$ GeV.

Another recent search at the {Tevatron} involved the use of the
 SQUID technique by the E882 experiment to detect monopoles bound
  in
detector material at the D0 and CDF
experiments\cite{Kalbfleisch:2000iz,Kalbfleisch:2003yt}. In these
works three sets of samples were taken from the D0 and CDF detectors
during their upgrades for {Tevatron} Run-2 data-taking: the Be
beampipe and Al extension cylinders from D0, Pb from the CDF Forward
Electromagnetic Calorimeters, and one half of an Al cylinder
supporting the CDF Central Tracking Chamber. These detector pieces
were exposed to integrated luminosities of $172\pm 8$ pb$^{-1}$ (D0)
and $180 \pm 9$ pb$^{-1}$ (CDF). Three different assumptions were
made concerning the angular distribution of monopoles in the
monopole-antimonopole centre-of-mass system:
$\frac{d\sigma}{dcos\theta}\propto$ constant,
$\frac{d\sigma}{dcos\theta}\propto 1+cos^2\theta $, and
$\frac{d\sigma}{dcos\theta}\propto 1-cos^2\theta $. Upper limits on the
production cross section of monopoles with charge $g_d,2g_d,3g_D$
and $6g_D$ were found to be $0.6$,$0.2$, $0.07$, and $0.02$ pb,
respectively, at 90\% CL, when taking the isotropic
case. Using the modified Drell-Yan formalism as a cross-section
prediction, these correspond to mass limits of 265, 355, 410, and
375 GeV, respectively. Varying the angular distributions according
to the three assumed forms described above leads to a spread of
around 10 GeV in these mass values.

Taken together, the works
\cite{Kalbfleisch:2000iz,Kalbfleisch:2003yt} and
\cite{Abulencia:2005hb} represent the most comprehensive limits for
the production of monopoles in hadron-hadron collisions. The
searches each employ complementary methods, and together are
sensitive to the production of monopoles with charges between $g_D$
and $6g_D$ and masses up to around 900 GeV. Nevertheless, since both
CDF and D0 are expected to accumulate several fb$^{-1}$ of
luminosity at the completion of the {Tevatron} program, there
remains sensitivity to search for monopoles with a cross section
around an order-of-magnitude lower than that which has already been
excluded. In addition, a further search could address the
possibility of the production of monopoles with a charge less than
$g_D$.

Other searches\cite{Price:1987py,Price:1990in,Bertani:1990tq} at the
{Tevatron} based on substantially smaller samples of luminosity have
used plastic track detectors placed around the interaction point.
Again, no evidence for monopole production was found and the
exclusion limits on cross sections are typically several thousand
times larger, and the mass limits around three times smaller than
those obtained in
\cite{Kalbfleisch:2000iz,Kalbfleisch:2003yt,Abulencia:2005hb}.

Lower
energy hadron-hadron experiments have employed a variety of search
techniques including plastic track
detectors\cite{Hoffmann:1978mp,Aubert:1982zi} and searches for
trapped
monopoles\cite{Carrigan:1973mw,Carrigan:1974un,Carrigan:1977ku}. The
latter searches used external magnetic fields to accelerate stopped
monopoles and
 were sensitive to a very wide charge range (0.03-24$g_D$) albeit for masses less than 30 GeV.

\subsubsection{Summary and discussion of direct searches}
In Tab.~\ref{monosearches}, the upper cross-section
limits from monopole searches and their charge and mass sensitivities are given.
It is difficult to compare the mass and cross-section exclusion
limits extracted in the various searches because of the theoretical
uncertainties on monopole production.
What can be reliably
calculated is the region of finite acceptance for a monopole with a
specific charge and mass, and then, with some model dependence, the
derived upper limit on the production cross section. Mass limits,
extracted under the assumption of a
known production cross section, are the least reliable information
reported by experiments.

\begin{table}[t]\scriptsize \begin{tabular}{ccccccccc}
\hline $\sqrt{s}$ & Collision & Experiment & Charge & Mass & Charge & Cross-section  & Ref. \vspace{-.12cm}\\
(GeV) & &  & sensitivity & sensitivity  & ($g_D$)   & limit
& \vspace{-0.12cm}\\
 &  &     &  ($g_D$)    & (GeV)       &       &          (pb) & &  \\
\hline
       1960 & $p\bar{p}$  & CDF  & - & 100-900 & 1  & 0.2      &  \cite{Abulencia:2005hb}       \\
       1800 &  $p\bar{p}$  & E882   & - & - & 1  & 0.6    &
       \cite{Kalbfleisch:2000iz,Kalbfleisch:2003yt}\\
            &      &     & &  &   2  & 0.2    &       \\
            &      &     &  &   & 3  & 0.07    &       \\
            &      &     & &   & 6  & 0.2    &       \\
       1800 & $p\bar{p}$ & M.~Bertani {\it et al.}   & 0.4-5 & $<850$ & $>$ 0.5 & 200    &
        \cite{Bertani:1990tq}\\
       56 & $pp$ & Hoffmann {\it et al.}  & 0.3-7 & $<30$ & 0.3-3
& 0.1 &  \cite{Hoffmann:1978mp} \\
 63& $pp$  & Carrigan {\it et al.}  & 0.07-24& $<30$& 0.2-1.2 &
0.13 &
\cite{Carrigan:1977ku}  \\
         &    &      &                           &      & 1.2-24&       0.4 &   \\
25-28 & $pN$ & Carrigan {\it et al.}   &  0.03-24 & $<13$ & 0.03-24
& $5\times
10^{-6}$  &  \cite{Carrigan:1973mw,Carrigan:1974un} \\
  300& $ep$  & H1  &$>$0.1& $<140$& 1 & 2.2 &
\cite{h1mon}  \\
          &   & &           &      & 2 &0.18&       \\
          &   &  &          &      & 3 & 0.07&       \\
          &   &   &        &      & 6 & 0.04&       \\
        88-94 & $e^+e^-$ & J.L.~Pinfold {\it et al.}  & 0.9-3.6 & $<45$ &1  & 0.3 & \cite{Pinfold:1993mq} \\
             &           &      &  & $<41.6$ & 2  & 0.3    &        \\
 89-93 & $e^+e^-$ & MODAL  & 0.1-3.6 & $<44.9$ &1  & 70 & \cite{Kinoshita:1992wd} \\
 34 & $e^+e^-$ & P.~Musset {\it et al.} & 0.98-5.9 &  $<16$ & 0.98-5.9   & 0.04
& \cite{Musset:1983ii} \\
29  & $e^+e^-$ & D.~Fryberger {\it et al.}  &
0.29-2.9 & $<14$ & 0.29-2.9  & 0.03 & \cite{Kinoshita:1982mv,Fryberger:1983fa} \\
 35 & $e^+e^-$  & TASSO  & 0.15-1 & $\leq 17$ & 0.15   & 4  & \cite{Braunschweig:1988uc} \\
           &               &                 &        &      & .44 & 0.04  \\
           &               &            &             &  &   1  & 0.08 & \\
10.6 & $e^+e^-$  & CLEO  & 0.03-0.12 & $<5$ & 0.03
 & 0.8
& \cite{Gentile:1986sf}\\
     &           &         &  &       & 0.07      & 0.24     &  \\
     &           &         &  &       & 0.12      & 1.6   &  \\
\hline

\end{tabular}
\vspace{0.15cm} \caption{A summary of selected direct searches for
monopoles. The searches are categorised according to the collision
centre-of-mass energy, the colliding beams, the experiment, the
quoted acceptance regions in charge and mass, and the cross-section
limits for specific values of magnetic charges. The symbol ``-"
denotes that an experiment did not explicitly quote an acceptance
region. The cross section limit from Ref.~\cite{Abulencia:2005hb} is quoted 
for monopole masses between around 200 and 700 GeV, a region in which 
the limit is roughly constant. The cross section limits from Ref.~\cite{h1mon} are quoted
for a monopole mass of 100 GeV, which is approximately one half of
the average photon-proton centre-of-mass energy at the experiment.
The cross section limits for works \cite{Braunschweig:1988uc} and
\cite{Gentile:1986sf} are quoted for mass values of 8 and 2.5 GeV,
respectively. At these values cross-section limits are available for
the different charge hypotheses considered.}
\label{monosearches}
\end{table}





\begin{figure}[t!]
\begin{center}
\epsfig{file=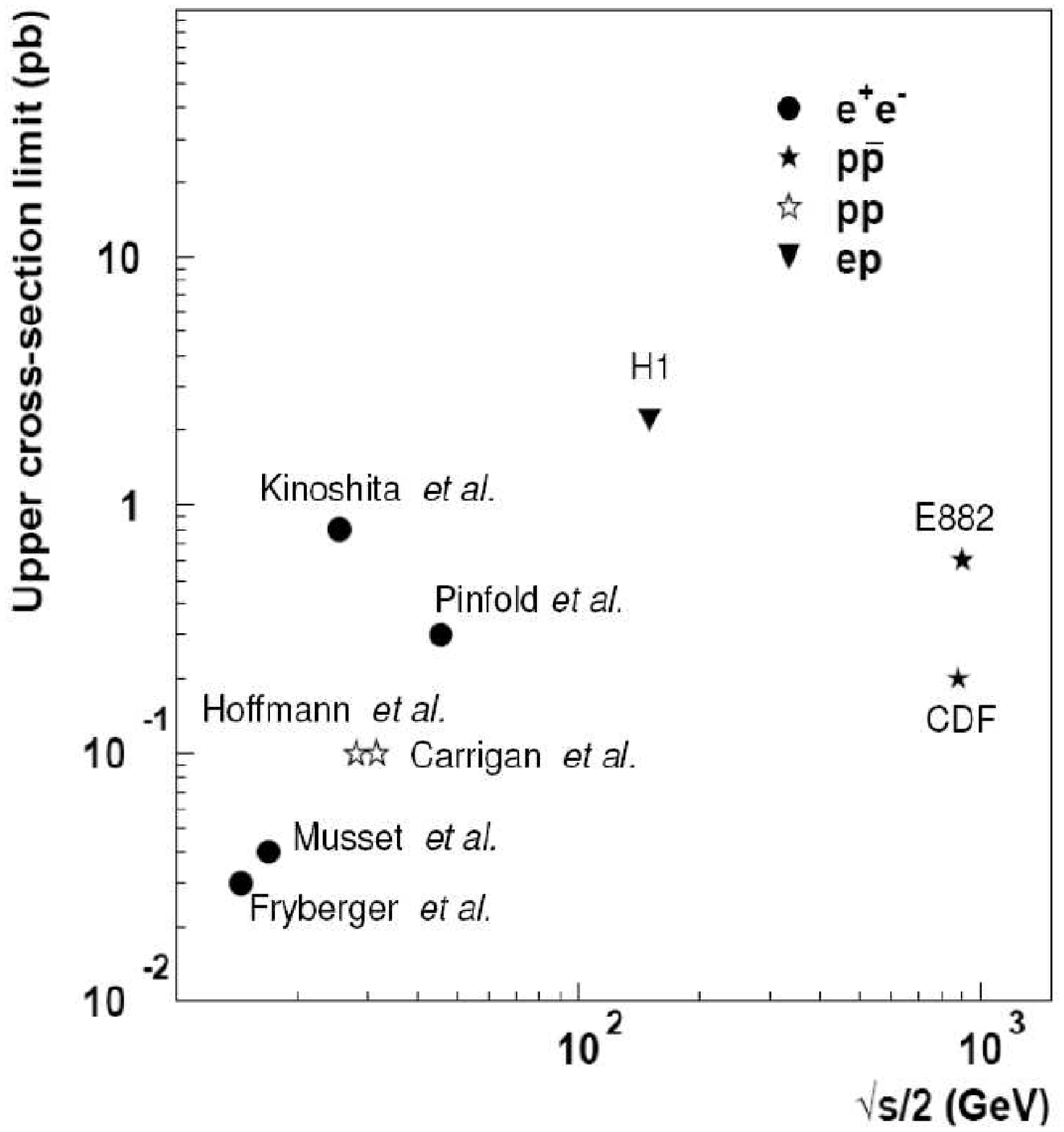,height=8cm,width=10cm} \caption{Summary
of upper limits on the monopole production cross section at a range
of experiments at different centre-of-mass
energies~\cite{Carrigan:1977ku,Hoffmann:1978mp,Musset:1983ii,
Fryberger:1983fa,Kinoshita:1988cn,Pinfold:1993mq,Kalbfleisch:2000iz,Kalbfleisch:2003yt,Abulencia:2005hb,h1mon}.
} \label{fig:magmon}
\end{center}
\end{figure}

As motivated by Dirac's argument (see Section~\ref{mmtheory}),
a large number of experiments have searched for
monopoles possessing the Dirac charge. Fig.~\ref{fig:magmon} shows the upper cross-section limits, which
have been obtained. The limits are shown versus one half of the centre-of-mass energy of the
experiments, which is indicative of the largest mass value to which
an experiment is sensitive, currently 900 GeV at the Tevatron\cite{Kalbfleisch:2003yt} .
There also exist theoretical arguments for
considering particles with charges up to $6g_D$. To reach these values it is usually
necessary to look for stopped monopoles owing to the vastly
increased energy loss associated with high-charge monopoles. Searches for monopoles with
magnetic charges up to $6g_D$ have so far only taken place at two high energy colliders: HERA\cite{h1mon} and the
{Tevatron}\cite{Kalbfleisch:2003yt}.

Sensitivity to low magnetic charges ($\gapprox \frac{1}{10}g_D$) was
provided by the H1~\cite{h1mon} and MODAL~\cite{Kinoshita:1992wd}
searches. Low-energy $e^+e^-$ experiments searching for parabolic
tracks~\cite{Braunschweig:1988uc,Gentile:1986sf} and hadron-hadron
searches in which external magnetic fields were used to accelerate
trapped
monopoles~\cite{Carrigan:1973mw,Carrigan:1974un,Carrigan:1977ku},
provided sensitivity to charges as low as $0.03g_D$. However, these searches were only sensitive
to monopoles of masses less than around 30 GeV.

There have been
no dedicated searches for monopoles with charges less
 than $0.03g_D$. Searches for electrically charged particles with
anomalous ionisation energy loss (Section~\ref{sec:esearch}) could be sensitive to monopoles in this low charge region. However, this depends on
the dynamic range and calibration of the read-out electronics of the tracking devices, which were used
in these searches. In addition, measurements of charged particles with high
transverse momentum could also provide sensitivity to low charge monopoles. Since a
monopole with a large mass ($\gapprox$
100 GeV) and low charge would not leave an identifiable parabolic
track, the signature would be a straight track consistent with a
very high-momentum electrically charged particle.

\subsubsection{Indirect searches}\label{mmin}
Mass sensitivity at colliders can be increased via indirect
searches, which rely on processes mediated by virtual monopole
loops, as shown in Fig.~\ref{fig:inloop} in Section~\ref{monocal}. Such processes
can lead to multi-photon final-states arising from electromagnetic radiation
from monopoles.
The D0 experiment were able to exclude monopole masses between 610
and 1580 GeV\cite{Abbott:1998mw}, while L3 reports a lower mass limit of $510$
GeV\cite{Acciarri:1994gb} using this approach. However, it should be
stressed that these cross-section calculations employ perturbative
techniques\cite{DeRujula:1994nf,Ginzburg:1999ej} with uncertainties
which are difficult to estimate and the derived cross-section limits
have been criticised for this\cite{Gamberg:1998xf}.

An indirect lower limit on the mass of Dirac monopoles of 120 GeV
has also been inferred from measurements of the anomalous magnetic
moment of the muon\cite{Graf:1991xe}. However, it should again be
emphasised that this result relies on perturbative techniques and
that the result is therefore questionable.
\section{SMPs at the LHC}
The LHC will allow searches for SMPs with masses up to several
TeV, representing an order of magnitude increase in
 mass sensitivity compared with earlier colliders. As explained in
 Section~\ref{sec:scen}, well motivated extensions of the SM predict
 the production of a range of SMPs at the
 LHC. However, it is important that the LHC experiments perform searches for SMPs
  which are motivated not only by what is currently theoretically favoured but also by
 what is experimentally feasible. Therefore, in this section
 we discuss the sensitivity of the two
high-luminosity and multi-purpose experiments,
ATLAS\cite{atlastdr1,atlastdr2} and CMS\cite{cmstdr1,cmstdr2}, to
SMPs with a variety of properties. In addition, we also briefly
discuss the planned MOEDAL experiment\cite{moedal,Pinfold:1999sp},
which aims to deploy plastic track detectors around the LHCb vertex
detector.

We use the findings of a number of studies which explore the
feasibility of discovering SMPs at the LHC. These have been made
both by interested phenomenologists and the experiments themselves
(see for
example~\cite{nisati,polesello2,cms,tarem,Ambrosanio:2000ik,Ambrosanio:2000zu,Allanach:2001sd,Hewett:2004nw,richardson,
aafkejorgen,milstead,johansen,Ellis:2006vu}). The former works tend
to apply simple ansatzes regarding detector performance whereas the
latter use sophisticated detector simulation packages. While the
latter studies provide a more realistic description of the detector
response it is pertinent to note that the detector simulation
programs are still being optimised and have yet to be confronted and
verified with collision data. Nevertheless, it is still possible
to use the feasibility studies to draw a number of conclusions
regarding the discovery potential for various SMPs and to highlight SMP-specific
experimental challenges.

\subsection{Detector sensitivity}
\subsubsection{Timing issues}
One of the principal constraints on the design of a collider
experiment and its ability to measure slow-moving SMPs is the bunch
crossing time. The LHC crossing time of 25 ns, will be considerably
shorter than that at LEP (25 $\mu$s), the Tevatron (396 ns) or HERA
(96 ns). Therefore, for an SMP at the LHC to be detected or
triggered in a certain detector system and be associated to the
correct bunch crossing, it should arrive at most 25 ns after the
default arrival time of a particle travelling at the speed of
light~\cite{Hauser:2004nd,aafkejorgen}. Later arrival would imply
triggering or detection within the next crossing time window. The
large size of the ATLAS and CMS detectors (the central ATLAS and CMS
muon chambers extend to 10 and 7 m, respectively) ensures that this
will be an important source of inefficiency in detecting SMPs. For
example, it is only possible to reconstruct the track of a slowly
moving SMP in the ATLAS central muon chambers within the correct
bunch crossing window if $\beta \gapprox 0.5$\cite{aafkejorgen}.
Even if the SMP travels within the appropriate timing window,
additional problems may arise from its slowness. The sampling time
and reconstruction software of each sub-detector is optimised
assuming particles travelling at luminal speed. Hence, the quality
of the read-out signal or reconstructed track or cluster may be
degraded for an SMP, especially for sub-systems far away from the
interaction point.
 Detector simulations so far suggest that it will still be
possible to trigger and measure slowly moving particles at ATLAS and
CMS\cite{cms,aafkejorgen}. However, this is an area which must
continue to be studied as the simulation programs are further
developed and the detectors better understood.

\subsubsection{Important sub-detectors}
The principal detector systems needed for SMP searches are the
tracking systems. ATLAS and CMS both contain inner tracking systems
capable of precision measurements of the momenta of tracks of
charged SMPs\cite{atlastdr1,cmstdr1}. These devices can potentially
be used to search for the 'classic' signature of SMPs losing
anomalous amounts of ionisation energy in an inner detector, albeit
probably with poorer precision than has been obtained in previous
experiments. The ATLAS inner tracking system comprises a
semiconductor tracker system and a transition radiation tracker
(TRT). The precision with which
$\frac{dE}{dx}$ can be exploited is limited since the TRT will only record
hits satisfying high
and low ionisation thresholds. The total number of high/low
threshold hits and the time in which a channel is read out over a
specific threshold can be used in SMP searches\cite{aafkejorgen}.
The CMS inner tracking system consists of silicon pixel and silicon
strip detectors. As described in Section~\ref{sig}, the resolution
in $\frac{dE}{dx}$ of such devices is poorer than with gaseous
detectors. Work is ongoing to estimate the sensitivity of the CMS tracking system to SMPs~\cite{Adam:2005pz}.

Arguably the most important detector component in an SMP search at
the LHC is the muon system since most proposed SMPs leave a
signature of a slow-moving penetrating particle (see Sec.~\ref{searchtechniques}). The
ATLAS\cite{atlastdr1} and CMS\cite{cmstdr1} experiments employ muon
chambers which will be able to trigger on SMPs efficiently. They also
offer considerably greater precision in track reconstruction than
has been previously available at current colliders
experiments\cite{d0muon,Abazov:2005uk,cdfmuon,Ginsburg:2004fa}. The
excellent momentum resolution implies that charge misidentification
will be typically less than 1\%\cite{atlastdr1}, and that $R$-hadrons
undergoing the
charge exchange reactions described in Section~\ref{sec:hadscat} can
therefore be identified~\cite{milstead}. In addition to providing
the reconstruction of the track of an SMP, the muon systems will be
able to provide time of flight measurements with an accuracy of
around 1ns. The time of flight is likely to be one of the most
powerful discriminants between SMPs and
muons~\cite{cms,aafkejorgen}. Furthermore, in the event of a
discovery, it will allow the reconstruction of the mass of an SMP.
It has been shown that a mass resolution of around 20 GeV is
possible for an $R$-hadron of mass $500$ GeV at
ATLAS~\cite{johansen} using this method. An understanding of the SMP
mass resolution is important in both probing the mass hierarchy of
particles in any exotic scenario and also in establishing the
existence of different types of SMPs with similar masses. In
addition to the time of flight information, the muon systems employ
gaseous detectors which can provide $\frac{dE}{dx}$ measurements.
These may also be used as discriminants in an SMP search and in
measurements of SMP properties.

Since direct searches for most species of SMPs are based on charged
particle signatures, calorimeters are less crucial, though still
important, detector components. The ATLAS and CMS calorimeters can
be employed in a number of ways in SMP searches. Measurements of
electromagnetic and hadronic energy loss can distinguish between
hadronic and leptonic SMP species. Furthermore, the electromagnetic calorimeters can also provide supplementary $\frac{dE}{dx}$ information.
Calorimeter triggers based on jet multiplicity or global event properties, such as visible energy, can also be used to select events containing
SMPs~\cite{aafkejorgen}. These same observables could be used in
offline analyses for SMPs predicted to occur in specific event
topologies.

A complementary way to detect SMPs at the LHC is via passive
detectors. The MOEDAL collaboration\cite{moedal,Pinfold:1999sp} proposes
to deploy a plastic track-etch detector comprising three layers of
CR39/lexan plastic around the LHCb vertex detector.  The
major systematic
uncertainty in any search with MOEDAL will be background tracks from the radiation
environment of the LHC, and this is presently under study.


\subsection{Specific types of SMPs at the LHC}
\subsubsection{Expected rates}

Using {\sc Pythia} we have calculated the expected yields for
different types of SMPs at the LHC, following the accumulation of
10fb$^{-1}$. This corresponds to one year of low-luminosity
operation at the LHC.
Table~\ref{tabrates} shows the expected yield of pair-production
processes giving rise to different types of SMPs: gluinos, stops,
fourth-generation quarks with charge $\pm \frac{2}{3}e$, heavy
leptons and superpartners of left and right-handed leptons. Since predictions of
this type are necessarily scenario-dependent, we have made
conservative calculations assuming that the exotic particle is
produced via SM couplings and that the influence of other exotic
particles is minimal. The predictions do not include contributions
from the cascade decays of other exotic particles.


Studies indicate that the acceptance of the ATLAS and CMS detectors
following SMP selection cuts typically vary between 5\% and 80\%
depending on the mass and type of
SMP~\cite{nisati,polesello2,cms,Ambrosanio:2000ik,Ambrosanio:2000zu,tarem,aafkejorgen,milstead}. A
detailed simulation for each type of SMP of detector inefficiencies
and systematic uncertainties, such as the effects of nuclear
scattering and delayed reconstruction, is ongoing and therefore
beyond the scope of this paper. Furthermore, many systematic
uncertainties, particularly for searches for hadronic SMPs, remain
difficult to quantify even with detector simulations. This is
discussed below, where we describe experimental issues related to
specific types of SMPs which could be sought at the LHC.

\begin{table}[b]
\begin{center}
\begin{tabular}{cccccc}
\hline
Particle & $M=200$ GeV  & $M=500$ GeV & $M=1.0$ TeV & $M=1.5$ TeV & $M=2.0$ TeV\\
\hline $\tilde{g}$ &   $2.3 \times 10^7$  & $1.8 \times 10^5$ & $2.3 \times 10^3$ & 100  & 8  \\
$\tilde{t}$ &  $4.3\times 10^5$  & $3.7 \times 10^3$ & 47 & 2.1 & 0.1    \\
$t'$ &  $2.5 \times 10^6$ & $2.5 \times 10^4$ & 440 & 22 & 2 \\
$\tilde{l}_L$ & 190 & 4.9 & 0.14 & 0.009 & $7\times10^{-4}$ \\
$\tilde{l}_R$ & 82  & 2.1 & 0.06 & 0.004 & $3\times 10^{-4}$ \\
 $l'$ &  1220 & 34 & 1.1 & 0.08 & 0.008  \\
\hline
\end{tabular}
\end{center}
\vspace{0.06cm} \caption{{\sc Pythia} predictions of the total
number of SMP pairs expected at the LHC following the accumulation
of 10fb$^{-1}$ of integrated luminosity. }\label{tabrates}
\end{table}

\subsubsection{Generic SMPs}
Generic SMP's which possess electric charge and penetrate through the whole
detector will leave a clear signature of a high-$p_T$, slow,
muon-like track. Since no assumption, other than non-zero electric
charge, is made regarding the properties of a generic SMP it is not
possible to calculate its expected production cross section.
However, background is expected to be highly suppressed when making use of the time of flight technique~\cite{aafkejorgen}.
Thus, penetrating SMPs with appropriate production cross sections
would be easily observed. A null observation would also allow
stringent upper limits on these cross sections. For SMPs which stop
in the calorimeter and do not traverse the muon system, searches
could rely on observing an excess of high-$p_T$ tracks in the inner
detectors. In conjunction with this, observables such as a charged
particle leaving a signature of anomalous ionisation in the tracking
chambers together with a characteristic energy deposition profile in
the calorimeters could also be used. However, as described earlier
in this section, the precision with which ionisation energy loss can
be measured is likely to be poorer than obtained in earlier
searches. This could make it difficult in particular to identify
fractionally charged objects such as for example free quarks.

The planned MOEDAL experiment would be able to complement the above
search strategies by allowing searches for electrically charged SMPs
with $\frac{Z}{\beta}>5$, which implies a maximum electric
charge sensitivity of up to around $220e$~\cite{pinfoldpriv}. The present upper limit
on the SMP electric charge to which ATLAS and CMS will be sensitive is still unclear.
However, it is likely to be substantially below the reach of MOEDAL.

While it can be expected that the ATLAS and CMS tracking systems
will be able to reconstruct particles with charges down to at least
$\sim 0.5e$, studies have not been performed to establish this and
to estimate the reconstruction efficiency. Measurements of SMPs with
extremely low charge (millicharged-charged particles) would not be
directly recorded in the LHC tracking chambers.

\subsubsection{Heavy leptons and sleptons}
The possibility of stable heavy leptons and sleptons at the LHC hs been
considered in
Refs.~\cite{Allanach:2001sd,polesello2,cms,Ambrosanio:2000ik,Ambrosanio:2000zu,Allanach:2001sd,tarem,Ellis:2006vu}.
As these studies illustrated, the detection of these particles presents the fewest detector difficulties of all of the SMPs
considered, although their direct cross section would be suppressed in relation to coloured SMPs.
Heavy leptons and sleptons would manifest themselves as delayed, muon-like
particles leaving little energy deposition in the hadronic
calorimeters.
Since these particles would not necessarily be
associated with a jet, isolation criteria could also be used to
identify them. In an extremely minimal scenario in which only the
properties of the slepton or lepton would play a role in the
production cross section, modelling these process is relatively
straightforward. However, the presence of other exotic particles
complicates the picture. In SUSY scenarios accommodating stable sleptons, e.g. GMSB,
stable sleptons would be expected to be dominantly produced
via the cascade decays of copiously produced squarks and gluinos at the LHC. Studies of
stable sleptons have focused on their detection and the exploration of the parameter space of
GMSB and GMSB-like models~\cite{polesello2,cms,Ambrosanio:2000ik,Ambrosanio:2000zu,Allanach:2001sd,tarem,Ellis:2006vu}.
For conservative estimates of rates, Tab.~\ref{tabrates} presents the
expected yields of the stable superpartners of left and right handed leptons, which are
produced directly and not via cascade decays. The latter quantities depend on masses and couplings of many SUSY sparticles so are extremely model-dependent. Making use of the time of flight technique (see Section~\ref{sec:tof}), the discovery of leptons or sleptons possessing masses up to
several hundred GeV should be possible in early LHC running with the ATLAS detector~\cite{polesello2,cms,Ambrosanio:2000ik,Ambrosanio:2000zu}. In the optimistic situation of an SMP-discovery, the angular
distribution of the pair-produced SMPs could allow discrimination between a lepton or slepton hypothesis, as outlined
in Ref.~\cite{Allanach:2001sd}.



\subsubsection{Gluinos}
Gluinos are produced through strong interactions and would thus be
copiously produced at the LHC, as is shown in Tab.~\ref{tabrates}. A
number of studies exploring the possibility of gluinos at the LHC
have taken place~\cite{Hewett:2004nw,richardson,
aafkejorgen,milstead}, which were mainly inspired by the recent
split SUSY model. These studies indicate that stable gluinos
possessing masses up to 1 TeV could be discovered in early LHC
running. However, there are special issues associated with hadronic
SMPs which must be addressed in a stable gluino search. For example,
poorly understood nuclear interactions could potentially inhibit a
discovery. It is possible, though extremely unlikely, that charged
gluino $R$-hadrons could dominantly be stopped or convert into
neutral states in the calorimeter and thus escape detection in the
muon chambers. Furthermore, even if the $R$-hadrons remain charged
then events could be produced containing tracks with different signs
of charge in the inner and muon systems (so-called 'flippers'),
possibly challenging the track reconstruction software. Current
simulations suggest that the ATLAS reconstruction programs can deal
with this situation, and it may in fact be used as a means of
searching for $R$-hadrons~\cite{milstead}. However, in the absence
of collision data to confirm this, a conservative approach would be
to consider the most straightforward signature in which the
$R$-hadron maintains the same non-zero charge in the inner and muon
chambers. The string fragmentation model with default parameters
(Section~\ref{hadro}) and the geometric scattering model
(Section~\ref{sec:aafke}) predict that the proportion of gluino
$R$-hadrons produced at the LHC which possess the same non-zero
electric charge at the production vertex and after their passage
through a typical calorimeter is around 10\%. In this situation,
gluino $R$-hadrons could be found with masses up to at least 1 TeV
in early LHC running when using measurements of muon-like tracks and
topological variables, and significantly beyond when also making use
of the time of flight technique~\cite{aafkejorgen}. However, it
should be noted that calculations of nuclear scattering processes
for $R$-hadrons are extremely uncertain and any search must take
this into account.

Another pessimistic scenario can be envisaged if the fraction of
gluino-gluon states formed in the hadronisation process is around
unity, leading to no track in the inner detector. In this unlikely,
but not inconceivable scenario, neutral $R$-hadrons which convert
into charged states via nuclear interactions may be visible as
tracks in the muon systems. However, as for the charge 'flippers'
this signature may again challenge reconstruction software and the
event filtered away with background removal algorithms.

The most experimentally difficult scenario occurs in the case of
gluino $R$-hadrons being produced as neutral particles and remaining neutral
during their passage through the detector. Here, a signature of
$R$-hadron production could be manifest through an anomalously high
jet rate\cite{Hewett:2004nw}.

\subsubsection{Stable squarks and quarks}
Stable squarks and quarks at the LHC have been
postulated, as described in Section~\ref{sec:scen} although no
detector simulations of these have been published. Searches for
stable heavy colour-triplet states would benefit from the strong
production cross section and exploit the same techniques as for the
stable gluino searches described above. Similarly, they would also
be sensitive to uncertain nuclear interactions which could
potentially reduce their discovery potential. The nuclear
interactions of stable colour triplets have received far less
attention than gluinos and searches would benefit from more
phenomenology in this area. One model which treats both colour-octet
and triplet states~\cite{Kraan:2004tz} predicts comparable energy
loss and charge exchange rates for both types of particles. Should
this model reproduce the data, it can therefore be expected that a
similar sensitivity is obtainable as for stable gluinos, i.e. stable squark and quarks with masses up to at least 1 TeV could be discovered at early stages of the running of the LHC. Again, the angular distribution of the pair-produced SMPs could allow discrimination between a quark or squark hypothesis in case of a discovery.

\subsubsection{Monopoles}
Magnetic monopoles at the LHC will leave a number of striking signatures.
However, little work has been done estimating the sensitivity of
ATLAS and CMS to these particles. Those studies which do exist
concern indirect searches in which magnetic monopoles are produced as
internal loops~\cite{atlastdr2}. Furthermore, as discussed in
Section~\ref{monocal}, calculations of these processes suffer from
unquantifiable uncertainties\cite{DeRujula:1994nf,Ginzburg:1999ej}.
Even in the absence of detailed studies of direct magnetic monopole
production it is possible to make several remarks concerning their
detection. As described in Section~\ref{dedxMM}, a Dirac monopole
will typically lose several thousand times as much ionisation energy
as a MIP. A detailed $\frac{dE}{dx}$ calibration is therefore not
necessary to observe them. However, electronic saturation effects
due to the enormous ionisation energy must be carefully studied in
any search. Furthermore, customised track-finding algorithms may
have to be written to account for the parabolic track trajectory
followed by a magnetic monopole in a magnetic field. It is also important to
employ GEANT simulations to calculate the stopping of magnetic monopoles in
detector material and the energy deposition of profiles of magnetic monopoles
which progress to the calorimeters. A comprehensive search must
consider the full mass and magnetic charge range which is
experimentally available.

The MOEDAL experiment bypasses the problems described above of detecting
magnetic monopoles with active detectors by using the plastic track
technique. Current acceptance calculations indicate that MOEDAL will be sensitive to
monopoles with masses up to around 6 TeV and charges up to around
3$g_D$~\cite{pinfoldpriv}. MOEDAL may offer the most promising method of hunting
magnetic monopoles at the LHC. However, this depends on the magnetic monopole production
cross section since MOEDAL would be exposed to integrated luminosity around 100 times
lower than that collected by ATLAS and CMS.
\section{Summary}
Searches for Stable Massive Particles have been made at colliders
for several decades. So far, no particles beyond those accommodated
within the Standard Model have been observed. The searches are
motivated by a number of theories which address key issues in modern
physics. A review has been given of theoretical scenarios predicting
SMPs, the phenomenology needed to model their production at
colliders, the experimental techniques used to find SMPs and the
searches which have been made to date. The interplay between
collider searches and open cosmological questions has also been
addressed.

We look forward to the extension of these searches in the coming
years at the LHC and at cosmic ray facilities. The discovery of an
SMP would change our view of particle physics.
\section*{Acknowledgements}
We gratefully acknowledge the following people who have helped us in
the preparation of this paper by providing useful comments and
suggestions to various sections: Lars Bergstr\"om, Wilfried Buchm\"uller, Bogdan Dobrescu, Beate Heinemann, Barry King, Maxim Perelstein, Jim Pinfold, Christoph Rembser, Peter Richardson, Pietro Slavich, and Tim Tait.

David Milstead is Royal Swedish Academy Research Fellow supported by a 
grant from the Knut and Alice Wallenberg Foundation.











\end{document}